\documentclass[a4paper,11pt]{article}
\pdfoutput=1

\usepackage{jheppub}

\usepackage{amsmath}
\usepackage{booktabs}
\usepackage{bm}
\usepackage{slashed}

\usepackage{tikz}
\usepackage{physics}
\usepackage{tikz-feynman}
\usetikzlibrary{decorations.pathmorphing,decorations.markings,trees,shapes}
\usetikzlibrary{arrows.meta}
\usepackage[force]{feynmp-auto}
\tikzset{
    photon/.style={decorate, decoration={snake}},
    electron/.style={draw=black, postaction={decorate},decoration={markings,mark=at position .55 with {\arrow[draw=black]{>}}}},
        antielectron/.style={draw=black, postaction={decorate},decoration={markings,mark=at position .55 with {\arrow[draw=black]{<}}}},
    gluon/.style={decorate, draw=magenta,
        decoration={coil,amplitude=4pt, segment length=5pt}} 
}
\PassOptionsToPackage{colorlinks=true, allcolors=magenta}{hyperref}

\usepackage{graphicx}
\usepackage{slashed}
\usepackage{cancel}
\usepackage{xcolor} 

\usepackage{multirow} 
\usepackage{subcaption}
\usepackage{amssymb}
\usepackage{amsmath}

\usepackage{comment}
\usetikzlibrary{arrows.meta}

\newcommand{\gutw}{\theta^{\rm GUT}_W}
\newcommand{\w}{\theta_W}

\newcommand{\dm}{\nu'_{R}}

\newcommand{\TeV}{\text{TeV}}
\newcommand{\zp}{Z'}
\newcommand{\swp}{\theta'}

\newcommand{\gzp}{g_{Z'}}
\newcommand{\mzp}{m_{Z'}}
\newcommand{\mdm}{m_{\chi}}

\newcommand{\neu}{\mathcal{N}}

\usepackage{pifont}
\usepackage{array}

\hypersetup{
  colorlinks = true,
  allcolors  = magenta
}

\usepackage{tikz,xcolor,hyperref}
\definecolor{lime}{HTML}{A6CE39}
\DeclareRobustCommand{\orcidicon}{%
	\begin{tikzpicture}
	\draw[lime, fill=lime] (0,0) 
	circle [radius=0.16] 
	node[white] {{\fontfamily{qag}\selectfont \tiny ID}};	\draw[white, fill=white] (-0.0625,0.095) 
	circle [radius=0.007];	\end{tikzpicture}
	\hspace{-2mm}}
\foreach \x in {A,...,Z}{%
	\expandafter\xdef\csname orcid\x\endcsname{\noexpand\href{https://orcid.org/\csname orcidauthor\x\endcsname}{\noexpand\orcidicon}}
}

\title{Discovering Warped Extra Dimensions at \\ Next Generation Dark Matter Detectors}

\author[1]{Kaustubh Agashe\orcidA{}\,,}
\author[2,1]{Abhishek Banerjee\orcidB{}\,,}
\author[1]{Kushan Panchal\orcidC{}\,,}
\author[3,4]{and Nicholas L. Rodd\orcidD{}\,}

\affiliation[1]{\footnotesize Maryland Center for Fundamental Physics, University of Maryland, College Park, MD 20742, USA}
\affiliation[2]{\footnotesize Deutsches Elektronen-Synchrotron DESY, Notkestr. 85, 22607 Hamburg, Germany}
\affiliation[3]{\footnotesize Theory Group, Lawrence Berkeley National Laboratory, Berkeley, CA 94720, USA}
\affiliation[4]{\footnotesize The Leinweber Institute for Theoretical Physics, University of California, Berkeley, CA 94720, USA}

\emailAdd{kagashe@umd.edu}
\emailAdd{abhishek.banerjee@desy.de}
\emailAdd{kushu23@umd.edu}
\emailAdd{nrodd@lbl.gov}

\abstract{The coming years will see what could be the final push in the search for WIMP dark matter (DM), with direct detection reaching into the neutrino fog and indirect detection reaching TeV-scale thermal relics.
We study a class of models that these experiments could discover: DM emerging from a warped extra dimension.
In the IR, these scenarios appear generically as a $Z'$-type portal. 
We show that DM with electroweak couplings and TeV scale masses remains viable and discoverable within such a bottom-up context.
However, the models require a 20\% tuning between the DM and $Z'$ masses, a tuning that is relaxed if the DM and $Z'$ are strongly coupled and heavier ($\sim$$10$\,TeV).
In fact, it is the {\em strongly}-coupled $Z'$-type portal which is UV-completed by the framework of a warped extra-dimensional Grand Unified Theory (GUT), where DM stability is a spin-off of suppressing proton decay, resulting in a $\sim$$10$\,TeV Kaluza-Klein (KK) scale.
We identify a variant of this framework where all gauge fields except a baryon-number gauge boson have suppressed couplings to the DM; this model significantly suppresses the direct detection rate, but it remains discoverable through indirect detection.
Further, to address the little hierarchy problem associated with $\sim$$10$\,TeV KK scale, inspired by the Super-Composite Higgs, we augment our extra-dimensional GUT setup with supersymmetry.
We show that this model also possesses a DM candidate, now with a weakly-coupled and $\sim$${\rm TeV}$ mass $Z'$ mediator.
In all cases, a near-term discovery remains possible, for instance with the now online Cherenkov Telescope Array Observatory (CTAO), suggesting that our first hint of extra dimensions could emerge from the high-energy sky.
}

\preprint{\begin{tabular}[t]{@{}r@{}}
  DESY-26-164\\
  UMD-PP-026-10
\end{tabular}}

\begin{document}

\maketitle

\section{Introduction}

The idea of dark matter (DM) being composed of Weakly Interacting Massive Particles (WIMPs) remains a highly attractive paradigm.
The WIMP miracle -- that a particle with a weak scale annihilation cross section naturally gives the correct DM thermal relic density -- is compelling. 
The WIMP further has deep UV motivations.
It can arise naturally in scenarios that aim to resolve the electroweak hierarchy, for instance supersymmetry (SUSY) as well as warped extra dimensions. 
Arguably, the most important aspect of the WIMP is that it is eminently testable.
Indeed, many sharp realisations of the framework are now excluded through the classic DM trident: colliders, direct, and indirect detection. For the example of SUSY, a pure thermal wino has been essentially excluded with indirect detection~\cite{Fan:2013faa,Cohen:2013ama,Rinchiuso:2018ajn,Rodd:2024qsi,Safdi:2025sfs}.
Other scenarios, like a Higgsino, have been identified as a key benchmark that is eminently detectable at the next generation of instruments, including future colliders~\cite{Low:2014cba,Capdevilla:2021fmj,Chigusa:2025azs,Bhattiprolu:2025zwt} or direct detection~\cite{Co:2021ion,Martin:2024ytt,Bloch:2024suj}, and possibly even within the next few years at the Cherenkov Telescope Array Observatory (CTAO)~\cite{Rinchiuso:2020skh,Rodd:2024qsi,Abe:2025lci}.

This raises an important broader question: as we enter the next era of WIMP searches, what other scenarios sit poised for detection? To answer this question, we first study a generic class of infrared (IR) theories of WIMP DM models. This bottom-up analysis helps us set benchmarks for various constraints and signals for WIMP DM. Perhaps the most natural IR description is one consisting of a secluded dark sector that interacts with the Standard Model (SM) through a weakly coupled vector mediator, a scenario generically referred to as $Z'$-portal DM.
Such models have been studied broadly, see e.g. Refs.~\cite{Langacker:2008yv,Alves:2015pea,Alves:2015mua,Blanco:2019hah}, although we confront these scenarios with present constraints.
Within these scenarios, we find we are often pushed to consider strong coupling between the DM and the mediator due to the following argument. For models with electroweak sized couplings, achieving the correct relic abundance whilst evading direct detection and collider constraints requires a mass tuning $\mzp \simeq 2 m_\chi$ of ${\cal O}(10\%)$, with $\mathcal{O}(10)$\,TeV $Z'$ masses. This tuning can be alleviated if we couple the $Z'$ more strongly to DM (with couplings to nucleons being either electroweak-sized or strong). However, while the tuning gets relaxed, the $Z'$ mass gets pushed above $\mathcal{O}(20)$\,TeV.
Another aspect of these models not always considered, is the possibility of a $Z'$ coupling to the SM Higgs.
We show that including this coupling can essentially remove all sensitivity of direct detection, but an almost equivalent bound arises from electroweak precision.
On the front of future signals, we demonstrate that the relevant parameter space for these models is testable at CTAO, but also partly within reach of other experimental searches, such as next generation direct detection. 

The $Z'$-portal models described above suffer from arbitrary choices of couplings and masses, compelling us to look for a UV model. The case for UV completions is even stronger for the strongly coupled $Z'$-portal, as a large $Z'$ coupling at the TeV scale leads to a Landau pole much below the Planck scale. A hint for where to look for UV completions is provided by the proximity of the DM and $Z'$ masses to the weak scale. This suggests that the UV completion could arise in theories addressing issues like the Planck-weak hierarchy.
One such class of theories, which also addresses the flavour puzzle, is warped extra dimensions or Randall-Sundrum (RS1) models \cite{Randall:1999ee}. In these models, the DM and mediator can emerge as Kaluza-Klein (KK) excitations. As the KK scales are $\mathcal{O}$(TeV), and the KK modes are strongly coupled, this framework can readily incorporate strongly coupled $Z'$ models.

Serendipitously, the model outlined in Refs.~\cite{Agashe:2004ci,Agashe:2004bm} offers a suitable framework for studying strongly coupled DM.  In this setup, an $SO(10)$ Grand Unified Theory (GUT), when augmented with warped extra dimensions, predicts DM in the form of a  TeV scale Dirac fermion. In this model, like every SM particle, the right handed top quark ($t_R^c$), has several KK fermions as GUT partners. One of these, $\nu'^c_R$, has $SO(10)$ quantum numbers of a right-handed neutrino. The model has gauged baryon-number (or $U(1)_B$) to ensure proton stability, which imparts the same $U(1)_B$ charge to $t_R^c$ and $\nu'^c_R$. This implies $\nu'^c_R$ is an SM colour singlet with $U(1)_B$ charge. This makes it an exotic particle, forbidding its decay to purely SM states. Being the lightest exotic, $\nu'^c_R$ is stable and thus, can be identified as DM. Therefore, DM is obtained as a spin-off of suppressing proton decay (akin to $R$-parity in SUSY). While it is an SM singlet, the DM couples strongly to non-SM GUT gauge fields.

The $SO(10)$ gauge group contains two $U(1)$ subgroups, with the DM charged under one of them. This $U(1)$ furnishes a $Z'$ gauge boson coupled to DM. Due to its $U(1)_B$ charge, the DM also couples to the baryon number gauge boson. This setup of DM and neutral gauge bosons from extra $U(1)$s is reminiscent of the $Z'$-portal. However, there are several distinguishing features. For instance, the DM here also couples to the charged and coloured $X_s$ gauge boson. Furthermore, the $X_s,Z'$ and the baryon number gauge boson have flavour non-universal couplings with the SM fermions, due to extra dimensional geography. In particular, couplings to $t^c_R$ are much larger than the couplings to lighter fermions. Due to these features, this model is more appropriately termed as a generalised $Z'$-portal model. Apart from $t^c_R$, the $Z'$ above also couples strongly to the SM Higgs field. Unlike the IR $Z'$-portal where it was a choice, the coupling of $Z'$ to the SM Higgs is now unavoidable. We find that this coupling is one of the primary contributors to the direct detection amplitude.

We first study and update the bounds on the original $SO(10)$ model from Ref.~\cite{Agashe:2004bm} (cf. Ref.~\cite{Belanger:2007dx}) where $SO(10)$ is unbroken in the 5D bulk and on the IR brane. The DM along with $X_s,Z'$ and the baryon number gauge boson all have masses given by the KK scale. The DM primarily annihilates to a pair of top quarks, along with pairs of longitudinal $W^\pm/Z$s and Higgs bosons. To our knowledge, this is the first analysis of this model that includes the contribution of the baryon-number gauge boson to DM annihilation. Due to the GUT gauge fields being strongly coupled with DM, obtaining the correct relic abundance, while being safe from direct detection, only requires a mass tuning (between the DM and mediator masses) of roughly $40\%$. However, direct detection still pushes the KK scale to $\sim$$20$\,TeV. These results are in line with our analysis of the strongly coupled $Z'$ portal.

The high KK scale opens the possibility to construct variations of the original model by breaking $SO(10)$, specifically the $SU(2)_R$ subgroup, on the IR brane. This incurs bounds from custodial isospin violation, or the $\rho$ parameter. In \cite{Agashe:2004ci,Agashe:2004bm}, the bound on the KK scale from direct detection was \(\mathcal O(\mathrm{TeV})\) or lower. Thus, maintaining a natural \(\mathcal O(\mathrm{TeV})\) KK scale required custodial isospin to be consistent with the stronger bounds from the \(\rho\) parameter. However, this argument is now rendered moot; the KK scales are already pushed to $\mathcal{O}(20)$\,TeV by direct detection, and hence, are expected to be consistent with bounds from the $\rho$ parameter even without custodial isospin. 
Thus, with the freedom to break $SU(2)_R$, we explore various breaking patterns for $SO(10)$.
This is achieved by changing the IR brane boundary conditions (BCs) of the GUT gauge fields. For these variations, we find that direct detection bounds are comparable to, but slightly weaker than the custodial isospin bounds. Similar to the original model, we maintain a low mass tuning of $\mathcal{O}(30\%)$ while being pushed to $\mathcal{O}(20)$\,TeV KK scales.
Particularly interesting among the models here, is the variation where $SO(10)$ is broken to the SM gauge group on the IR brane. For this case, all the GUT gauge bosons have a suppressed coupling with DM. However, the baryon number gauge boson retains its original coupling, thus providing sufficient DM annihilation. As it turns out, with GUT gauge and Yukawa couplings suppressed in this model, we also suppress the direct detection cross section. However, the CTAO reach still covers the viable parameter space, hence cementing its place as the leading signal. Overall, we show that CTAO proves to be a promising avenue for discovery for all the models above, as it can probe KK scales up to $\mathcal{O}(100)$\,TeV.

While these warped GUT models offer rich phenomenology, they suffer from a little hierarchy between the electroweak and KK scales, as the KK scales do not go any lower than $10\,$TeV. 
It was recently shown in Ref.~\cite{Agashe:2026ggy} that further augmenting the model with SUSY can help ameliorate this issue. 
Inspired by this framework, we study a warped SUSY model, with gaugino-mediated SUSY breaking. The model has separate IR and Higgs branes. The gauge and gaugino fields propagate till the final IR brane. The SM Higgs, together with the SM and GUT fermions, including the DM, only propagate till the Higgs brane. As DM is sequestered from GUT gauge fields, its gauge couplings are highly suppressed if the gauge fields retain any of the BCs from the models discussed above. This can lead to an overabundance of DM. Remarkably, we are rescued by yet another BC change, this time on the UV brane, for the $Z'$ gauge boson. With this new BC, the $Z'$ now couples to the SM as well as DM with electroweak strength. Thus, the setup closely resembles a weakly coupled \(Z'\)-portal DM model, with similar collider and direct-detection bounds. We find the phenomenologically allowed DM and $Z'$ masses of this model to be greater than about $5$\,TeV and $10$\,TeV respectively, with an $\mathcal{O}(10 \%)$ mass tuning. Even though this translates to a KK scale of roughly 80 TeV, no hierarchy problem arises for the Higgs mass. This is because corrections to the Higgs mass are sensitive only to SUSY breaking masses, such as the stop mass, which remain $\mathcal{O}$(TeV). 
CTAO remains a promising discovery channel for this model.

We organise our discussion of $Z'$-portal WIMP DM and its attractive UV completions as follows.
Starting in Sec.~\ref{sec:IRZprime} we consider the IR story, studying the present status of $Z'$-portal DM and what parameter space can be covered at the next generation DM detectors.
The section also reviews the various experimental probes of WIMP DM relevant for the present work. From here we turn to the UV, and in Sec.~\ref{sec:ED_GUT} study a range of DM scenarios where the UV is a warped extra dimension with a GUT.
As noted, generically these scenarios are pushed by present constraints to have a new physics scale at the $\sim$$10$\,TeV range, introducing a tension in their ability to explain the electroweak hierarchy problem.
To address this, in Sec.~\ref{sec:warped_susy} we explore how SUSY can alleviate this tension.
Lastly, we outline our conclusions in Sec.~\ref{sec:conclusion}, followed by a series of appendices that include technical details of calculations we eschew from the main text.

\section{Lessons from the IR: $Z'$-Portal DM}
\label{sec:IRZprime}

We begin by exploring the IR landscape motivated by extra-dimensional DM models: the $Z'$ portal.
This provides us an opportunity to introduce the experimental searches that are most promising and constraining for these models, and in doing so update the constraints on widely discussed realisations of the $Z'$ scenario.
Although in the IR (and UV) there are many model variants that can be explored, we show that quite generically the well motivated parameter space in these scenarios is within reach of next generation DM searches.

As emphasised already, the UV models we consider at low energies take the form of DM coupled to the standard model through a $Z'$ portal.
This is a framework rather than a model; for that we need to be more specific.
The UV is rich and there are many realisations of the $Z'$ portal one can take; indeed, the literature contains a variety of versions of this scenario, see e.g. Refs.~\cite{Dudas:2009uq,Alves:2015pea,Alves:2015mua,Kahlhoefer:2015bea,Blanco:2019hah,Arcadi:2024ukq} and the review in Ref.~\cite{Langacker:2008yv}.
Motivated by the fact that in our extra-dimensional models the DM takes the form of a heavy partner of the right-handed neutrino, we take DM to be a right-handed Dirac fermion, $\chi_R$.
(For related constructions where a right-handed neutrino is the DM in gauged $U(1)$ extensions, see e.g. Ref.~\cite{Okada:2016gsh}.)
Further, we assume that the pattern of $Z'$-standard model fermion couplings is identical to those of the $Z$ boson, the so-called sequential scenario~\cite{Barger:1980dx,Altarelli:1989ff}.
Sequential couplings have been adopted for $Z'$-portal DM in Refs.~\cite{Alves:2015pea,Arcadi:2017hfi}, in the former as a limit of a more general parameterisation; other realisations instead gauge $B-L$ or more general $U(1)$ charges~\cite{Alves:2015mua,Blanco:2019hah,Okada:2016gsh}.

With the above consideration in mind, we explore the following model,\footnote{As written, the model is anomalous, containing no state to cancel the anomaly associated with $\chi_R$.
The anomaly is removed in the UV completions we consider, yet as the compensating state need not be relevant for the IR physics we do not consider it here.}
\begin{equation} \label{eq:L-Zprime}
{\cal L} \supset g_{\chi} Z'_{\mu} \Big[ \bar{\chi}_R \gamma^{\mu} \chi_R + \bar{f} \big(c_L P_L + c_R P_R \big) \gamma^{\mu} f \Big].
\end{equation}
Here $f$ represents all fermions in the standard model, which have distinct left and right handed couplings -- $c_L = (I^3_W - Q s_W^2)/c_W$ and $c_R = - Q s_W^2/c_W$ -- with $P$ the corresponding projectors.
(If we fix the $Z'$ coupling to the weak coupling, $g_{\chi} = g_W \simeq 0.65$, the fermion interactions become identical to that of the standard model $Z$-boson.)
This model is inspired by the $Z'$-scenario considered in Ref.~\cite{Alves:2015pea}, which we return to in Sec.~\ref{sec:alves+}.

\begin{figure}
\centering
\includegraphics[width=0.48\linewidth]{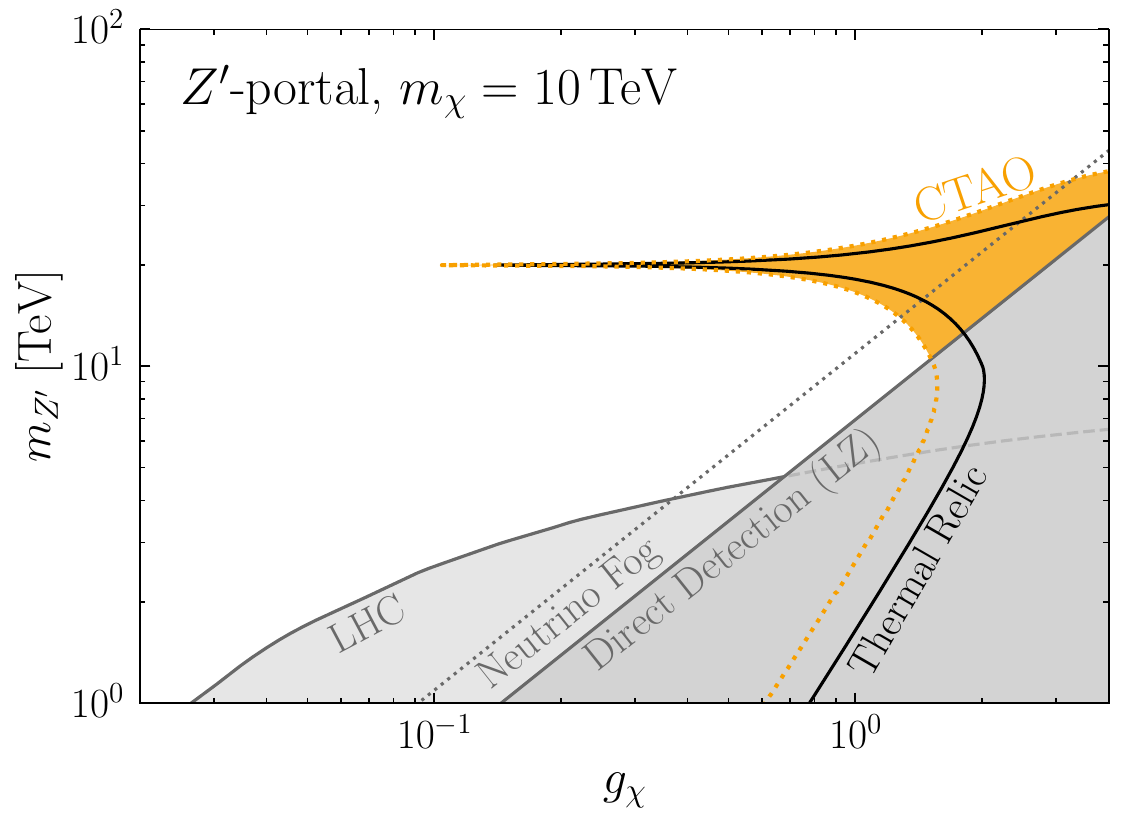}
\hspace{0.2cm}
\includegraphics[width=0.48\linewidth]{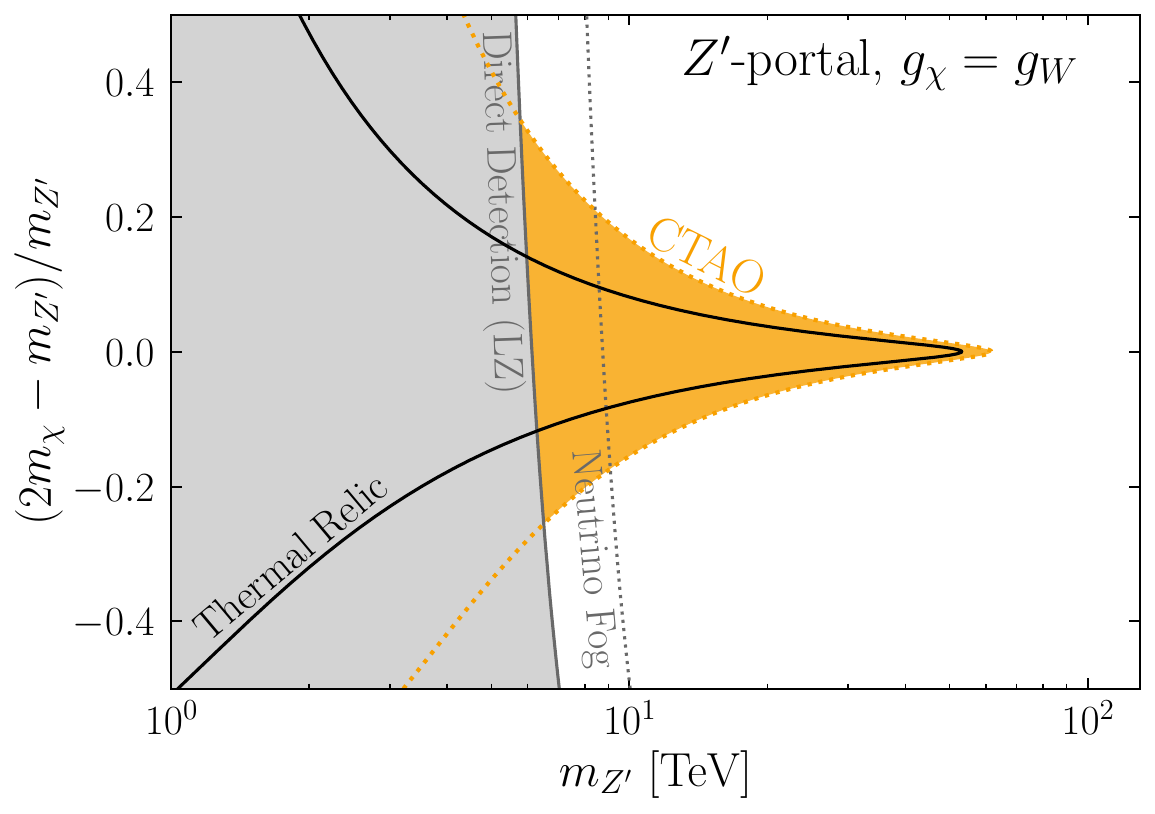}
\caption{The parameter space of the $Z'$ model in Eq.~\eqref{eq:L-Zprime}.
The model has three parameters, $(g_\chi,\mzp,m_\chi)$, and we show slices through the space at fixed mass (left) and coupling (right).
Generically, evading direct detection whilst maintaining the correct relic abundance requires $2m_{\chi} \sim \mzp$, although that parameter space appears entirely within reach of CTAO.
(The LHC constraints are entirely subdominant to direct detection for the results at fixed coupling.)
See text for details.}
\label{fig:Zprime1}
\end{figure}

To fully define the model, we must specify values for the three free parameters: the mass of the DM and $Z'$, and the coupling between the sectors $g_\chi$.
We show two slices through this parameter space in Fig.~\ref{fig:Zprime1}, showing already the existing (solid) and future (dotted) constraints on the model.
The specific details of the constraints are described in the remainder of this section.
We can already see that as this DM behaves in many ways as a canonical WIMP, the time-honoured trident of colliders, direct, and indirect detection are relevant.
Being a WIMP, the most compelling part of parameter space occurs where the model predicts an annihilation of DM to standard model final states with cross section $\langle \sigma v \rangle \simeq 4.4 \times 10^{-26}\,\mathrm{cm}^3/\mathrm{s}$~\cite{Steigman:2012nb}:\footnote{This is the annihilation rate required for a Dirac fermion, the Majorana value that appears more commonly is a factor of two smaller.
For Dirac fermions, the indirect detection flux given in Eq.~\eqref{eq:ID_flux} is also a factor of two smaller than its Majorana counterpart, so that the two factors cancel in the total rate.
For details, see for example Ref.~\cite{Cirelli:2024ssz}.} with this annihilation rate, the DM realises the observed relic abundance through the WIMP miracle.
We see immediately that the parameter space where DM obtains the correct relic abundance -- labelled \textit{Thermal Relic} -- and evades current constraints (shaded grey) has several common features.
Both the $Z'$ and DM must be heavy, with masses $\mathcal{O}(10\,\mathrm{TeV})$, and further be relatively strongly coupled with $g_\chi \sim 0.1$--$1$.
Both features are important to understanding the UV completions studied in Secs.~\ref{sec:ED_GUT} and \ref{sec:warped_susy}.

Figure~\ref{fig:Zprime1} reveals a related feature of the model that proves a common theme of this work.
In particular, we observe that the thermal relic parameter space is already in tension with the incredibly strong bounds from direct detection unless we tune $\mzp \sim 2 m_{\chi}$.
To be quantitative, we define the tuning as
\begin{equation}
\Delta = \frac{2m_{\chi}-\mzp}{\mzp},
\end{equation}
which is plotted on the right of Fig.~\ref{fig:Zprime1}, revealing the scenario requires $\mathcal{O}(10\%)$ tuning.
Intuitively, the tuning arises because of an inherent tension between requiring a relatively large annihilation cross section to set the relic abundance and a sufficiently small DM-nucleon cross section to evade direct detection.
However, as the annihilation can proceed through an $s$-channel $Z'$ exchange -- direct detection occurs via the $t$-channel -- taking $\Delta \to 0$ resonantly enhances this channel~\cite{Griest:1990kh}.
Accordingly, $|\Delta| \lesssim 20\%$ provides a refuge for this model from direct detection.
There is no hiding, however, from indirect detection.
This process searches for the late universe signature of the same annihilation processes setting the relic abundance---boosting the latter enhances the former.
CTAO is poised to realise that potential, and as shown can probe essentially the entire open parameter space for the $Z'$-portal.

In the following subsections we discuss the constraints on these DM scenarios, focusing on direct detection, indirect detection, and colliders.
We then comment on the impact of including Higgs couplings, before ending with a brief discussion of an alternate realisation of the $Z'$-portal framework, demonstrating that although the details vary the qualitative conclusions do not.
Indeed, many of these lessons pass over to the UV.

\subsection{Direct Detection}

The DM in Eq.~\eqref{eq:L-Zprime} couples to \textit{all} standard model fermions and therefore can scatter off atomic nuclei via the exchange of the heavy $Z'$ mediator.
The resulting nuclear recoils can be measured with high precision for inert elements with a large number of nucleons, such as xenon. 
Assuming elastic scattering, the recoil energy of the nucleus is roughly $E_R \simeq \mu v^2$, with $\mu$ the reduced DM-xenon mass, and $v \sim 10^{-3}$ the characteristic DM speed.
For TeV WIMPs, $\mu \simeq m_\mathrm{Xe}$ implying $E_R \sim 100\,$keV, which is a range of energies exquisitely probed by LZ~\cite{LZ:2024zvo}, XENONnT~\cite{XENON:2025vwd}, and PandaX~\cite{PandaX:2024qfu}.

The $Z'$ coupling allows the DM to undergo both spin-independent (SI) and spin-dependent (SD) nuclear scattering; we focus on SI scattering owing to the stronger constraints arising from the coherent mass-number enhancement.
The per-nucleon SI cross section for DM scattering with a nucleus of mass number $A$ and atomic number $Z$ is then given by (see e.g. Ref.~\cite{Alves:2015pea})
\begin{equation} \label{eq:dd_master_eqn}
\sigma^{\rm SI} \simeq \frac{g^2_{\chi}\mu_N^2}{4\pi \mzp^4}\left[\frac{Zg_p +(A-Z)g_n}{A}\right]^2\!,
\end{equation}
where $\mu_N$ is the reduced DM-nucleon mass and $g_{p,n}$ the coupling of the $Z'$ to the proton and neutron, which from Eq.~\eqref{eq:L-Zprime} is given by
\begin{equation}
g_p = 2 g_u + g_d = \frac{g_{\chi}}{c_W} \left( \frac{1}{4} - s_W^2 \right)\!,\qquad
g_n = 2g_d + g_u = - \frac{g_{\chi}}{4c_W}.
\end{equation}

The direct detection event rate is proportional to the DM number density multiplied by $\sigma^{\rm SI}$, or for a fixed local DM mass density, $\sigma^{\rm SI}/m_{\chi}$.
The current strongest constraints at these masses come from the LZ collaboration~\cite{LZ:2024zvo}, which requires $\sigma^{\rm SI}/m_\chi \le 3 \times 10^{-47}\, {\rm cm}^2/{\rm TeV}$ for $m_\chi \gtrsim 100$\,GeV.
Implementing their full constraints leads to the bounds shown in Fig.~\ref{fig:Zprime1}.
For fixed $m_{\chi}$, the bound translates to $\mzp/g_{\chi} \gtrsim 7$\,TeV.
For fixed $g_{\chi}= g_W$, direct detection along with the requirement of obtaining the observed thermal-relic density, translates to a tuning of $|\Delta|\lesssim 20\%$.
Broadly, direct detection requires the DM to be heavy, $m_{\chi} \gtrsim 10\,$TeV, and mildly tuned $m_\chi \sim \mzp/2$.
If future direct detection instruments are able to push all the way to the neutrino fog, as defined in Ref.~\cite{OHare:2021utq}, then the constraints would marginally improve, although the general lesson that these scenarios survive in the funnel region would remain.
(Note for $g_{\chi} \sim 1$ direct-detection bounds subsume those from the LHC discussed below, which is why those do not appear in the right panel of Fig.~\ref{fig:Zprime1}.)

\subsection{Indirect Detection}

We next look to the sky and in particular, indirect detection.
As we focus exclusively on DM in the TeV range, from Eq.~\eqref{eq:L-Zprime} DM can annihilate to all standard model states through $\chi \bar{\chi} \to f \bar{f}$.
A further channel, $\chi \bar{\chi} \to Z' Z'$ opens for $m_{\chi} \geq \mzp$.
The tree level cross section for both is a simple modification to those given in Ref.~\cite{Alves:2015pea}, and we do not reproduce it here.
Adjusting parameters such that
\begin{equation}
\langle \sigma v \rangle_{\chi \bar{\chi} \to Z' Z'} + \sum_{f \in {\rm SM}} \langle \sigma v \rangle_{\chi \bar{\chi} \to f \bar{f}} = 4.4 \times 10^{-26}\,\mathrm{cm}^3/\mathrm{s},
\end{equation}
defines the thermal-relic curve shown in Fig.~\ref{fig:Zprime1}.

The promise of indirect detection is that we can probe the same physics that set the relic abundance in the early universe using astrophysical observations of our own galaxy today.
The rate at which the annihilation of Dirac fermions produces photons that reach our detector is given by the canonical expression for indirect detection,
\begin{equation} \label{eq:ID_flux}
\frac{d\Phi}{dE} = \frac{\langle \sigma v \rangle}{16 \pi m_{\chi}^2} \frac{dN}{dE}\,J.
\end{equation}
Here $\Phi$ is the flux measured in $[\mathrm{photons}/\mathrm{cm}^2/\mathrm{s}]$.
We break down the different components in this expression next.

The combination of the cross section and spectrum in Eq.~\eqref{eq:ID_flux} is schematic.
Formally, the differential flux depends on the differential cross section, although it is convention to decompose this as $d\langle \sigma v \rangle/dE = \langle \sigma v \rangle \times dN/dE$.
Separated out, $\langle \sigma v \rangle$ is simply the Born cross section, which then defines $dN/dE$ as the spectrum per annihilation.
For SM fermion final states, $f\bar{f}$, we obtain the spectrum directly from \texttt{HDMSpectra}~\cite{Bauer:2020jay} (the results are comparable to those obtained from \texttt{PPPC4DMID}~\cite{Cirelli:2010xx}).

The spectrum for the $Z'Z'$ final state needs to be treated with additional care as the $Z'$ states themselves can decay to SM final states.
The spectrum of $Z' \to f \bar{f}$ in the $Z'$ rest frame can again be computed using \texttt{HDMSpectra}.
To then boost this to the DM rest frame, we use the formalism of cascade annihilations outlined in Refs.~\cite{Mardon:2009rc,Elor:2015tva}.\footnote{As discussed in Ref.~\cite{Baldes:2020hwx}, the polarisation of the $Z'$ also impacts the spectrum and as such a full calculation requires weighting the different polarisations of the new vector in the process.
This only has a minor impact on the spectra, and an even smaller one on the indirect detection sensitivity, and therefore we have not included it here.}
We start from the photon spectrum in the $Z'$ frame, $dN/dx_0$, with $x_0 = 2E_0/\mzp$ and $E_0$ the photon energy in the $Z'$ rest frame.
Moving frames, we take $x_1 = E_1/m_\chi$, with $E_1$ the photon energy in the rest frame of the DM and approximately the experiment.
If we define $\epsilon_1 = \mzp/m_{\chi}$, the kinematics of the process restrict the observed photon energy to the range,
\begin{equation}
0 \leq x_1 \leq \frac{1}{2} \left( 1 + \sqrt{1-\epsilon_1^2} \right)\!.
\end{equation}
The boosted spectrum in the DM frame can then be computed as~\cite{Mardon:2009rc,Elor:2015tva},
\begin{equation}
\frac{dN}{dx_1} = 2 \int_{x_{\rm min}}^{x_{\rm max}} \frac{dx_0}{x_0 \sqrt{1-\epsilon_1^2}} \frac{dN}{dx_0},
\end{equation}
with
\begin{equation}
x_{\rm min} = \frac{2x_1}{\epsilon_1^2} \left( 1 - \sqrt{1-\epsilon_1^2} \right)\!, \qquad
x_{\rm max} = \mathrm{min} \!\left[ 1,\,\frac{2x_1}{\epsilon_1^2} \left( 1 + \sqrt{1-\epsilon_1^2} \right) \right]\!.
\end{equation}

The next ingredient in Eq.~\eqref{eq:ID_flux} is the $J$-factor~\cite{Bergstrom:1997fh},
\begin{equation}
J = \int d\Omega\,ds\,\rho^2_{\chi}(s,\Omega).
\end{equation}
This is an integral over the DM-density squared -- for annihilation -- along the line of sight, $s$, and over the region of the sky viewed, $\Omega$.
The most sensitive region for DM annihilation is expected to be the centre of our own galaxy, and therefore to evaluate this integral we need a model for $\rho_{\chi}$ in the Milky Way.
An oft-used ansatz for this is the Einasto profile
\begin{equation}
\rho(r) \propto \exp \!\left\{ - \frac{2}{\alpha_s} \left( \left( \frac{r}{r_s} \right)^{\alpha_s} - 1 \right) \right\}\!,
\end{equation}
with $r$ the distance from the Galactic centre (we take our own distance from the centre as $8.5\,$kpc) and the density normalised to the local value that we take as $\rho_{\odot} = 0.39$\,GeV/cm$^3$.
We adopt this profile by default, taking $r_s = 20\,$kpc and $\alpha_s = 0.17$.
Of course, the actual distribution of DM in the galaxy could deviate significantly from the Einasto profile.
This is the central systematic of indirect detection and we comment on it shortly.

Having assembled the ingredients in Eq.~\eqref{eq:ID_flux}, we can ask what scenarios existing limits on the flux of high-energy photons could probe.
More exciting, however, is to consider the revolutionary sensitivity improvements that are about to be opened by CTAO~\cite{CTAConsortium:2010umy,CTAConsortium:2017dvg}.
For this we study the expected reach of the southern CTAO array that should begin operations at the start of the next decade -- specifically the Alpha configuration -- as with the Galactic centre rising overhead, this is expected to be the most sensitive array for DM.
(There is also a northern CTAO array, which although it is expected to be less sensitive to DM has the advantage of coming online \textit{now}; it could drive a near term discovery of the scenarios we consider~\cite{Abe:2025lci}, but here we choose to focus on the ultimate reach of CTAO.)
Our CTAO projections closely follow those in Ref.~\cite{Baumgart:2025dov} and we refer there for further details.
We assume 500\,hrs of Galactic centre observations, which would correspond to approximately 5\,years of operations; see, e.g., Ref.~\cite{Montanari:2022buj}.
We assume this time is divided between observations around the Galactic centre to provide an effectively uniform exposure within our region of interest, taken to be angles within $2^\circ$ of the centre, although we also mask the region within $0.3^\circ$ of the Galactic plane.
We model the expected primary background of CTAO which is misidentified cosmic rays that arrive uniformly over the region.
We do not include a model for the potential Galactic diffuse emission, even though that was studied in Ref.~\cite{Baumgart:2025dov}, as we find its presence has a negligible impact on our projections.
Finally, we note that although the observation strategy and performance of CTAO could be different to what we assume here, studies that have adopted different modelling choices for the instrument generally find comparable reach, see e.g. Refs.~\cite{Rodd:2024qsi,CTAO:2024wvb}.
As such, we do not expect our conclusions to qualitatively depend on the choices made here.

\begin{figure}
\centering
\includegraphics[width=0.48\linewidth]{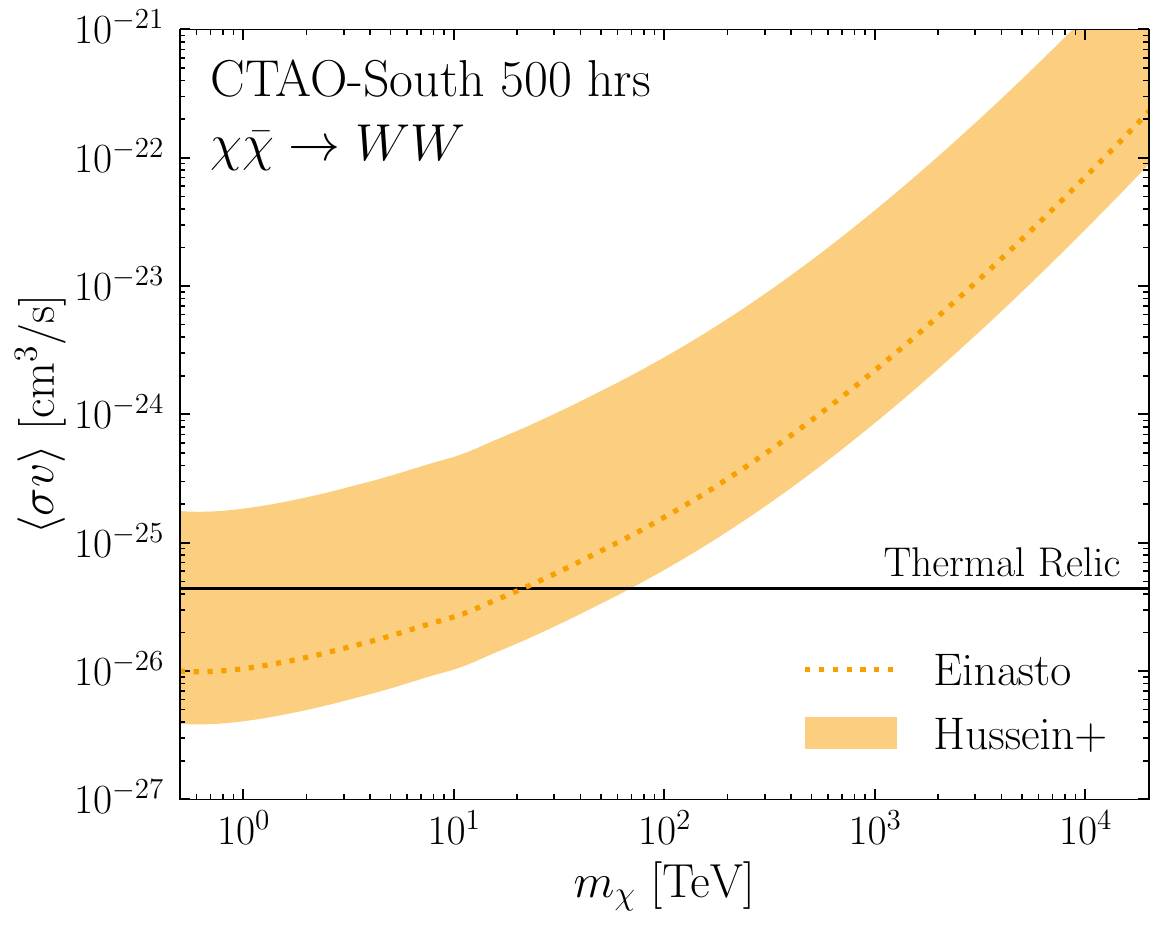}
\caption{The expected reach of CTAO for a generic Dirac fermionic WIMP annihilating to $WW$.
These results highlight that for an Einasto profile, CTAO could probe the thermal relic predictions well into the TeV range of DM masses.
The results also show the variation in sensitivity with the unknown $J$-factor of the Galactic centre; the band corresponds to the uncertainty range proposed in Ref.~\cite{Hussein:2025xwm}.}
\label{fig:CTAO_WW}
\end{figure}

Taken together, assuming a fiducial Einasto profile, the CTAO sensitivity to the $Z'$-portal DM is shown in Fig.~\ref{fig:Zprime1}.
For these choices, we can see CTAO is able to cover essentially the entire thermal relic parameter space of the models.
This is a reflection of the fact that CTAO is the first instrument poised to test a broad range of thermal WIMPs with TeV masses.
To highlight this, in Fig.~\ref{fig:CTAO_WW} we show the sensitivity of CTAO under our same modelling assumptions to a generic Dirac fermionic WIMP that annihilates to $W$-boson final states.
For the Einasto profile, CTAO can probe DM masses up to tens of TeV.
We also show there the potential uncertainty band associated with the unknown $J$-factor.
For this we draw on Ref.~\cite{Hussein:2025xwm}, which provided an uncertainty band on $\rho_{\chi}$ for the DM distribution of the Milky Way.
As we move across that band, the $J$-factor varies, and so too do the DM sensitivities.
Einasto is towards the upper end of the range, which spans roughly a factor of fifty, although we note it is possible for an even larger $J$-factor than Einasto.
The presence of this uncertainty applies for all CTAO projections shown in this work.

\subsection{Collider Searches}

In addition to the bounds pertaining to DM phenomenology, we also expect bounds on the $Z'$ gauge boson solely due to its coupling to SM particles.
As the $Z'$ couples to charged SM leptons, the model can be tested with LHC di-lepton searches~\cite{ParticleDataGroup:2024cfk}. 
The constrained quantity here is $\sigma(p p \to Z')\times {\rm BR}(Z' \to \ell \bar{\ell})$, which has the parametric dependence
\begin{equation}
\sigma(p p \to Z')\times {\rm BR}(Z' \to \ell \bar{\ell}) \sim \frac{g_\chi^2}{m_{Z'}^2} {\rm BR}(Z' \to \ell \bar{\ell}) .
\end{equation}
Recasting the bounds on a sequential $Z'$ from Ref.~\cite{CMS:2021ctt}, we obtain the grey curve in Fig.~\ref{fig:Zprime1}. For the plot with $m_\chi$ fixed, we focus on a $\TeV$ scale $Z'$ with a mass $\mzp < 2m_\chi$. As the $Z'$ is much heavier than any SM particle, the branching ratio to charged leptons is roughly constant.
Therefore, the constrained quantity scales only with the model parameters as $g_\chi^2/\mzp^2$.
On the other hand, for the plots where we fix $g_\chi$, we allow $\Delta<0$ or $m_{\zp}>2m_\chi$.
In this region of parameter space, the decay $\zp\to \chi \bar{\chi}$ opens up, resulting in a reduction of the branching ratio to leptons, and therefore a weakening of the collider bound.
However, as $\chi$ couples to $\zp$ with comparable strength to the SM, having one additional decay channel does not relax the bound substantially.   
As $g_\chi\to 1$, the direct detection bound subsumes this collider bound, which is why it is absent in the fixed coupling results.

\subsection{Higgs Couplings}

A guiding principle of the model in Eq.~\eqref{eq:L-Zprime} is the sequential structure: the $Z'$-fermion couplings to the SM mirror those of the $Z$.
This raises a natural question: what if the $Z'$ couples to the Higgs?
Of course, looking up from the IR all these questions are ad hoc; the $Z$ does not couple to DM, whilst the $Z'$ does.
Nevertheless, here we show that sequential Higgs couplings can substantially impact the phenomenology, and leave questions about what these couplings should be to our discussion of the UV.

If we assume the sequential structure extends to the Higgs sector, then the covariant derivative becomes
\begin{equation}
D_{\mu} H \supset \big(- i g_W Q_Z Z_{\mu} - i g_\chi Q_Z Z'_{\mu} \big) H,
\qquad Q_Z = \frac{1}{2c_W} \mathrm{diag}\big( 1-2s_W^2,\,-1 \big).
\end{equation}
Expanding out $|D_{\mu} H|^2$, we have $Z-Z'$ mixing.
This shifts the mass eigenstates away from the bare $m_Z$ and $\mzp$ as follows,
\begin{equation}
m_1^2 \simeq m_Z^2 \left( 1 - \frac{g_\chi^2 m_Z^2}{g_W^2 \mzp^2} \right)\!, \qquad
m_2^2 \simeq \mzp^2 \left( 1 + \frac{g_\chi^2 m_Z^2}{g_W^2 \mzp^2} \right)\!,
\end{equation}
which holds for $\mzp \gg m_Z$ and $\mzp \gg g_\chi v$.
The resulting fractional shift in the $Z$-boson mass is given by
\begin{equation}
\frac{\Delta m^2_Z}{m_Z^2} \simeq \frac{g_\chi^2}{\mzp^2} \frac{v^2}{4 c_W^2}.
\end{equation}
This mass shift signals custodial isospin violation, and thus faces bounds from the measurement of the $\rho$-parameter~\cite{ParticleDataGroup:2024cfk}, which numerically is approximately ${\mzp}/{g_\chi}\gtrsim 7\,\mathrm{TeV}$.

This $\rho$-parameter constraint is almost identical to the direct detection bound shown in Fig.~\ref{fig:Zprime1}.
Importantly, however, direct detection sensitivity is \textit{severely} suppressed in this scenario. After accounting for the exchange of the two mass eigenstates, the spin-independent cross section is
\begin{equation}
\sigma^{\rm SI} \simeq \frac{g^2_{\chi}\mu_N^2}{4\pi \mzp^4}\left[\frac{Zg_p +(A-Z)g_n}{A}\right]^2 \left( \frac{|\mathbf{q}|}{m_Z} \right)^4\!,
\end{equation}
where $\mathbf{q}$ is the momentum exchange (in terms of the Mandelstam variable, $|\mathbf{q}|^2 = -t$).
For TeV DM, $|\mathbf{q}| \simeq m_\mathrm{Xe} v \simeq 100\,\mathrm{MeV} \ll m_Z$.
Consequently, the direct detection cross section is suppressed by twelve orders of magnitude as compared with Eq.~\eqref{eq:dd_master_eqn} and is completely irrelevant.

The above represents the most dramatic impact of including a Higgs coupling: direct detection is essentially killed, whilst an almost equivalent bound from the $\rho$-parameter arises in its place.
There are other modifications to the model in this case also.
For example, new channels open in indirect detection and the $Z-Z'$ mixing allows DM to annihilate to $Zh$ and $W^+ W^-$.
This does not qualitatively change the thermal relic or indirect detection calculations, however.

Finally, we note one peculiarity of neglecting a Higgs coupling is that the SM Yukawas would appear to violate $U(1)_{\zp}$.
This demands a UV explanation; for example, a model which is $U(1)_{Z'}$ invariant, that reduces to the SM after $U(1)_{\zp}$ is broken and heavy degrees of freedom are integrated out. 
In this sense $Z'$-portal models, coupled with various phenomenological bounds, are a playground for constructing interesting UV completions.
This proves a constant theme throughout our work.

\subsection{Alternative Realisations of the $Z'$ Portal}
\label{sec:alves+}

\begin{figure}
\centering
\includegraphics[width=0.48\linewidth]{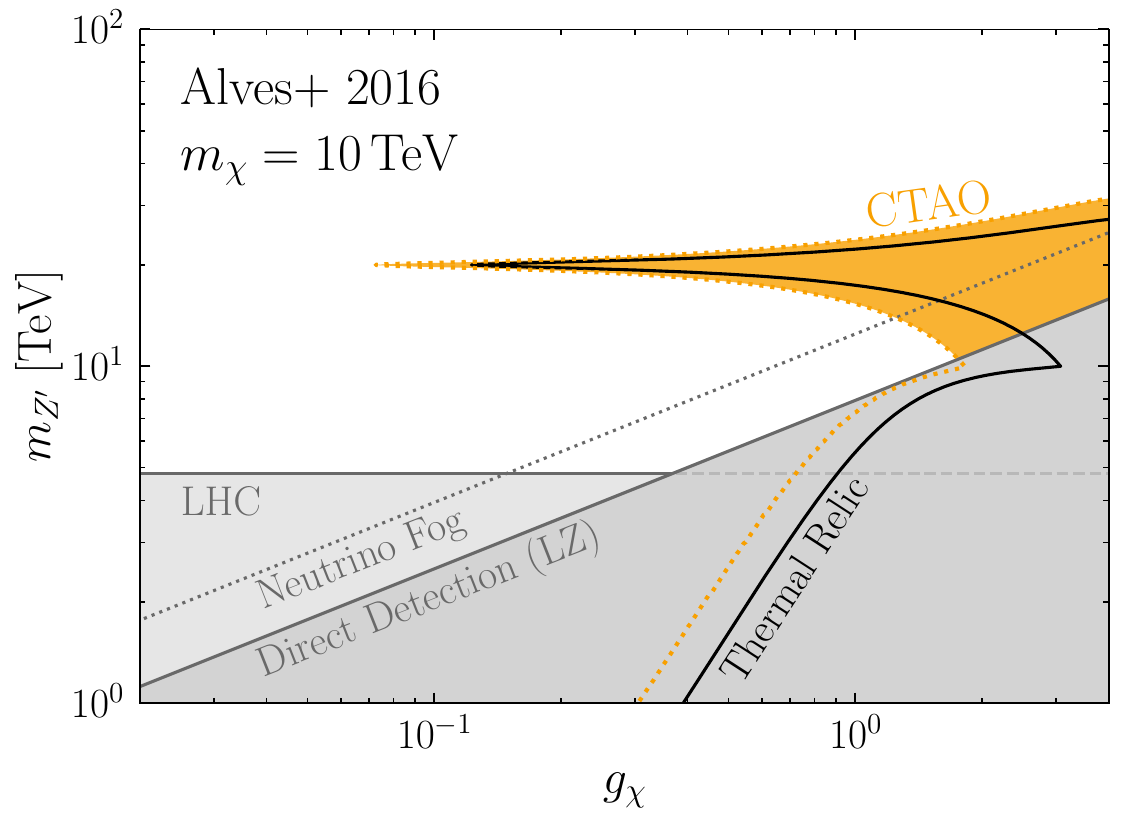}
\hspace{0.2cm}
\includegraphics[width=0.48\linewidth]{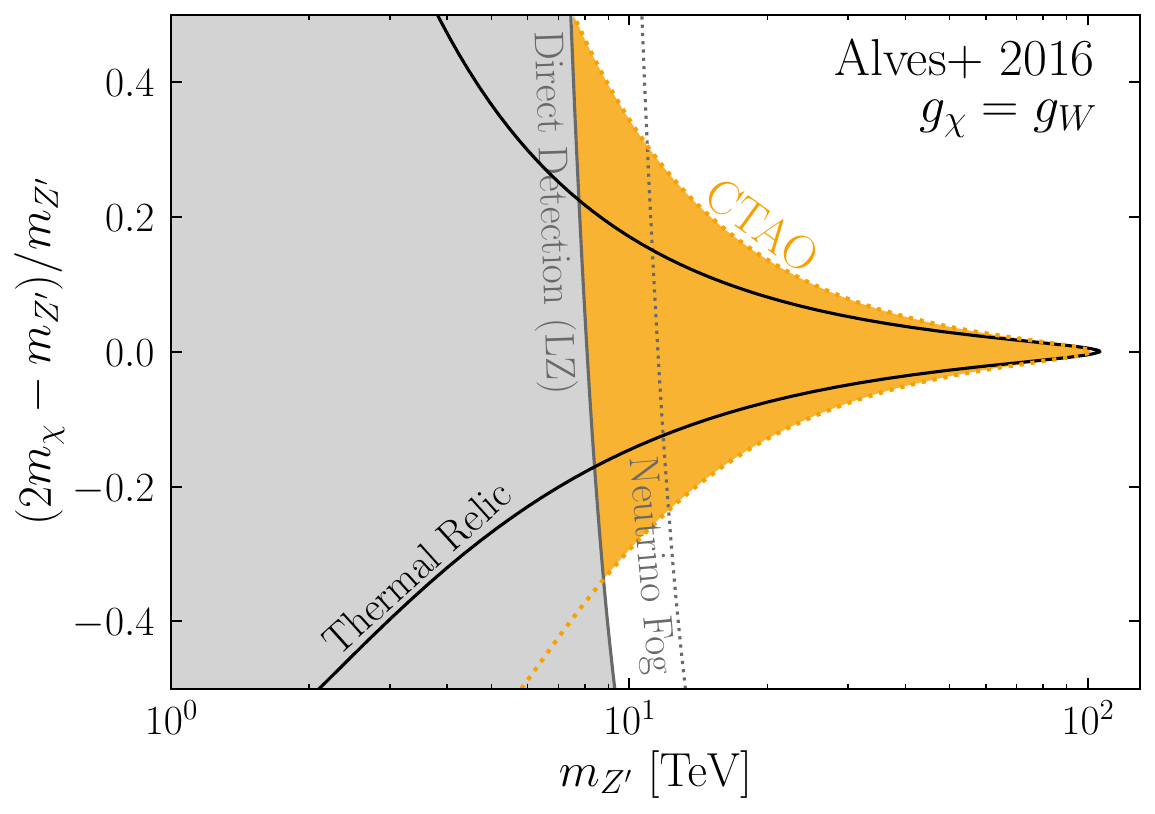}
\caption{As in Fig.~\ref{fig:Zprime1} but for the model of Ref.~\cite{Alves:2015pea} defined in Eq.~\eqref{eq:L-Zprime2}.
Here only the DM-$Z'$ coupling can vary; standard model states have an identical coupling to the $Z$.}
\label{fig:Zprime2}
\end{figure}

As emphasised already, Eq.~\eqref{eq:L-Zprime} is a single realisation of $Z'$-portal DM.
As another example, the influential study in Ref.~\cite{Alves:2015pea} considered a variant where the coupling of the $Z'$ to the standard model was essentially fixed to exactly that of the $Z$, whereas the coupling to DM could vary.
The resulting Lagrangian is given by
\begin{equation} \label{eq:L-Zprime2}
{\cal L} \supset Z'_{\mu} \Big[ g_{\chi}\, \bar{\chi}_R \gamma^{\mu} \chi_R + g_W\, \bar{f} \big(c_L P_L + c_R P_R \big) \gamma^{\mu} f \Big].
\end{equation}
(In the notation of Ref.~\cite{Alves:2015pea}, we have taken $g_{\chi v} = g_{\chi a} = 1/2$ and $a=1$.)
Recomputing the full sensitivity, we find the results in Fig.~\ref{fig:Zprime2}.
Whilst details differ, the qualitative conclusions do not: direct detection pushes us to higher masses and into a funnel with $10-20\%$ tuning, a region essentially entirely accessible to CTAO.

\section{UV Completions: Extra Dimensional GUTs}
\label{sec:ED_GUT}

Our analysis of the IR phenomenology of WIMP DM coupled to a $Z'$ gauge boson revealed that a thermal relic consistent with current direct detection bounds requires $m_{Z'}\sim \mathcal{O}(10)\,\TeV$. 
Since models solving the Planck-weak hierarchy generically lead to new states at the $\TeV$ scale~\cite{Giudice:1998bp,Agashe:2004rs,Martin:1997ns}, they can also be utilised as UV completions for WIMP DM satisfying these requirements~\cite{Agashe:2026ggy}.

A salient feature of the UV completion discussed in this paper, first proposed in Refs.~\cite{Agashe:2004ci,Agashe:2004bm}, is the connection between proton and DM stability. 
The stability of the proton can be a result of an underlying symmetry of the UV theory. 
One such symmetry, which is an accidental symmetry of the SM, is baryon number, or $U(1)_B$. 
This symmetry could also give rise to a stable, SM singlet particle as follows. 
We consider a 5D extra-dimensional theory with a gauged $U(1)_B$ symmetry, and define a $Z_3$ transformation of every field in our theory given by
\begin{equation}
    \Psi \to e^{2\pi i (B-\frac{n_c-n_{\bar{c}}}{3})} \Psi\,.
\label{eq:z3_symmetry}
\end{equation}
Here, $n_c(n_{\bar{c}} )$ is the number of $SU(3)_c$ coloured (anti-coloured) fields that make up $\Psi$. 
We follow the normalisation for $B$ under which $B=1/3$ for SM quarks. 
As the theory is invariant under $SU(3)_c$ and $U(1)_B$, this leads to a conserved $Z_3$ charge. 
Since the states in the SM that carry baryon charge are also always coloured, the entire SM is a singlet under the $Z_3$ symmetry. 
However, the UV theory will have states charged under the $Z_3$ symmetry. 
The lightest $Z_3$ charged particle will be identified as DM. 
Thus, a WIMP DM candidate is obtained as a spin-off of proton stability.

In the following subsections, we review some of the key details on how GUTs, together with extra dimensions, can realise such a DM candidate with a baryon charge.

\subsection{A Brief Review of Extra Dimensions}
\label{subsec:extra_dim_review}

As shown in the previous section, $\Delta$ is a key parameter in obtaining the observed thermal relic cross section, and is tuned to be $\mathcal{O}(10\%)$ for a generic weakly coupled model. 
This tuning can be relaxed if the WIMP and the $Z'$ are strongly interacting. 
However, if the coupling to the light quarks is also made stronger, the direct searches by LHC may rule out such a model.

Therefore, we need hierarchically different couplings for DM vs light SM particles. 
Extra dimensions offer explanations for hierarchies in mass scales (Planck and weak scales) as well as couplings (Yukawa couplings of SM fermions). 
Therefore, extra dimensions could also lead to a model of a WIMP and $Z'$ with small couplings with light SM fermions. 
We review some of the basics of extra-dimensions below, with most details deferred to App.~\ref{app:5D_specifics}.
We consider the following space-time metric, which is a slice of $\text{AdS}_5$,
\begin{equation}
   ds^2= e^{-2k|y|}dx_\mu dx^\mu - dy^2\,.
\label{eq:rs_metric} 
\end{equation}
Here, the curvature $k$ is related to the bulk 5D cosmological constant and the 5D Planck mass by $k^2=-\Lambda_{5D}/24 M^3_{5}$. 
The extra dimension is an $S^1/Z_2$ orbifold with boundaries at $y=0$ and $y=\pi R$. 
These boundaries are called the UV and IR branes respectively. 
A complex 5D scalar $\Phi$, a 5D fermion $\Psi$ and a 5D gauge field $A_M~(M=1,\ldots,5)$, will have the free-field action,
\begin{equation}\begin{aligned}
    \mathcal{S}^{5D}_{\rm free} = \int_{}{} d^4x dy\sqrt{-g} ~&\bigg[(D_M \Phi)^\dagger (D^M \Phi) - m^2_{5\Phi}|\Phi|^2+ i \bar{\Psi} \Gamma^M D_M \Psi \\
    &\hspace{2.5cm}- ck\, {\rm sgn}(y) \bar{\Psi}\Psi-\frac{1}{4}F_{MN}F^{MN}\bigg],
\label{eq:free_field_5D}
\end{aligned}\end{equation}
where $D_M$ denotes the 5D covariant derivative, and $m_{5\Phi}$ and $c$ are the scalar and fermion mass parameters. 
Each of these fields will have a KK decomposition into a 4D field times a profile in the fifth dimension. 
For instance, for the fermion field it is given by
\begin{align}
    \Psi(x,y)=\sum_n\psi_n (x) f_n(y).
\end{align}
The profiles of various modes depend on the BCs imposed on the UV/IR branes, as well as the 5D mass parameters $c$ and $m_{5\Phi}$, as well as brane-localised parameters (see App.~\ref{app:5D_specifics} for details). 
Just $\mathcal{O}(1)$ variations in these parameters can lead to significant changes in the profiles. 
The couplings of the 4D fields depend on the overlap of these profiles. 
For example, SM Yukawa couplings can be derived from the 5D Yukawa coupling as follows
\begin{align}
   \mathcal{L}_{5D}=y_5 H Q_iu_j^c \to  \mathcal{L}_{4D}= y^{u}_{ij} Q_iHu_j^c ,~ y^{u}_{ij}=y_5 \int_{0}^{\pi R}\sqrt{-g} ~dy~f_H(y) f^{i}_Q(y) f^{j}_u(y)\,. 
\label{eq:Yukawa_4D-definition}
\end{align}
The exact profiles for various SM fields can be found in App.~\ref{app:5D_specifics}. 
The zero modes of gauge fields, which correspond to the SM gauge fields, have a flat profile, meaning all SM fields have the same gauge couplings. 
The SM Higgs is peaked near the IR brane. All chiral fermions of the SM have profiles peaked near the UV brane, except for $Q_3$ and $t_R^c$. 
The doublet $Q_3$ has an almost flat profile whereas $t_R^c$ is peaked near the IR brane. These profiles together with Eq.~\eqref{eq:Yukawa_4D-definition} give the hierarchies in the SM Yukawas.

\subsection{DM from an Extra-dimensional GUT}

The freedom of boundary conditions and the localised nature of profiles make an appealing case for studying GUTs in the framework of RS1. 
A major issue with GUTs is the tuning between different Higgs vevs. 
This tuning is necessary to obtain a light SM Higgs while simultaneously making the extra Higgs fields (such as the triplet Higgs in $SU(5)$ GUTs) heavy. 
This tuning can be circumvented using BCs in a warped extra dimension. 
Another bottleneck for a generic GUT is the presence of X/Y type gauge bosons, which are charged under $SU(3)_c$. 
These leptoquarks can mediate interactions between protons and leptons, possibly leading to proton decay~\cite{Weinberg:1979sa,Dorsner:2024jiy,Super-Kamiokande:2020wjk}. 
The lower bound on the proton lifetime pushes the mass of these gauge bosons to $M_{\rm GUT} \gtrsim 10^{15}\,\rm{GeV}$. 
As we review here, splitting the GUT multiplets and assigning different types of BCs provides a way to get these masses down to $O(10)$\,TeV. 
Thus, these GUT models become more phenomenologically viable.

We study a warped $SO(10) \times U(1)_B$ GUT gauge theory, where the breaking of $SO(10)$ to the standard model gauge group $\mathcal{G}_{\rm SM}=SU(3)_c \times SU(2)_L\times U(1)_Y$ is achieved using boundary conditions~\cite{Csaki:2003dt,Csaki:2003sh}. 
We take $SO(10) \times U(1)_B$ to be a gauge symmetry in the bulk of the 5D spacetime. 
We break $SO(10) \to \mathcal{G}_{\rm SM}$ on the UV brane. 
This is tantamount to assigning a UV brane BC that is Neumann or $(+)$ BC for the SM gauge field, and Dirichlet or $(-)$ boundary condition for the other $SO(10)$ gauge bosons. 
We assign $(+)$ BCs to all gauge bosons on the IR brane. 
After carrying out the KK decompositions, the resulting 4D theory has only $\mathcal{G}_{\rm SM}$ as the unbroken gauge symmetry. 
The GUT gauge bosons, including the $X/Y$ leptoquarks, are at $\mathcal{O}(10)$\,TeV. 
However, despite these low mass scales for the leptoquarks, the model evades proton decay due to how matter fields are set up.

All the matter field representations of the SM can be accommodated in a $\bf{16}$ of $SO(10)$. 
However, since we require $U(1)_B$ to commute with the $SO(10)$ GUT gauge group, we cannot have all the SM fermions in one $\bf{16}$. 
Instead, each $\bf{16}$ only has one left-handed (LH) chiral fermion with $(++)$ BCs, which is identified as the SM fermion. 
The other 15 fermions are KK, Dirac fermions with $(-+)$ BCs for their LH chirality, and $(+-)$ BCs for the right-handed (RH) chirality. 
For instance, the multiplets containing the doublet $Q_3=(t,b)_L$ and $t_R^c$, are as follows:
\begin{align}  \bf{16}_{t_R}=\begin{pmatrix}
        t^c_R ~(++)\\
        d_R^{'c}\\
        Q'\\
        L'\\
        e^{'c}_R \\
        \nu^{'c}_R
    \end{pmatrix}\!,
    \qquad
    \bf{16}_{Q_3}=\begin{pmatrix}
        \tilde{u}_R\\
        \tilde{d}_R\\
        Q (++)\\
        \tilde{L}\\
        \tilde{e}_R \\
        \tilde{\nu}_R 
    \end{pmatrix}
    \label{eq:16_split}\!.
\end{align}
Here, all fields with a prime or a tilde are Dirac fermions with $(-+)$ BCs for their LH chirality. 
Every multiplet is assigned a common $U(1)_B$ charge, and that is the $U(1)_B$ charge of the SM field inside it. 
This split-multiplet setup ensures that $X/Y$ bosons only couple to fermions from the same $\bf{16}$. 
As this is a baryon number conserving interaction, it does not lead to proton decay.

\begin{figure}
    \centering    \includegraphics[width=0.65\linewidth]{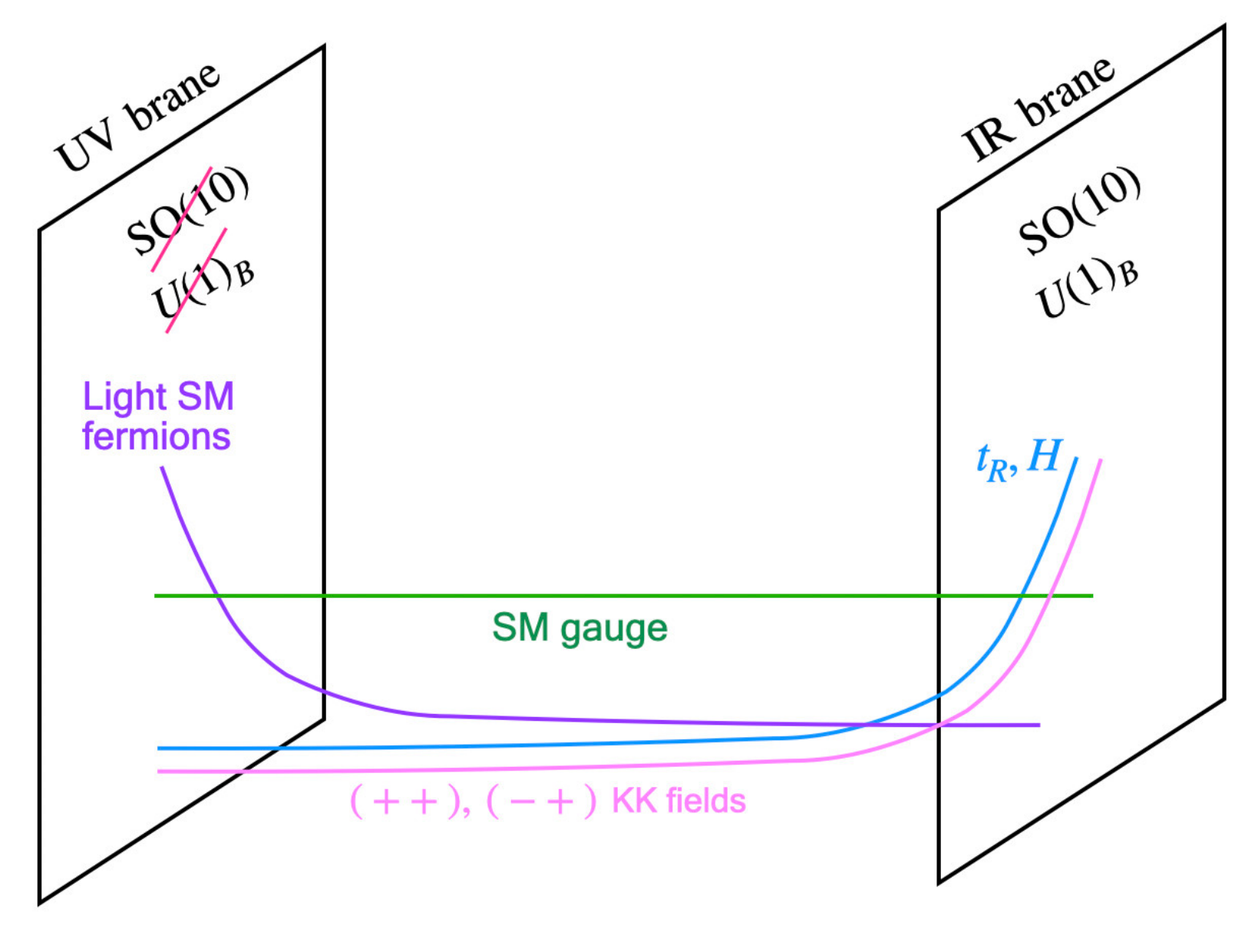}
    \caption{Basic set-up of the warped $SO(10)$ model. The curves show the 5D profiles of the various fields.}
    \label{fig:non_susy_setup}
\end{figure}

From Eq.~\eqref{eq:16_split}, we observe that we can have KK fermions that possess a $U(1)_B$ charge, but are $SU(3)_c$ singlets or vice-versa. 
In other words, we have states charged under the $Z_3$ symmetry of Eq.~\eqref{eq:z3_symmetry}. 
To identify the lightest $Z_3$ charged state, we note that the free field Lagrangian for a multiplet is given by
\begin{equation}
    \mathcal{S} \supset \int_{}{}d^4 x d y\sqrt{-g} \left[i\bar{\Psi}_{\bf{16}} \Gamma^M D_M  \Psi_{\bf{16}} - c k~ \text{sgn} (y) \bar{\Psi}_{\bf{16}}\Psi_{\bf{16}} \right],
\label{eq:ferm_action_so10}
\end{equation}
where the mass of the $(-+)$ KK fermion is given by
\begin{equation}
    m \simeq \begin{cases}
    \frac{\pi}{2}(1+c)ke^{-k\pi R},~~~ &c \gtrsim-1/4,\\
2\sqrt{c+\frac{1}{2}} ke^{-k\pi R}, &-1/2 \lesssim c \lesssim -1/4, \\    2\sqrt{|c+\frac{1}{2}|\left(|c+\frac{1}{2}|+1\right)}\left(e^{-k\pi R}\right)^{|c+\frac{1}{2}|}ke^{-k\pi R}, ~~&c \lesssim -1/2 .
    \end{cases}
\label{eq:mp_fermion_mass}
\end{equation}
It is evident from the relation above, that the multiplet with the smallest $c$ has the lightest $(-+)$ fermion. 
That multiplet is $\bf{16}_{t_R}$, as the large Yukawa coupling of the top requires $c_{t_R}\leq 0$. 
The only SM singlet field in $\bf{16}_{t_R}$ is $\nu'_R$, which we identify as our DM candidate. 
Since the DM stability is due to a $Z_3$ symmetry, we call it the lightest \(Z_3\)-charged particle (LZP). 

The complete setup of the model is summarised in Fig.~\ref{fig:non_susy_setup}. 
The profiles of the LH chirality of the GUT fermions, that is the chiralities with $(-+)$ BCs, are all peaked near the IR brane. 
This includes the LH  chirality of the DM, $\nu'_R$. 
The KK gauge fields and the Higgs are also peaked near the IR brane, while the SM fermions and gauge bosons retain their profiles from the usual RS1 setup as discussed in Sec.~\ref{subsec:extra_dim_review}.

\subsection{Symmetry Breaking and Dark-Matter Couplings}\label{sec:so10_specifics}

\subsubsection{$SO(10)$ Gauge Structure and Breaking}\label{subsec:gauge_structure}

In this section, we give the breaking pattern of $SO(10)$ in the model and obtain the relevant charges and couplings of various gauge bosons. 
Starting with the gauge sector, we have $SO(10)$ unbroken in the bulk as well as the IR brane, while 
$SO(10)$ is broken to the SM gauge group $SU(3)_c \times SU(2)_L \times U(1)_Y$ on the UV brane. 
This is achieved by giving $(++)$ BCs to the SM gauge bosons, and $(-+)$ to all other $SO(10)$ gauge fields.

For the remainder of this paper, we primarily discuss the gauge bosons of the Pati-Salam subgroup $SU(4)_c \times SU(2)_R \times SU(2)_L$ of $SO(10)$. 
This is because the gauge interactions from this subgroup are the ones relevant for the DM phenomenology. 
Further details about the gauge interactions for the full $SO(10)$ model can be found in Ref.~\cite{Agashe:2004bm}. 
We consider the following breaking pattern for Pati-Salam:\\
{\bf(i):} We break $SU(4)_c\to SU(3)_c\times U(1)_{B-L}$, which is done by imparting $(-+)$ boundary conditions to the $X_s$ gauge boson.\\
{\bf(ii):} We break $SU(2)_R\to U(1)_{R}$, which is done by imparting $(-+)$ boundary conditions to the charged $W^{\pm}_{R}$ gauge bosons.\\
{\bf(iii):} Finally, we break $U(1)_{R}\times U(1)_{B-L}\to U(1)_{Y}$. We achieve this with a scalar $\Phi$, that is localised on the UV brane. 
Under $SU(4)_c \times SU(2)_R \times SU(2)_L$, $\Phi$ transforms as a $(\bf{4,2,1})$, giving it a $B-L$ charge of $-1$. 
We assume that the component of $\Phi$ with +1/2 $U(1)_R$ charge gets a vev $\langle \Phi\rangle \sim M_{\rm pl}$, breaking $U(1)_R \times U(1)_{B-L} \to U(1)_Y$. 
This breaking via a Planckian vev is tantamount to assigning $(++)$ BCs to the hypercharge gauge boson, and $(-+)$ BCs to the orthogonal gauge boson $Z'$. 
We thus get the $U(1)_Y,U(1)_{Z'}$ charges to be
\begin{equation}
    Y= T_{3R}+ \frac{1}{2} (B-L),\qquad
    Q_{Z'}=T_{3R} - Y\sin^2 \theta'\,.
\end{equation}
The 5D gauge couplings of the subgroups $SU(4)_c$, $SU(2)_R$, and $SU(2)_L$ are all equal to the $SO(10)$ gauge coupling $g^{(5)}_{4c}=g^{(5)}_{2R}=g^{(5)}_{2L}=g_5$, while $SU(4)$ normalisation fixes the coupling $g^{(5)}_{B-L}=\sqrt{3/2}g_5$. 
Denoting the gauge bosons for $U(1)_R$ and $U(1)_{B-L}$ by $W_{3R ,\mu}$ and $(B-L)_\mu$, we have the two orthogonal combinations
\begin{align}
    Z'_\mu&=W_{3R ,\mu}\cos \theta'- \frac{1}{2}(B-L)_\mu \sin \theta', ~~~~g_{Z'}=\sqrt{g_{2R}^2+g_{B-L}^2}=\sqrt{\frac{5}{2}} g_5\,,\\ 
    B_\mu&= W_{3R ,\mu} \sin\theta'+ \frac{1}{2}(B-L)_\mu \cos \theta',~~~~\sin^2 \theta' =\frac{g^2_{B-L}}{g_{B-L}^2+g^2_{2R}}=\frac{3}{5}\,.
\end{align}
Finally, we have the relation $m_{Z'}=m_{X_s}=m_{W_R}=m_{\rm KK}\sim \mathcal{O} (10 \,\rm TeV)$ due to the gauge bosons having identical BCs. The profiles of $X_s, W^{\pm}_R$ and $Z'$ are peaked near the IR brane, as shown in Fig.~\ref{fig:non_susy_setup}.  
From here on, we report all the gauge couplings above in terms of two 4D couplings. 
The first is $g_{10}=g_5/\sqrt{\pi R}$, which is the coupling of zero modes or SM gauge bosons. 
The second is the strong coupling $g_{*} = g_5 \sqrt{k}=g_{10}\sqrt{k \pi R}$, which is the typical interaction strength between fields peaked near the IR brane. 

\subsubsection{$U(1)_B$ and Further Breaking of $SO(10)$} \label{subsec:breaking_so10_u1b}

On top of the $SO(10)$ breaking patterns discussed above, we need to include some more breakings in order to get a viable DM model. 
Since $U(1)_B$ is a gauge symmetry, the baryon number gauge boson (denoted by $\mathcal{B}$) has its own set of boundary conditions. 
Having $U(1)_B$ unbroken everywhere, or assigning $(++)$ BCs to $\mathcal{B}$, would imply a massless $\mathcal{B}$, which is experimentally ruled out. 
Therefore, $U(1)_B$ has to be broken on one of the branes. 
Breaking $U(1)_B$ on the IR brane can give rise to dimension six operators that lead to fast proton decay. 
This is because the scale suppressing these operators is only $\mathcal{O}(10)$\,TeV. 
This fixes the BC on the IR brane to be $(+)$. 
We note that there cannot be a Higgs peaked on the IR brane (akin to the SM Higgs) that breaks $U(1)_B$ as that would also lead to fast proton decay. 
Therefore, the only remaining option is to break $U(1)_B$ on the UV brane. 
This implies the presence of UV brane localised higher dimensional operators that can lead to proton decay. 
However, as these operators are typically suppressed by Planckian scales, the resulting proton decay is slow. 
Thus, $\mathcal{B}$ gets $(-+)$ BCs and an IR brane peaked profile like the gauge KK modes in Fig.~\ref{fig:non_susy_setup}.

While we are safe from proton decay, the breaking of $U(1)_B$ on the UV brane should be treated carefully, as it can lead to the violation of the $Z_3$ symmetry in Eq.~\eqref{eq:z3_symmetry}. 
However, we see that we can break $U(1)_B$ and still conserve $Z_3$ charges, as long as the breaking $\Delta B$ takes integer values. 
This can be realised by breaking $U(1)_B$ on the UV brane via the vev of a scalar that has an integer baryon number.

With the BCs for $\mathcal{B}$ fixed, we note that the 5D gauge coupling for $U(1)_B$, denoted by $g^{(5)}_B$, is a free parameter. 
For simplicity we choose $g^{(5)}_B=g_5$ for our analysis.

Having fixed the breaking patterns of $SO(10)$ in the previous subsection, DM phenomenology forces us to consider slight variations. 
From Eq.~\eqref{eq:ferm_action_so10}, we see that $\nu'_R$ has the same mass as its GUT partners at leading order. 
While, SM gauge loops result in a slightly larger mass for the GUT partners, the splitting is small, given parametrically by 
\begin{align}
    \delta m/m \simeq \frac{g^2_{\rm SM}}{4\pi^2} \lesssim 4\%.
\end{align}
This small difference means co-annihilation effects become important. 
In Sec.~\ref{sec:quark_mass}, we show that breaking $SO(10)$ on the IR brane can yield a $10\%$ mass-splitting between $\nu'_R$ and its GUT partners, which is sufficient to prevent co-annihilation. 
However, in the original model from Ref.~\cite{Agashe:2004bm}, motivated by theoretical reasons such as better gauge coupling unification, a bulk breaking of SO(10) was invoked. 
In particular, this was done by adding a mass term for the GUT partners
\begin{equation}
    \mathcal{L} \supset \int d^4 x dy  ~\epsilon(y)\Delta c \sum_{\Psi \neq t_R,\nu'_R} \bar{\Psi}\Psi. \label{eq:so(10)_bulk_breaking}
\end{equation}
To obtain a mass difference of $10 \%$, we need a splitting $\Delta c \simeq 0.1$. 
This small $c$ splitting will be assumed till Sec.~\ref{sec:quark_mass}.

\subsubsection{Couplings of Dark Matter} \label{subsec:couplings_of_dm}

In this section, we discuss the phenomenological implications of the DM model described in the previous subsections. 
For simplicity and consistency of notation, we use $\chi$ in place of $\nu'_R$ for the DM fermion. 
For $\chi$, the profiles of both the LH  (with $(-+)$ BCs) and RH (with $(+-)$ BCs) chiralities are peaked close to the IR brane. 
The peak for the LH chirality is sharper and closer to the IR brane. 
Thus, the LH chirality has stronger couplings with the KK gauge fields.

In order to specify the gauge interactions of $\chi$, we need to focus on the fermions of the $\bf{16_{t_R}}$ multiplet. 
The only two fermions relevant for the gauge interactions are $t^c_R$ and the $SU(2)_R$ partner of $\chi$ (the LH chirality of which is denoted by $e_R^{c'}$ in Eq.~\eqref{eq:16_split}). 
For simplicity, we denote this fermion by $e'$. 
With this notation, we have the following GUT gauge interactions of $\chi$, $t_R^c$ and the Higgs 
\begin{align}
\mathcal{L}_{\rm gauge} \supset -& g_* \left[ \sqrt{\frac{5}{2}}Q^{\chi}_{Z'}Z'_{\mu}\bar{\chi} \gamma^\mu (P_R a_{\chi_R}+P_L a_{\chi_L}) \chi+ \frac{1}{3} \mathcal{B}_{\mu}\bar{\chi} \gamma^\mu (P_R a_{\chi_R}+P_L a_{\chi_L}) \chi \right] \nonumber\\
&-g_*\left[ \frac{1}{\sqrt{2}} W_{R,\mu}\bar{e'} \gamma^\mu (P_R a_{\chi_R}+P_L a_{\chi_L}) \chi  \right ] -g_* a_{X_s} \frac{1}{\sqrt{2}}X_{s,\mu}\bar{t^c_R}\gamma^\mu P_L \chi \nonumber\\
&-g_* a_{t_R} \left[ \sqrt{\frac{5}{2}}Q^{t^c_R}_{Z'}Z'_{\mu}{t^c_R}^\dagger {\bar{\sigma}}^{\mu} t^c_R- \frac{1}{3} \mathcal{B}_{\mu}{t^c_R}^\dagger {\bar{\sigma}}^{\mu} t^c_R \right]\nonumber\\
&-g_* a_H \sqrt{\frac{5}{2}}Q^H_{Z'} Z'_\mu \left( D^\mu H^\dagger H- H^\dagger D^\mu H \right)\!,
\end{align}
where $P_R=\frac{1}{2}(1+\gamma^5)$, and $a_{\chi_R},a_{\chi_L},a_{t_R}, a_{X_s}$ and $a_H$ are $\mathcal{O}(1)$ factors arising from the overlap of the relevant 5D profiles. 
The factors $a_{X_s},a_{t_R}$ in particular, are important for DM phenomenology. 
This is because, as we vary the DM mass, the $c$ parameter from Eq.~\eqref{eq:ferm_action_so10} also changes. 
This in turn changes the profile of $t_R^c$, and results in $O(1)$ different overlaps of $t_R^c$ with other IR peaked fields, such as KK gauge fields. 
These variations of overlaps are captured entirely by $a_{X_s}$ and $a_{t_R}$. 
As shown in App.~\ref{app:holography}, the extra dimensional picture is holographic to a strongly interacting CFT. 
In this picture, the factors $a_{X_s}, a_{t_R}$ capture the degree of compositeness of $t_R^c$.

Here, we remind the reader that we choose a profile for the Higgs field which is that of an $A_5$ zero-mode of a 5D gauge field, for which $a_H \simeq 1$. 
For the estimates in the main text, we shall put $a_{X_s}=a_{t_R}=1$, reserving the exact numbers for the final results.

Apart from these interactions, there are a number of sub-leading effects. 
These couplings, while irrelevant for the annihilation of DM, can be very important for the direct detection cross section. 
To specify these interactions, we need an additional GUT fermion, namely the lepton doublet of the $(t,b)_L$ multiplet $\bf{16_{Q_3}}$ (that is, $\tilde{L}$ in Eq.~\eqref{eq:16_split}). 
We call this doublet $\ell'$ for simplicity. 
Furthermore, we call the neutral lepton inside the doublet $\mathcal{N}$. 
The various sub-leading interactions are as follows:\\ 
\textbf{(i) Coupling of DM to SM Higgs:} We have the $SO(10)$ invariant Yukawa interaction for the DM, 
\begin{align}
    \mathcal{L}^{5 \rm D}_{\rm Yukawa}&\supset \mathbf{16_Q16_{t_R} 10_H} \supset y_5 \chi \ell' H.
\label{eq:nul_nur_Higgs}
\end{align}
Since all the fields here are peaked close to the IR brane, we get the following Yukawa couplings after electroweak symmetry breaking
\begin{align}
    \mathcal{L}^{4 \rm D}_{\rm Yukawa}&= y_{A} \chi_L \neu_R \left(\frac{v+h}{\sqrt{2}}\right)+ y_{B} \chi_R \neu_L \left(\frac{v+h}{\sqrt{2}}\right),\\
    y_{A}&=y_5 \int_{}dy \sqrt{-g} f_H f_{\chi_L}f_{\neu_R},\qquad
    y_{B}=y_5 \int_{}dy \sqrt{-g} f_H f_{\chi_R}f_{\neu_L}.
\label{eq:Yukawas}
\end{align}
as well as mass mixings between $\neu$ and $\chi$ post electroweak symmetry breaking,
\begin{align}
    m_A=y_A v/\sqrt{2},\qquad
    m_B= y_B v/\sqrt{2}\,.
\label{eq:mass_mixings}
\end{align}

The mass mixings and Yukawas can be estimated by taking the ratio of profile overlaps of $\chi$ and $\mathcal{N}$ with the profile overlap of $Q_3,\,t^c_R$. 
The ratio of profile overlaps only depends on the 5D mass parameter $c$, and therefore the ratio of profile overlaps is fixed. 
We fix $0.3 \lesssim  c_{t_L} \lesssim 0.4$, and this gives us
\begin{align}
    \frac{m_A}{m_t}=\frac{y_A}{y_t}\simeq 2,\qquad\frac{m_B}{m_t}=\frac{y_B}{y_t}\simeq \frac{1}{2}~.
\label{eq:mmix_nul_nur}
\end{align}

The expression above is approximate since the top quark profile can shift as we vary the $c$ parameter. 
Due to this variation, the right hand side of Eq.~\eqref{eq:mmix_nul_nur} can be smaller by an $O(1)$ factor. 
We use Eq.~\eqref{eq:mmix_nul_nur} for estimates and reserve exact computations for the final phenomenological discussion. 
Upon diagonalisation of the mass matrix, the mass eigenstates (shown in bold) are related to the original basis as follows:
\begin{align}
    ~~~~N_R &= \sin \theta_1 {\bf  \chi_R} + \cos\theta_1 {\bf N_R},&\sin\theta_1 \simeq \frac{m_A m_{\chi}+m_Bm_\neu}{m^2_{\neu}-m^2_{\chi}},\nonumber\\
    ~~~~N_L &= \cos \theta_2 {\bf N_L} + \sin \theta_2 {\bf \chi_L},&\sin\theta_2 \simeq \frac{m_B m_{\chi}+m_Am_\neu}{m^2_{\neu}-m^2_{\chi}},
\label{eq:basis_change}
\end{align}
where, if we use $c_{t_L}=0.3$ and Eq.~\eqref{eq:mp_fermion_mass},  $m_\neu\simeq 0.85 m_{\rm KK}$. 
This mixing can induce a Higgs DM coupling
\begin{align}
\mathcal{L}_{\rm int} &\supset y_\chi h \bar{\chi}\chi=(y_A \sin\theta_1+y_B\sin \theta_2) \frac{h}{\sqrt{2}}\bar{\chi}\chi \nonumber\\
&\simeq\left[\frac{(y_A^2+y_B^2)m_\chi+2y_Ay_Bm_\neu }{m^2_{\neu}-m^2_{\chi}} \right]\frac{v}{2} h\bar{\chi}\chi\,.
\label{eq:Higgs_dm}
\end{align}
\textbf{(ii) Coupling of DM to SM Z via $\nu'_L-\nu'_R$ mixing:} Noting that $\nu'_L$ is charged under the SM $Z$, the diagonalisation in Eq.~\eqref{eq:basis_change} induces a coupling of the DM to the SM $Z$, 
\begin{align}
    &g^{\chi_R}_{Z,\rm mix} \simeq g_Z \times \left(\frac{m_Am_\chi+m_Bm_\neu }{m^2_{\neu}-m^2_{\chi}}\right)^2 \times \frac{1}{2} \,,\\
    &g^{\chi_L}_{Z,\rm mix} \simeq g_Z \times \left(\frac{m_Bm_\chi+m_Am_\neu}{m^2_{\neu}-m^2_{\chi}}\right)^2 \times \frac{1}{2} \,,
\label{eq:g_nul_nur}
\end{align}
where the factor of $1/2$ is the charge of $\nu'_L$ under the SM $Z$. \\
\textbf{(iii) Coupling of DM to SM $Z$ via $Z-Z'$ mixing:} Another sub-leading effect arises from the  Higgs coupling to both SM $Z$ and $Z'$. 
This leads to a  $Z-Z^{ \prime }$ mass mixing term given by (using $( B - L )$ charge of Higgs being 0)
\begin{align}
\Delta m^2 = \gzp g_{Z} v^2 T^H_{ 3 L } Y_H\cos^2 \theta^{ \prime }=-\frac{\gzp g_Z v^2 \cos^2 \swp}{4}\,,
\end{align}
where $\gzp = \sqrt{5/2}g_* $ is the 4D coupling of $Z'$. 
The squared mass matrix in the gauge basis can be diagonalised (assuming small mixing and $m_{ Z^{ \prime } } \gg m_Z $) giving the mass eigenstate for the $Z$ boson: $ {\rm SM }~Z ~\simeq Z- \frac{\Delta m^2}{M_Z'^2}Z' $,
resulting in $\nu^{ \prime }_R$ acquiring a coupling to the SM $Z$ boson of
\begin{align} 
g^{ \chi_R }_{ Z-Z' }\sim g^{ \chi_L }_{ Z-Z' } \simeq
g_Z \left(\frac{ v^2 \cos^2 \theta^{ \prime } g^2_{Z'} Q^{\dm}_{ Z^{ \prime } } }{ 4 m^{2}_{ \rm KK } }\right) \mathcal{O}(1)\,,
\label{eq: g_z_zprime_mixing}
\end{align}
where the $\mathcal{O}(1)$ factor comes from the overlap of $Z'$ and $\chi_L/\chi_R$. 
The total coupling of DM to SM $Z$, is therefore given by
\begin{align}
    \mathcal{L_{\rm int}} \supset (g^{\chi_R}_Z \bar{\chi_R} \bar{\sigma}^\mu \chi_R+g^{\chi_L}_Z \bar{\chi_L} \bar{\sigma}^\mu \chi_L )Z_{\mu}\,,
    \qquad g^{\chi_R,\chi_L}_Z&=g^{\chi_R,\chi_L}_{Z,\rm mix}+g^{\chi_R,\chi_L}_{Z-Z'}\,.
\label{eq:total_smz_coupling}
\end{align}
\textbf{(iv) Kinetic mixings with SM hypercharge:} Both $Z'$ and $\mathcal{B}$ can have a kinetic mixing with the hypercharge gauge boson,
\begin{equation}
    \mathcal{L_{\rm mix}} \supset -\frac{\epsilon_{Z'-Y}}{2} Z'_{\mu \nu} Y^{\mu \nu}-\frac{\epsilon_{\mathcal{B}-Y}}{2} \mathcal{B}_{\mu \nu} Y^{\mu \nu}\,.
\end{equation}
This mixing is mediated by loops of heavy fermions and Higgs fields, and is given by
\begin{align}
    &\epsilon_{Z'-Y} \simeq \frac{g_Yg_{Z'}}{24\pi^2} \sum_f Q^{Z'}_f Y_f\log(\frac{m^2_{KK}}{m_f^2})+\frac{g_Yg_{Z'}}{48\pi^2} \sum_\phi Q^{Z'}_\phi Y_\phi\log(\frac{m^2_{KK}}{m_\phi^2}),\\
&\epsilon_{\mathcal{B}-Y} \simeq \frac{g_Yg_{\mathcal{B}}}{24\pi^2} \sum_f \frac{1}{3} Y_f\log(\frac{m^2_{KK}}{m_f^2}).
\end{align}
Here, $g_Y$ is the hypercharge gauge coupling, the index $f$ represents the fermions in the $t^c_R$ GUT multiplet, while the index $\phi$ represents the (complex) scalar fields in the theory, namely the doublet and triplet Higgs. 
Light quarks, such as $u,d$, are effectively decoupled from $Z'$ and $\mathcal{B}$ due to the $(-)$ boundary condition of these gauge bosons on the UV brane. 
However, the kinetic mixing above induces the following couplings
\begin{equation}
    \mathcal{L} \supset- g_Y \left(Y_u\bar{u}\gamma^\mu u+ Y_d\bar{d}\gamma^\mu d \right) \left(  \epsilon_{Z'-Y} Z'_\mu+\epsilon_{\mathcal{B}-Y} \mathcal{B}_\mu\right)\!.
\end{equation}  
Estimating the direct detection amplitude, we find that the contribution from the kinetic mixing above is negligible. 
For instance, comparing it with the $Z-Z'$ mixing from Eq.~\eqref{eq: g_z_zprime_mixing}, we find that the direct detection amplitude is suppressed roughly by a loop factor of ${g_Y^2}/{16\pi^2}~$. 
Therefore, we ignore kinetic mixing for the final analyses.

We now have all the relevant couplings of the DM and we discuss its phenomenology in the next subsection. 

\subsection{Results for the Original Model}\label{subsec: results_unbroken_so10}
%
\begin{figure}[b]
  \centering \includegraphics[width=0.9\linewidth]{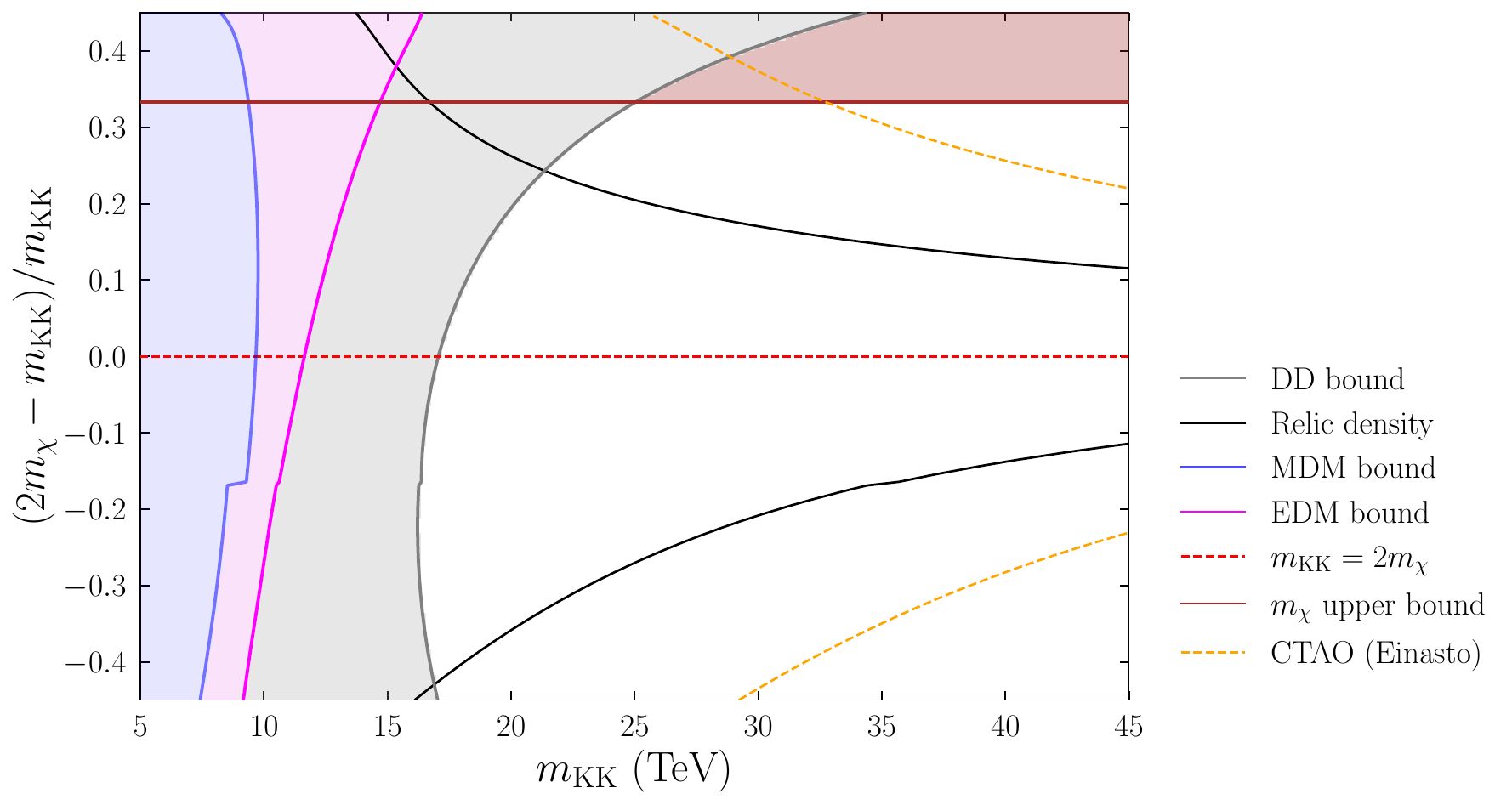}
   \caption{Parameter space for $\chi$ mass tuning vs. KK mass, for $SO(10) \times U(1)_B$ unbroken on the IR brane. Here, DD stands for Direct Detection, MDM for Magnetic Dipole Moment of $\chi$, and EDM for Electric Dipole Moment of $\chi$. As $c_{t_R}\leq0$, Eq.~\eqref{eq:mp_fermion_mass} leads to an upper bound on the mass of DM, as depicted by the brown line above. The CTAO projection is made assuming the Einasto profile for $\chi$ at the Galactic centre.}
   \label{fig:results_so10_unbroken}
\end{figure}
\begin{figure}[t]
\centering

\resizebox{\textwidth}{!}{%
\begin{tikzpicture}

\begin{scope}[xshift=7.5cm,yshift=7cm]
\draw[
  -{Triangle[length=3mm,width=3mm]},
  line width=1pt,
  draw=black,
  opacity=0.3
]
    (-1.5,-4) -- (1.5,-4)
    node[right, text=black, text opacity=0.4] {\small $\mathrm{Time}$};
\end{scope}

\begin{scope}[xshift=3.1cm]

\draw[thick,antielectron]
(-1.0,0) to[out=120,in=-60] (-2.0,1.9);

\draw[thick,electron]
(-1.0,0) to[out=-120,in=60] (-2.0,-1.9);

\draw[thick,photon]
(-1.0,0) -- (1.0,0);

\draw[thick,electron]
(1.0,0) to[out=60,in=-120] (2.0,1.9);

\draw[thick,antielectron]
(1.0,0) to[out=-60,in=120] (2.0,-1.9);

\node at (0,-0.45) {$Z'$};

\node[left] at (-2.0,1.9)
{\large $\chi$};

\node[left] at (-2.0,-1.9)
{\large $\bar{\chi}$};

\node[right] at (1.85,1.65)
{\Large $\substack{t_R\\ W_{\rm long}^+\\ h}$};

\node[right] at (1.85,-1.65)
{\Large $\substack{\bar t_R\\ W_{\rm long}^-\\ Z_{\rm long}}$};

\end{scope}

\begin{scope}[xshift=8.7cm]

\draw[thick,antielectron]
(-1.0,0) to[out=120,in=-60] (-2.0,1.9);

\draw[thick,electron]
(-1.0,0) to[out=-120,in=60] (-2.0,-1.9);

\draw[thick,photon]
(-1.0,0) -- (1.0,0);

\draw[thick,electron]
(1.0,0) to[out=60,in=-120] (2.0,1.9);

\draw[thick,antielectron]
(1.0,0) to[out=-60,in=120] (2.0,-1.9);

\node at (0,-0.45) {$\mathcal{B}$};

\node[left] at (-2.0,1.9)
{\large $\chi$};

\node[left] at (-2.0,-1.9)
{\large $\bar{\chi}$};

\node[right] at (2.0,1.9)
{\large $t_R$};

\node[right] at (2.0,-1.9)
{\large $\bar t_R$};

\end{scope}

\begin{scope}[xshift=14cm]

\draw[thick,antielectron]
(0,0.75) -- (-2.0,1.9);

\draw[thick,electron]
(0,0.75) -- (2.0,1.9);

\draw[thick,photon]
(0,0.75) -- (0,-0.75);

\draw[thick,electron]
(0,-0.75) -- (-2.0,-1.9);

\draw[thick,antielectron]
(0,-0.75) -- (2.0,-1.9);

\node[right] at (0.25,0)
{$X_s$};

\node[left] at (-2.0,1.9)
{\large $\chi$};

\node[right] at (2.0,1.9)
{\large $t_R$};

\node[left] at (-2.0,-1.9)
{\large $\bar{\chi}$};

\node[right] at (2.0,-1.9)
{\large $\bar t_R$};

\end{scope}

\end{tikzpicture}%
}

\caption{Annihilation channels for the LZP DM, $\chi$.}
\label{fig:lzp_annihilation}

\end{figure}
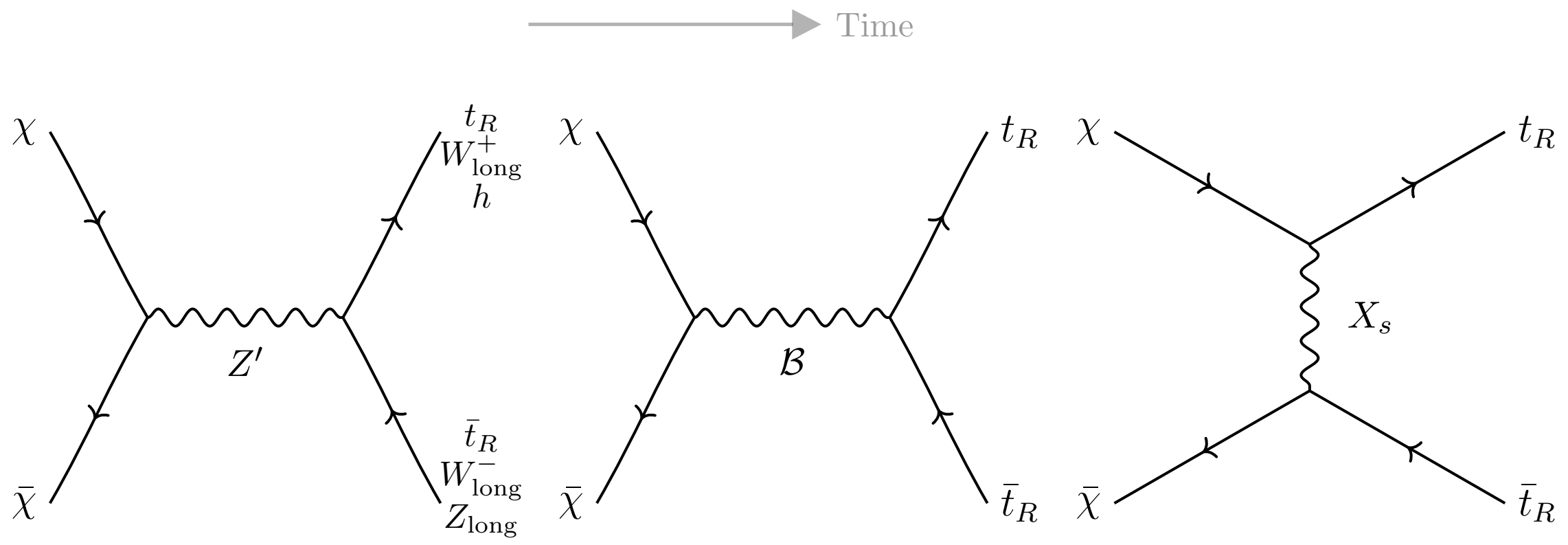

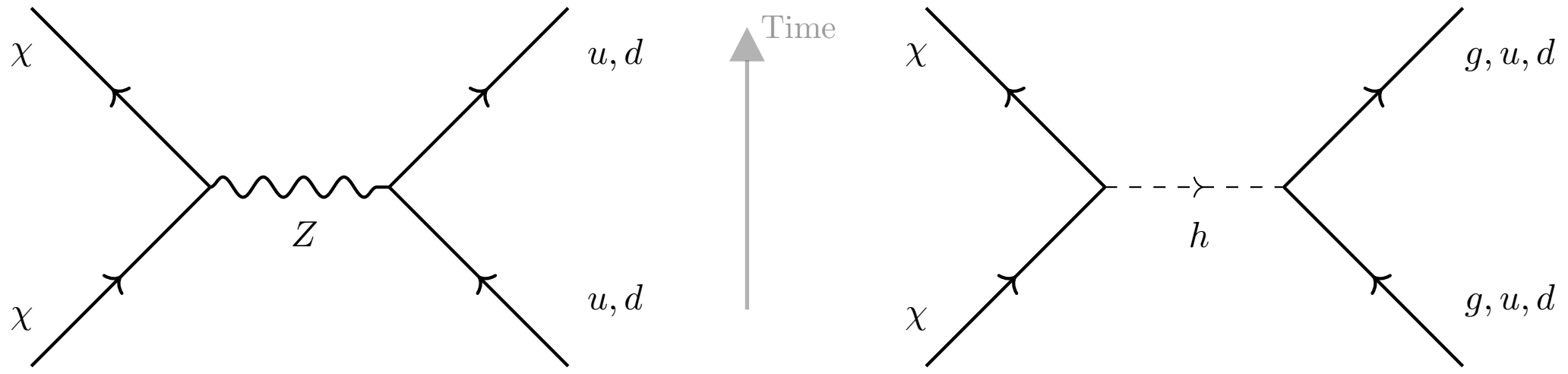
\begin{figure}[t]
\centering

\begin{tikzpicture}[scale=0.82,transform shape]

\begin{scope}[xshift=6cm]

\draw[thick,electron]
(-1.2,-1.2) to[out=135,in=-45] (-3.1,0.7);

\draw[thick,antielectron]
(-1.2,-1.2) to[out=-135,in=45] (-3.1,-3.1);

\draw[thick,photon]
(-1.2,-1.2) to[out=0,in=-180] (0.7,-1.2);

\draw[thick,electron]
(0.7,-1.2) to[out=45,in=-135] (2.6,0.7);

\draw[thick,antielectron]
(0.7,-1.2) to[out=-45,in=135] (2.6,-3.1);

\node at (-0.2,-1.7) {$Z$};

\node at (-3.2,0.2) {$\chi$};
\node at (-3.2,-2.6) {$\chi$};

\node at (3.1,0.2) {$u,d$};
\node at (3.1,-2.4) {$u,d$};

\end{scope}

\begin{scope}[xshift=7.5cm,yshift=-1cm]
\draw[
  -{Triangle[length=3mm,width=3mm]},
  line width=1pt,
  draw=black,
  opacity=0.3
]
    (3,-1.5) -- (3,1.5)
    node[right, text=black, text opacity=0.4] {\small $\mathrm{Time}$};
\end{scope}
\begin{scope}[xshift=15.5cm]

\draw[thick,electron]
(-1.2,-1.2) to[out=135,in=-45] (-3.1,0.7);

\draw[thick,antielectron]
(-1.2,-1.2) to[out=-135,in=45] (-3.1,-3.1);

\draw[dashed,electron]
(-1.2,-1.2) to[out=0,in=-180] (0.7,-1.2);

\draw[thick,electron]
(0.7,-1.2) to[out=45,in=-135] (2.6,0.7);

\draw[thick,antielectron]
(0.7,-1.2) to[out=-45,in=135] (2.6,-3.1);

\node at (-0.2,-1.7) {$h$};

\node at (-3.2,0.2) {$\chi$};
\node at (-3.2,-2.6) {$\chi$};

\node at (3.1,0.2) {$g,u,d$};
\node at (3.1,-2.4) {$g,u,d$};

\end{scope}

\end{tikzpicture}

\caption{Direct detection channels for the LZP DM, $\chi$. DM scatters off nucleons, therefore the quarks and gluon final states
are all understood as constituents of nucleons.}
\label{fig:lzp_dd}
\end{figure}

In this section, we discuss the phenomenology of the LZP DM, $\chi$, and update the bounds on the original model from Ref.~\cite{Agashe:2004bm} and the subsequent work of Ref.~\cite{Belanger:2007dx}. 
As discussed in the previous section, we assume $SO(10)$ is unbroken on the IR brane. 
We plot the various constraints in Fig.~\ref{fig:results_so10_unbroken}. 
The parameters we scan over are the DM mass tuning and the KK scale $m_{\rm KK}$, which is the mass of all the mediators ($Z',X_s$ and $\mathcal{B}$). 
For simplicity, we fix the parameters $g_{10} =g_W$ and $c_{t_L}=0.3$, which in turn fixes all the other couplings of the model. 
We describe the various features of the plot below.

\textbf{Annihilation cross section: }As in Sec.~\ref{sec:IRZprime}, we study the case of non-relativistic WIMP annihilation ($\beta_{\rm DM}\sim 0.1-0.2$). 
The dominant contribution to the annihilation of DM is from SM final states that are peaked near the IR brane. 
We show the various channels in Fig.~\ref{fig:lzp_annihilation}. 
In this model, DM primarily annihilates into $t_R^c$, because of the multiple annihilation channels mediated by $Z'$, $X_s$, and $\mathcal{B}$ gauge bosons. 
We find that this contribution is roughly $90\%$. 
Roughly $5\%$ of the total annihilation is to longitudinal $W^{\pm}$ as well as longitudinal $Z$ plus a Higgs boson.

\textbf{Direct Detection:} As is apparent from Fig.~\ref{fig:results_so10_unbroken}, direct detection places the strongest bound on the model. 
As discussed before, $Z'$, $X_s$, and $\mathcal{B}$ gauge bosons have negligible interactions with the light quarks due to their $(-+)$ BCs. 
Therefore, at leading order they do not mediate nucleon-DM scattering. 
However, we do have the sub-leading DM couplings to the Higgs and $Z$ from Eq.~\eqref{eq:Higgs_dm} and Eq.~\eqref{eq:total_smz_coupling} respectively. 
Therefore, we obtain the direct detection scattering amplitudes shown in Fig.~\ref{fig:lzp_dd}.

The SI direct detection amplitude takes the following form:
\begin{align}
\mathcal{M^{\rm SI}_{\rm DD}}\propto \frac{g^{\chi,V}_Z g^{N,V}_Z}{M_Z^2} - \frac{y_Ny_\chi }{M_h^2}\,,\qquad g^{N,V}_{Z}=\frac{Zf_p+(A-Z)f_n}{A}\,,
\label{eq:dd_lzp}
\end{align}
where, $V$ denotes the vectorial coupling of $Z$ and $f_p=2q^{u,V}_{Z}+q^{d,V}_{Z},~f_n=2q^{d,V}_{Z}+q^{u,V}_{Z}$. 
The couplings $y\chi$ and $g^{\chi,V}_Z$ can be obtained from Eq.~\eqref{eq:Higgs_dm} and Eq.~\eqref{eq:total_smz_coupling} respectively. 
Here, $y_N\sim m_{\rm nucleon}/v$ is the coupling of the Higgs boson to a nucleon, and is essentially the coupling of Higgs to gluons. 
In Eq.~\eqref{eq:dd_lzp}, the ratio of the Higgs amplitude to the $Z$ amplitude, can be estimated to be proportional to $m_{t} m_{\chi}/m^2_H$. 
Therefore, for this study, which involves TeV scale DM masses, the Higgs exchange contribution is significant. 
This is in stark contrast with the first rendition of the model in Ref.~\cite{Agashe:2004bm}, which focussed on $m_{\chi}\sim \mathcal{O}(100)\,\rm GeV$, for which the Higgs exchange is subdominant to the $Z$ exchange. 
Upon substituting all the relevant gauge charges in Eq.~\eqref{eq:dd_lzp}, we find that the Higgs exchange contribution has a sign opposite to that of the other contributions. 
This fortuitous (and partial) cancellation leads to a less severe direct detection bound. While this relative sign is indeed a coincidence, it should be noted that it cannot be changed as long as $SO(10)$ is respected by the model. 
This is because  Eq.~\eqref{eq:nul_nur_Higgs} ties $y_\chi$ to the Yukawa coupling of the top quark via $SO(10)$ invariance in the bulk and on the IR brane. 
The possibility of changing $y_\chi$, by breaking $SO(10)$ on the IR brane, is explored in later sections where we consider variations of the original model.

\textbf{Dipole interactions of DM:} Due to its couplings to charged states such as $e',t_R$ and $X_s$, $\chi$ may have magnetic as well as electric dipole moments. 
If a CP symmetry is assumed for the UV theory, the electric dipole moment vanishes. 
However, we make no such assumptions here, and we calculate the DM electric dipole moment assuming $\mathcal{O}(1)$ CP violating phases in the UV model. 
The exact expressions for these dipole moments are given in App.~\ref{app:ew_obs}. 
The DM magnetic dipole moment, $\mu_\chi$, can interact with the magnetic dipole moment of the nucleons, $d_N$, leading to spin-dependent DM-nucleon scattering with a cross section: $\sigma^{\rm SD}_{\rm dipole}\propto \mu_N^2 \mu^2_\chi d^2_N$. 
The electric dipole moment, $d_\chi$ can interact with the charge of the nucleons, leading to spin-independent DM-nucleon scattering $\sigma^{\rm SI}_{\rm dipole}\propto \mu_N^2 d^2_\chi Q^2_N$. 
Thus, direct detection bounds can constrain these dipole moments~\cite{Pospelov:2000bq}.

Overall, for the best case, we can get the correct relic density for the DM for a KK scale $m_{\rm KK }\simeq 17$\,TeV with a DM mass tuning  of roughly $40\%$. 
The fact that this tuning is milder than that in the weakly coupled $Z'$ model of Sec.~\ref{sec:IRZprime}, can be parametrically justified as follows. 
The annihilation and  direct detection cross sections are given parametrically by
\begin{align}
\sigma_{\rm annih}\simeq \frac{g^4_*}{M^2_{Z'}\Delta^2},\qquad
\sigma_{\rm DD}\simeq \frac{g^4_*}{M^4_{Z'}}\mu_N^2,
\end{align}
giving the ratio ${\sigma^2_{\rm annih}}/{\sigma_{\rm DD}}\propto \left({g_*}/{\Delta}\right)^4$. This ratio is fixed, meaning we get a larger $\Delta$ for a stronger coupling.

{\bf Miscellaneous bounds:} In addition to the bounds above, there are several bounds from the collider searches of various KK particles. 
For instance, collider searches for KK $W/Z$ and KK gravitons, put a limit of $m_{\rm KK}\gtrsim \mathcal{O}(5)$\,TeV. 
However, as these limits are weaker than the bounds presented here, they are not shown in our plots. 
Furthermore, as discussed in Sec.~\ref{sec:warped_susy}, without additional flavour and CP symmetries, there are stringent bounds $m_{\rm KK} \gtrsim \mathcal{O}(1000)$\,TeV due to the electron EDM. 
Thus, we assume that the model has these symmetries.

\begin{figure}[t]
    \centering
\includegraphics[width=0.75\linewidth]{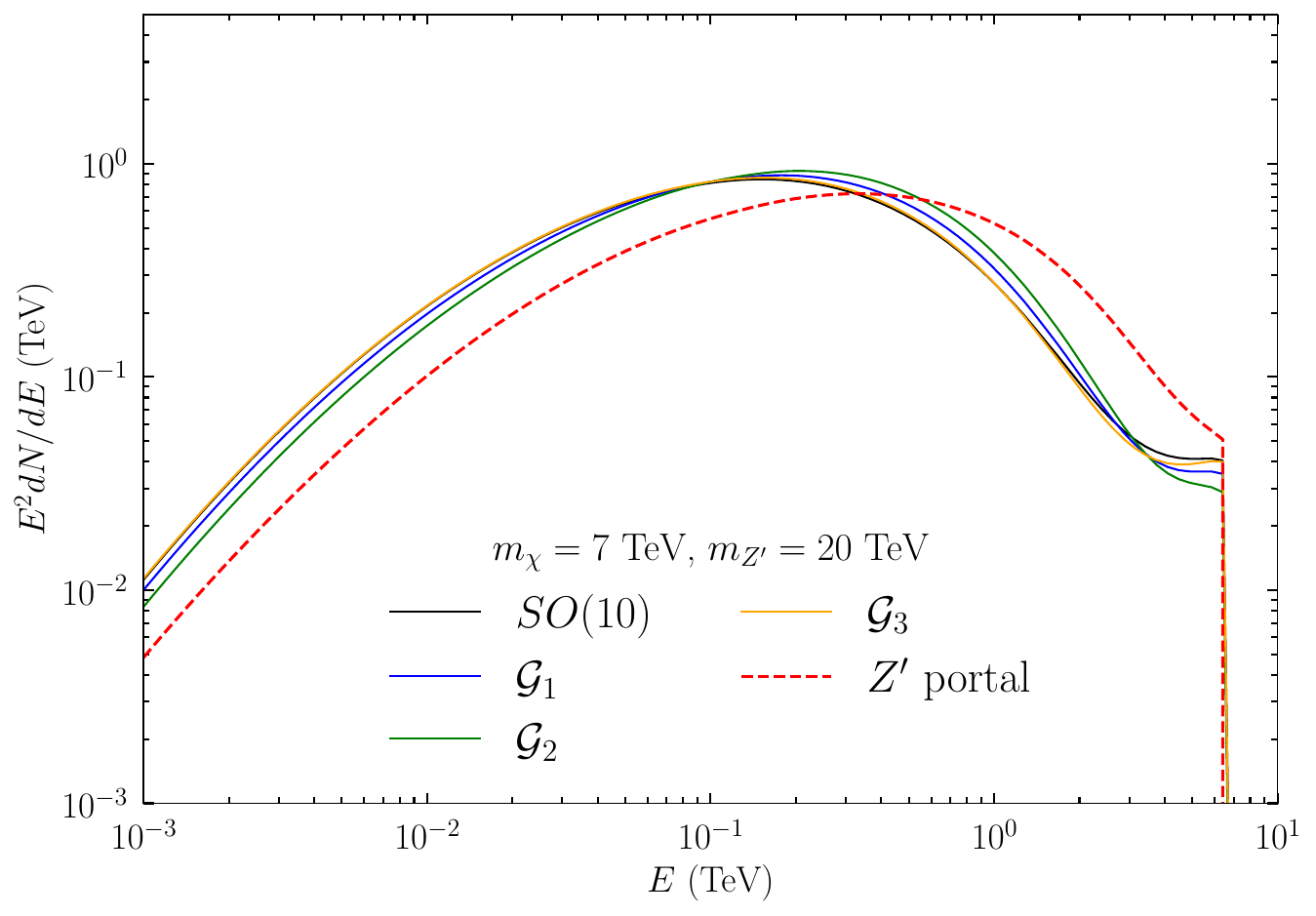}
    \caption{Photon spectra from DM annihilation for various WIMP DM + $Z'$ models. The $Z'$ portal is a benchmark for a generic bottom-up model, while the $SO(10)$, $\mathcal{G}_1$, $\mathcal{G}_2$, and $\mathcal{G}_3$ models are top-down models, motivated by GUTs and extra dimensions. For all the spectra, the mass of the DM and the $Z'$ are fixed to the values shown in the figure. For the extra-dimensional models, $m_{Z'}$ is the KK scale.}
    \label{fig:CTAO_spectra_all_models}
\end{figure}

The model above provides a UV completion for the top-philic DM and top-philic $Z'$ models, whose phenomenology has been extensively studied~\cite{PhysRevD.97.075002,Cox:2015afa,Jackson:2013tca}. 
A natural worry is the scope of discovery for such a model, especially since collider searches are blind to gauge bosons like $X_s, Z'$ and $\mathcal{B}$, due to their small couplings to the light quarks and leptons. 
However, for the parameter space on the edge of direct detection in Fig.~\ref{fig:results_so10_unbroken}, future improvements in various direct detection experiments can readily lead to a discovery. 
On the other hand, for high KK scales $m_{\rm KK}\gtrsim 25$\,TeV, indirect detection proves to be a promising avenue for discovery. 
From Fig. \ref{fig:results_so10_unbroken}, we see that the parameter space giving the correct relic density is well within the reach of CTAO. We have chosen not to show results in Fig.~\ref{fig:results_so10_unbroken} where $m_{\rm KK}$ becomes extremely large.
Nevertheless, the CTAO reach extends up to $\mathcal{O}(100)$\,TeV and can continue to probe the thermal mass in scenarios that would realise such a high KK scale. For this model, we show the expected photon spectrum from the annihilation of DM, which can be observed by CTAO, in Fig.~\ref{fig:CTAO_spectra_all_models} with a solid black line. 
As the DM annihilates into heavy SM states, like the top quark, the photon spectrum peaks at an $\mathcal{O}$(1) lower energy when compared to the $Z'$-portal DM from Sec.~\ref{sec:IRZprime}.

We thus conclude the analysis of the original model by noting that warped $SO(10)$ with split GUT multiplets gives us a viable DM candidate, with complementary direct and indirect detection signatures. 
This extensive phenomenology, along with various theoretical considerations, compels us to consider variations of the original model. 
In the next few subsections, we detail the model building and the associated phenomenology of these variations.
%

\subsection{Breaking $SO(10)$ on the IR Brane}\label{sec:quark_mass}

Motivated to get variations of the original model, in this section, we discuss the key aspects of the required model building.
As seen before, a central ingredient of the model building process is the BCs of KK fields. 
Therefore, as shown in this section, a simple way to obtain different variations of the original model is to modify BCs by breaking $SO(10)$ on the IR brane.

To study the most general warped GUT setup (and the associated DM phenomenology), we ought to explore all allowed variations and breaking patterns of $SO(10)$ on the IR brane. 
In addition to being more comprehensive, there are three main reasons to consider these variations: \\ 
\textbf{(i) High KK scale:} One of the most important effects of breaking $SO(10)$ on the IR brane is custodial isospin violation, due to the breaking of the $SU(2)_R$ subgroup. 
Due to stringent experimental constraints on the $\rho$ parameter, this can result in bounds of $\mathcal{O}(10)$\,TeV on the KK scale. 
In the previous renditions of the original model~\cite{Agashe:2004bm}, the bound on the KK scale from direct detection was only $\mathcal{O}(1)\,\TeV$, which motivated custodial isospin, and therefore $SO(10)$ invariance on the IR brane. 
However, as we have seen from the previous section, the bound from direct detection is now $\mathcal{O}(10)$\,TeV, affording us the possibility of breaking $SO(10)$.\\
\textbf{(ii) Maintaining bulk $SO(10)$ invariance:} As discussed in Sec.~\ref{subsec:breaking_so10_u1b}, bulk breaking mass terms are required to suppress co-annihilation of the GUT partners of our DM. 
As a result, due to other bulk $SO(10)$ breaking terms that could also be present in general, the theory is less controlled. 
Therefore, keeping $SO(10)$ unbroken in the bulk, we can instead break $SO(10)$ on the IR brane, which allows us to add IR brane-localised kinetic terms (BLKTs) for the LH chirality of fermions as follows
\begin{equation}
    \mathcal{L_{\rm IR}} \supset \begin{cases}
        \sqrt{-g}~ \delta ( y- \pi R) b_{\rm IR} k^{-1} i\bar{\Psi}\Gamma^\mu P_L D_\mu \Psi~,\Psi=t^c_R,\chi ~&\text{if} ~SU(4)_c ~\text{on IR brane,}  \\
\sqrt{-g}~ \delta ( y- \pi R) b_{\rm IR} k^{-1} i\bar{\Psi}\Gamma^\mu P_L D_\mu \Psi~,\Psi= \chi ~&\text{if} ~\cancel{SU(4)_c} ~\text{on IR brane.}
    \end{cases}
\end{equation}
These terms can decrease the mass of $\chi$ as follows
\begin{equation}
    m_{\chi} \simeq \begin{cases}
    \left[\frac{\pi}{2}(1+c)-\tan^{-1}(\frac{\pi}{2}(1+c) b_{\rm IR})\right]ke^{-k\pi R},~~~~& ~c\gtrsim -1/4,\\
    \frac{1}{{\sqrt{1+2b_{\rm IR}}}}2\sqrt{c+\frac{1}{2}}~ke^{-k\pi R}, ~~&-1/2 \lesssim c \lesssim -1/4 \,.
    \end{cases}
\label{eq:mp_fermion_mass_blkt}
\end{equation}
Thus, $b_{\rm IR} \sim \mathcal{O}(0.1)$ gives us the required $10\%$ difference to suppress co-annihilation.\\
\textbf{(iii) Validity of the 5D EFT:} The 5D gauge theory is non-renormalisable due to the negative mass dimension of the 5D gauge coupling $g_5$~\cite{Sundrum:2005jf}. 
The loop corrections to $g_5$ are linearly divergent, and are given by
\begin{equation}
    \frac{1}{g_{5}^2}= \frac{1}{g_{5,\,\rm tree}^2}+\frac{\Lambda_{\rm UV}}{16\pi^2}  \times \frac{2}{3} \times 2 \times N_f \times C (\textbf{16})-\frac{\Lambda'_{\rm UV}}{16\pi^2} \times \frac{11}{3} \times C_2 (\rm adj)\,,
\end{equation}
where the second and third terms on the right hand side come from fermion and gauge loops respectively. 
Here, $C(\bf{16})=2$ is the Dynkin index of the $\bf{16}$ of $SO(10)$, $C_2(\rm adj)=8$ is the quadratic Casimir of the adjoint of $SO(10)$, and $N_f$ is the number of fermion multiplets. 
The gauge and fermion loops have different UV regulators in general. 
Since the divergence is physical, this regulator dependence cannot be removed and it is difficult to determine if the two loops cancel each other, and if so, to what degree. 
Thus, being conservative, we demand for the validity of the EFT,
\begin{equation}
    \frac{\Lambda'_{\rm UV}}{16\pi^2}  \frac{88}{3} \lesssim \frac{1}{g_{5}^2}\,,~~~\frac{\Lambda_{\rm UV}}{16\pi^2} \frac{8}{3} N_f \lesssim \frac{1}{g_{5}^2}\,. 
\end{equation}
We have $N_f=18$ for $SO(10)$ unbroken on the IR brane. 
Substituting $g_5 =g_{10}\sqrt{\pi R}=g_W\sqrt{\pi R}$, gives $\Lambda'_{\rm UV} \simeq 0.4 k$ and $\Lambda_{\rm UV} \simeq 0.24 k$. 
The ideal scenario would be $\Lambda_{\rm UV}\gg k$, so that the warped IR cut-off $\Lambda_{\rm IR} \gg m_{\rm KK}$. 
This condition makes processes involving KK gauge bosons calculable in the EFT. 
Unfortunately, both $\Lambda_{\rm UV}$ and $\Lambda'_{\rm UV}$ are far from satisfying that condition.

A straightforward way to increase the UV cutoff, is to lower $g_5$. 
However, this lowers the DM gauge coupling, making it difficult to obtain its annihilation cross section. 
Therefore, we keep $g_5$ unchanged. 
The gauge cutoff $\Lambda'_{\rm UV}$ is fixed once $g_5$ and the GUT gauge group are fixed, and therefore, it cannot be altered. 
However, the fermionic cut-off $\Lambda_{\rm UV}$ can be increased by reducing $N_f$.

The reduction in $N_f$ can be achieved by putting multiple SM fields in the same multiplet. 
However, it should be done in a way that preserves $U(1)_B$. 
Keeping this in mind, we unify the SM lepton fields $L,e_R$ into one multiplet, while the right-handed neutrino $\nu_R$ is still kept in a separate multiplet (to obtain the correct neutrino masses). 
Similarly, we can unify $(u,d)_R$ into one multiplet. This reduces $N_f$ from 18 to 12, giving $\Lambda_{\rm UV}=0.36k$. 
As we see in the next section, this reduction in $N_f$ should be followed by breaking $SU(2)_R$ on the IR brane to get the correct quark masses.

We note that while breaking $SO(10)$ on the IR brane does help mitigate the damage from the EFT invalidity, we still have $\Lambda_{\rm UV}< k$. 
Thus, the 5D EFT remains, in principle, incalculable. Hence, the numbers presented for all the models, such as the cross sections, mass tunings, etc., should be treated as having $O(1)$ uncertainties. 

\subsection{Quark Masses, $SO(10)$ Breaking Routes and DM Couplings}\label{subsec:quark_masses}

As discussed in the previous section, the number of $SO(10)$ GUT multiplets needs to be reduced. 
However, it should be done in a way that preserves $U(1)_B$, that is, the baryon number for all the species in a multiplet is still the same. 
To start with, we consider putting the SM up and down type right handed fields in one multiplet. 
This implies quarks in the same multiplet have identical $c$ parameters, which implies the following incorrect mass relations: $m_u=m_d$, $m_c=m_s$, and $m_t=m_b$. 
To get the correct quark masses, we need additional ingredients. To begin, the $SU(2)_R$ symmetry needs to be broken on the IR brane to allow for different up and down type 5D Yukawas. 
The mass ratios are given by
\begin{align}
\frac{m_u}{m_d}=\frac{y_{5u}}{y_{5d}},\qquad
\frac{m_c}{m_s}=\frac{y_{5c}}{y_{5s}},\qquad
\frac{m_t}{m_b}=\frac{y_{5t}}{y_{5b}}\,,
\end{align}
implying that the right quark masses are obtained by choosing $y_{5d}\simeq 2y_{5u}$, $y_{5c}\simeq 11y_{5s}$, and $y_{5t}\simeq 60y_{5b}$. 
These couplings are quite hierarchical, especially for the top and bottom Yukawas. 
To reduce this hierarchy, we need two additional ingredients. 
First, we postulate a two Higgs doublet model, with the up and down Higgs vevs related by $v_u/v_d=\tan \beta$. 
Second, we take a UV BLKT for the RH up quark,
\begin{equation}
\mathcal{L} \supset \ k^{-1}\delta(y)b_{\rm UV} {u_R^c}^\dagger i\bar{\sigma}^\mu \partial_\mu u_R^c\,.
\label{eq:d_b_blkt}
\end{equation}
We thus have the following modified quark mass relations
\begin{equation}
\frac{m_u}{m_d}=\frac{y_{5u}}{y_{5d}} \frac{\tan \beta}{\sqrt{b_{\rm UV}(2c_{u_R}-1)+1}},\qquad
\frac{m_c}{m_s}=\frac{y_{5c}}{y_{5s}} \tan \beta, \qquad
\frac{m_t}{m_b}=\frac{y_{5t}}{y_{5b}}  \tan \beta.
\end{equation}
Taking $\tan \beta = 8$, the 5D Yukawa relations $y_{5d}\simeq 8y_{5u}$, $y_{5c}\simeq1.4y_{5s}$, and $y_{5t}\simeq8y_{5b}$ along with $b_{\rm UV} \sim  \mathcal{O}(10)$ gives the correct quark masses.

With the issue of quark masses resolved, we now discuss the different ways of breaking $SO(10)$ on the IR brane, while keeping it unbroken in the bulk. 
The breaking of $SO(10)$ on the UV brane is the same as discussed in Sec.~\ref{sec:so10_specifics}.
We consider three groups to which  $SO(10)$ can be broken:\\  
{\bf (i) Variant I: }$SO(10) \to \mathcal{G}_1=SU(4)_c \times U(1)_R \times SU(2)_L$. 
This can be achieved by assigning $W^{\pm}_R$ $(--)$ BCs while $X_s$ and $Z'$ retain $(-+)$ BCs.\\
{\bf (ii) Variant II: }$SO(10) \to \mathcal{G}_2 =SU(3)_c \times U(1)_{B-L} \times U(1)_R \times SU(2)_L$. 
In this case, $W_R, X_s$ get $(--)$ BCs while $Z'$ retains $(-+)$ BCs. \\
{\bf (iii) Variant III: }$SO(10) \to \mathcal{G}_3 =SU(3)_c \times U(1)_Y \times SU(2)_L$. 
This is achieved by further breaking $\mathcal{G}_2$ via an IR brane-localised Higgs that is a $\bf{(4,2,1)}$ of the Pati-Salam group. 
In this case, $W_R, X_s$ and $Z'$ all get $(--)$ BCs.

In all three cases, the baryon number gauge boson, $\mathcal{B}_\mu$, has $(-+)$ BCs, and the SM group as well as $U(1)_B$ is unbroken on the IR brane. 
Breaking $SO(10)$ on the IR brane can alter couplings of DM to $X_s,W^{\pm}_R$ and $Z'$ by $O(1)$ factors. 
Furthermore, a $(--)$ KK mode is heavier than a $(-+)$ KK mode. 
Denoting the mass of a $(--)$ gauge KK mode by $m_{(--)}$, we have the relation
\begin{align}
m_{(--)}\simeq 1.6\, m_{\rm KK}.
\label{eq:m_mm}
\end{align}
A derivation of this relation is given in App.~\ref{app:5D_specifics}. 
These changes in the couplings, and the masses of KK gauge bosons, can change the sub-leading effects discussed in Sec.~\ref{subsec:couplings_of_dm} as follows:\\
\textbf{(i) Coupling of DM to SM Higgs:~} If $SO(10)$ is broken to $\mathcal{G}_1$, $SU(4)_c$ remains unbroken which requires the 5D Yukawa of the top and the 5D Yukawa of Eq.~\eqref{eq:nul_nur_Higgs} to be the same. 
Thus, the mass mixing between $\neu$ and $\chi$ and the Higgs coupling to DM remains unchanged. 
However, if $SO(10)$ is broken to $\mathcal{G}_2$ or $\mathcal{G}_3$, additional brane-localised Yukawas of the form
\begin{equation}
    \mathcal{L}_{\rm IR} \supset \sqrt{-g}~ \delta(y-\pi R)~y_{\rm IR} k^{-1} H\ell' \chi + \mathrm{h.c.}\,,
\end{equation}
are allowed, where $y_{\rm IR}$ is a dimensionless parameter. 
This additional Yukawa coupling can change the Higgs coupling to DM. 
In other words, we may choose $y_{\rm IR}$ such that $y_A$ in Eq.~\eqref{eq:Yukawas} changes as $y_A \to \varepsilon y_A$. 
From Eq.~\eqref{eq:Higgs_dm}, in the limit $y_A\gg y_B$, we see that this suppression of $\varepsilon$ in $y_A$, results in a suppression of $\varepsilon^2$ of the Higgs DM coupling. 
For the purpose of illustration, we choose $\varepsilon = 1/2$ for our analysis.
\\
\textbf{(ii) Coupling of DM to SM $Z$ via $\neu-\chi$ mixing:~} This coupling also depends on the mass mixing between $\neu$ and $\chi$. 
Therefore, it remains unchanged when $SO(10)$ is broken to $\mathcal{G}_1$ and suppressed otherwise. 
From Eq.~\eqref{eq:g_nul_nur}, we see that both $g^{\chi_L}_{Z,\rm mix}$ and $g^{\chi_R}_{Z,\rm mix}$ are proportional to $y^2_A$, which means these couplings are suppressed by a factor of $\varepsilon^2$ when $y_A$ is suppressed by a factor of $\varepsilon$ as discussed above.\\
\textbf{(iii) Coupling of DM to SM $Z$ via $Z-Z'$ mixing:~} This coupling, as shown in Eq.~\eqref{eq: g_z_zprime_mixing},  only depends on the coupling of $Z'$ to the Higgs. 
Thus, it remains unchanged for Variants I and II, as $Z'$ retains its $(+)$ BC on the IR brane. 
For Variant III, we expect this coupling to be suppressed  by $O(1)(m_{\rm KK}/m_{(--)})^2 \simeq 1/3$, where the $\mathcal{O}(1)$ factor comes from the changing $Z'$ profile.

For the reader's convenience, we summarise all the couplings of the DM for the three variants in Tab.~\ref{tab:dm_couplings}.

\begingroup
\renewcommand{\arraystretch}{1.8}
\begin{table}[h]
\centering
\begin{tabular}{|l|
  >{\centering\arraybackslash}p{2.4cm}|
  >{\centering\arraybackslash}p{2.4cm}|
  >{\centering\arraybackslash}p{2.4cm}|}
\hline
\multicolumn{4}{|c|}{\textbf{Leading order Couplings}} \\ 
\hline
\multicolumn{1}{|c|}{\textbf{Coupling}} & 
\multicolumn{3}{c|}{\textbf{Gauge group on IR brane}} \\ 
\cline{2-4}
 & $\mathcal{G}_1 \times U(1)_B$ & $\mathcal{G}_2 \times U(1)_B$ & $\mathcal{G}_3 \times U(1)_B$ \\
\hline
$\bar{\chi} \gamma^\mu P_{L,R} ~\chi Z'_\mu $ &  $g_* \sqrt{5/8}~a_{\chi_L,\chi_R}$ & $g_* \sqrt{5/8}~a_{\chi_L,\chi_R}$ & $g_* \sqrt{5/8}~ b_{\chi_L,\chi_R}$ \\
$\bar{\chi} \gamma^\mu P_{L,R}~ \chi B_\mu $ &  $g_*/3 ~a_{\chi_L,\chi_R}$ & $g_* /3~a_{\chi_L,\chi_R}$ & $g_*/3~b_{\chi_L,\chi_R}$ \\
${t^c_R}^{\dagger} \gamma^\mu  P_L\chi X_{s,\mu} $ &  $g_* \sqrt{1/2} ~a_{X_s}$ & $g_* \sqrt{1/2} ~b_{X_s}$& $g_* \sqrt{1/2} ~b_{X_s}$\\
${t^c_R}^{\dagger} \bar{\sigma^\mu} t^c_R Z'_{\mu} $ &  $g_* \sqrt{1/40} ~a_{t_R}$ & $g_* \sqrt{1/40} ~a_{t_R}$& $g_* \sqrt{1/40} ~b_{t_R}$\\
${b^c_R}^{\dagger} \bar{\sigma^\mu}  b^c_R Z'_{\mu} $ &  $-g_* \sqrt{9/40} ~a_{t_R}$ & $-g_* \sqrt{9/40} ~a_{t_R}$& $-g_* \sqrt{9/40} ~b_{t_R}$\\
${t^c_R}^{\dagger} \bar{\sigma^\mu} t^c_R  \mathcal{B}_{\mu} $  and ${b^c_R}^{\dagger} \bar{\sigma^\mu} b^c_R \mathcal{B}_{\mu} $&  $g_* /3 ~a_{t_R}$ & $g_* /3 ~a_{t_R}$& $g_* /3 ~a_{t_R}$\\
$(D^\mu H ^\dagger H-H^\dagger D^\mu H ) Z_\mu'  $ &  $g_* /\sqrt{10} ~a_H$ &$g_* /\sqrt{10} ~a_H$& $g_* /\sqrt{10} ~b_{H}$\\
\hline
\multicolumn{4}{|c|}{Subleading Couplings} \\ 
\hline
Gauge group on IR brane & \multicolumn{2}{|c|} {$ \bar{\chi} \gamma^\mu \chi Z_\mu$}& $\bar{\chi}\chi h$\\
\hline
$\mathcal{G}_1 \times U(1)_B$ & \multicolumn{2}{|c|} { $g_Z \left(\frac{g^2_* v^2}{m_{\rm KK}^2}+\frac{y^2_* v^2}{m_{\rm KK}^2}\right) $ }&  $y_* \frac{ y_*v}{m_{\rm KK}} $ \\
$\mathcal{G}_2 \times U(1)_B$ & \multicolumn{2}{|c|} { $g_Z \left(\frac{g^2_* v^2}{m_{\rm KK}^2}+\varepsilon^2\frac{y^2_* v^2}{m_{\rm KK}^2}\right) $ } & $\varepsilon^2 y_* \frac{ y_*v}{m_{\rm KK}} $\\
$\mathcal{G}_3 \times U(1)_B$ & \multicolumn{2}{|c|}  { $g_Z \left(b_\chi b_H\frac{g^2_* v^2}{m_{(--)}^2}+\varepsilon^2\frac{y^2_* v^2}{m_{\rm KK}^2}\right) $ } & $\varepsilon^2 y_* \frac{ y_*v}{m_{\rm KK}} $\\
\hline
\end{tabular}
\caption{Couplings for the dark matter, $\chi$, with $SO(10)$ broken on IR brane. The gauge groups are $\mathcal{G}_1=SU(4)_c \times U(1)_R \times SU(2)_L$ , $\mathcal{G}_2=SU(3)_c \times U(1)_{B-L} \times U(1)_R \times SU(2)_L$  and $\mathcal{G}_3=SU(3)_c \times U(1)_Y \times SU(2)_L$ . Here, $g_{*}=g_{10}\sqrt{k\pi R}$, $y_* \sim 2 y_t$ and $m_{(--)}\simeq 1.6m_{\rm KK}$. The parameter $\epsilon \sim 1/2$ is the suppression to the Yukawa coupling $y_A$ in Eq.~\eqref{eq:Yukawas} when $SU(4)_c$ is broken on the IR brane. 
For the couplings in this table, we assume (parametrically) $y_A\gg y_B$. 
The $a$'s and $b$'s are $\mathcal{O}(1)$ numbers coming from the overlap of 5D profiles of the relevant fields. 
The entries for the sub-leading couplings are all parametric estimates, with exact values reserved for the final phenomenological discussion.}
\label{tab:dm_couplings}
\end{table}

\endgroup

\subsection{Results for $SO(10)$ Broken on IR Brane}\label{subsec: results_broken_so10}

In this section, we discuss the DM phenomenology for the three variants of the original model that were described in the previous section. 
For all the cases, we fix the GUT coupling $g_{10}=g_W$ and $c_{t_L}=0.3$. 
Furthermore, for all the variants discussed below, $U(1)_B$ is broken only on the UV brane. This means the BCs of $\mathcal{B} $ remain $(-+)$.

\begin{figure}[h!]
   \centering
\includegraphics[width=0.9\linewidth]{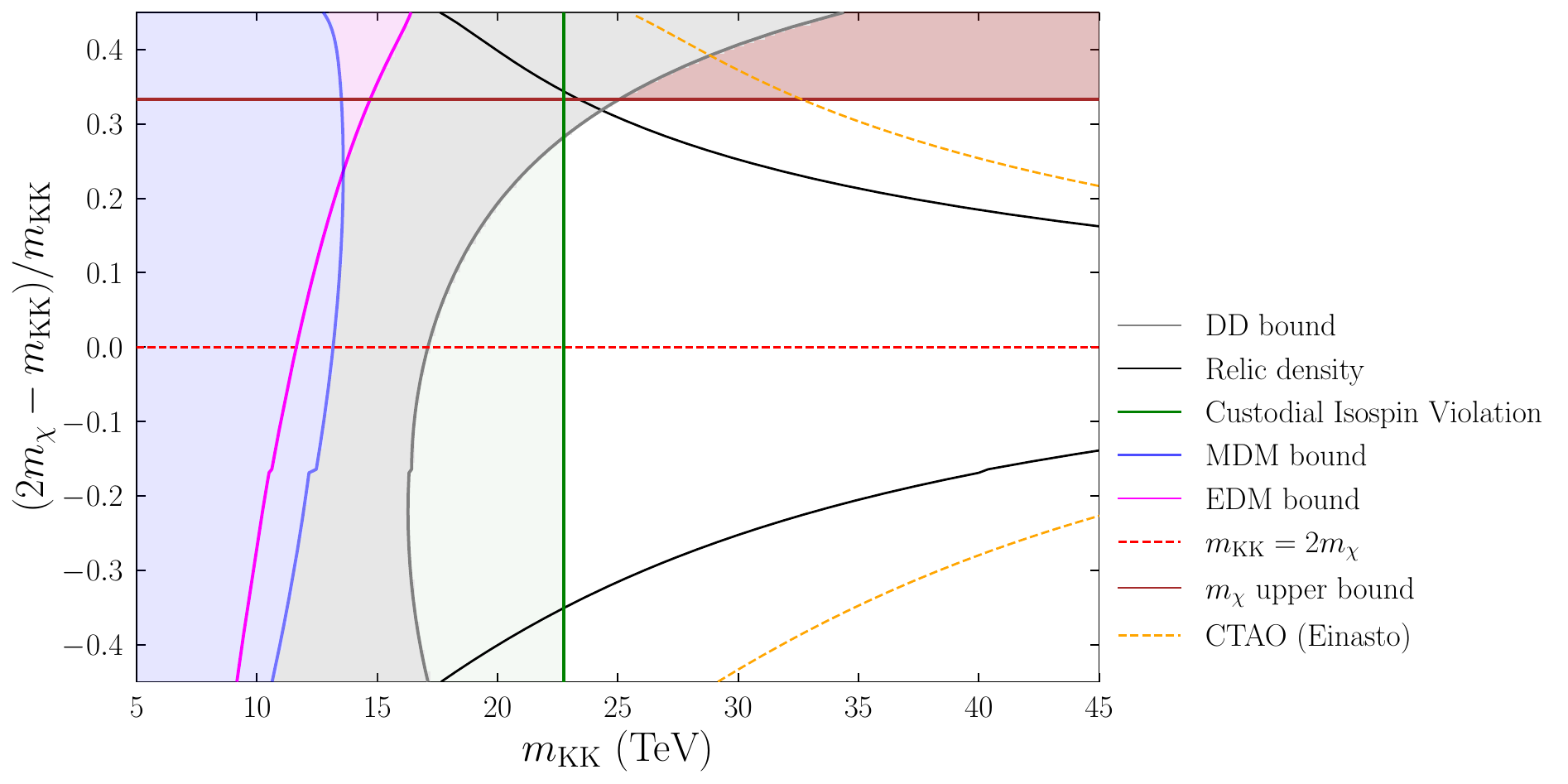}
   \caption{As in Fig.~\ref{fig:results_so10_unbroken}, here for $\mathcal{G}_1=SU(4)_c \times U(1)_R \times SU(2)_L \times U(1)_B$ unbroken on the IR brane.}
   \label{fig:results_mp_all}
\end{figure}

\noindent \textbf{Variant I:} We start with the variant with $SO(10)$ broken to $SU(4)_c \times U(1)_R \times SU(2)_L$ on the IR brane. The constraints on this model are shown in Fig.~\ref{fig:results_mp_all}. We retain all the annihilation channels from the original model. This is because the profiles/BCs of $X_s$, $Z'$, and $\mathcal{B}$ remain unchanged. In addition to these channels, as $b_R^c$ is now in the $\bf{16_{t_R}}$ multiplet, $\chi$ can annihilate into $b_R^c$ final states. 
As the boundary conditions on the gauge bosons and the $\chi$-Higgs coupling are the same as in Sec.~\ref{subsec: results_unbroken_so10}, the direct detection bound is essentially unchanged. 
However, we now face a bound from the $\rho$ parameter due to the breaking of $SU(2)_R$ (and hence custodial isospin) on the IR brane. 
The $\rho$ parameter bound, $m_{\rm KK} \gtrsim 23 \,\TeV$, is the strongest bound for this model. 
A detailed calculation of the $\rho$ parameter can be found in App.~\ref{app:ew_obs}. 
Overall, a mass tuning $\Delta\simeq 35\%$ is required to obtain the required thermal relic density.

\begin{figure}[h!]
   \centering
\includegraphics[width=0.9\linewidth]{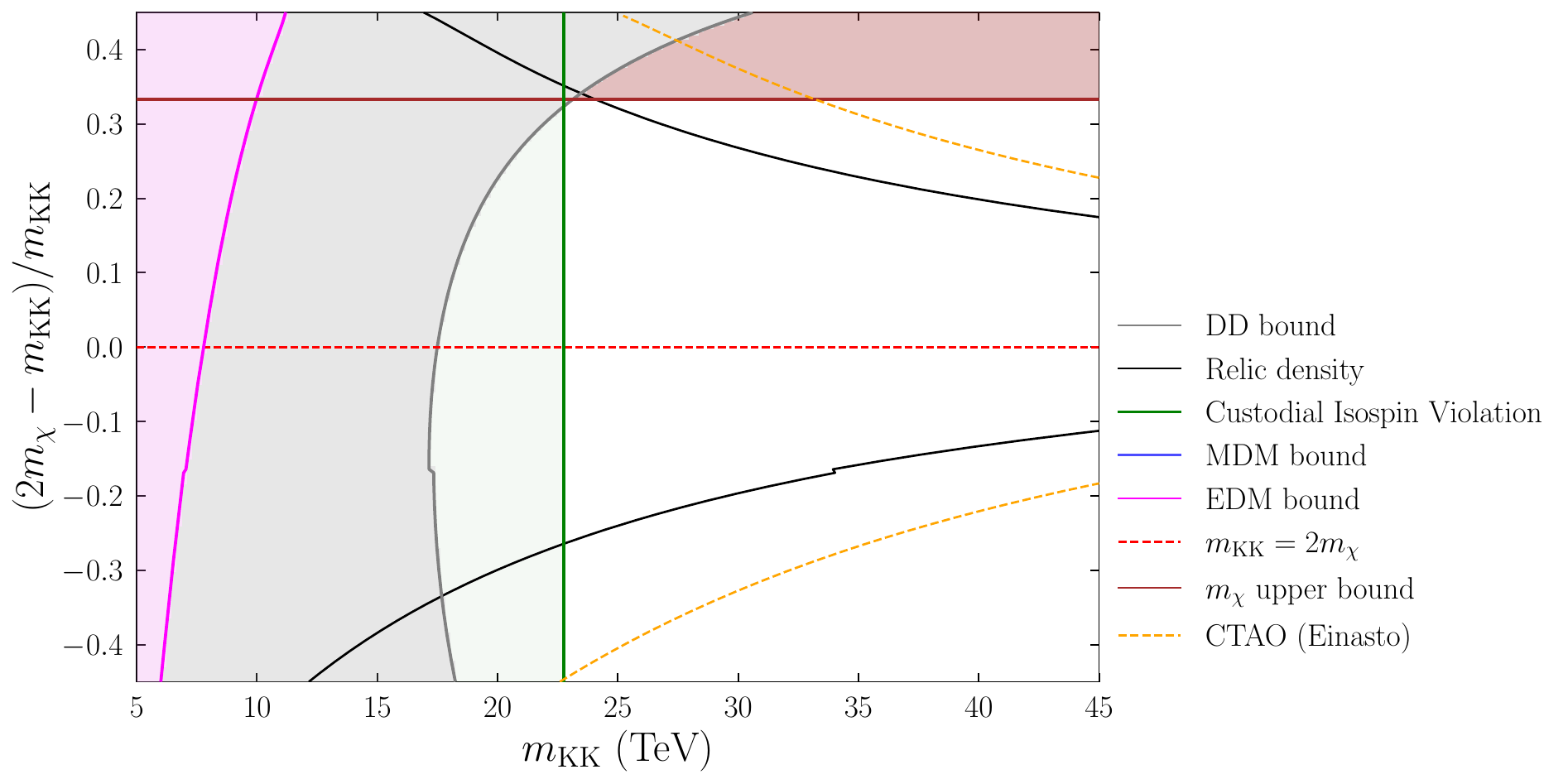}
   \caption{As in Fig.~\ref{fig:results_so10_unbroken}, here for $\mathcal{G}_2=SU(3)_c \times U(1)_{B-L}\times U(1)_R \times SU(2)_L \times U(1)_B$ unbroken on the IR brane.}
   \label{fig:results_mp_zp}
\end{figure}

\noindent \textbf{Variant II:} For this case, $SO(10)$ is broken to $SU(3)_c  \times U(1)_{B-L}\times U(1)_R \times SU(2)_L $ on the IR brane. The constraints on this model are shown in Fig.~\ref{fig:results_mp_zp}. 
Due to $X_s$ having $(--)$ BCs,  the coupling of $\chi$ with $X_s$ is suppressed, and $X_s$ now has a heavier mass $m_{(--)}$ defined in Eq.~\eqref{eq:m_mm}. 
Both the coupling suppression, and the mass increase are $\mathcal{O}$(1) effects. 
Therefore, the $t$-channel annihilation of DM to $t_R^c$ is considerably suppressed. 
Due to the $Z'$ charge of $b_R^c$ being larger than $t_R^c$ ($Q^{Z'}_{b_R}= -3\times Q^{Z'}_{t_R}$), the branching fractions of DM annihilating into $b^c_R$ and $t_R^c$ are now roughly the same. 
The direct detection bounds, from Eq.~\eqref{eq:dd_lzp}, depend on the coupling of $\chi$ to $Z$~($g^\chi_Z$) and the Higgs boson~($y_\chi$). As discussed in Sec.~\ref{subsec:quark_masses}, while the  $\chi$-$Z$ coupling remains unchanged for this variant, the $\chi$-Higgs coupling could be $O(1)$ different, which leads us to modify $y_\chi\to y_{\chi}/2$. 
Even after this modification, we see that the direct detection bound remains sub-leading compared to the $\rho$ parameter bound. 
As the BCs of $SU(2)_R$ gauge bosons are the same as Variant I, we obtain the same $\rho$ parameter bound $m_{\rm KK} \gtrsim 23\,$TeV. 
This finally leads to a tuning of $\Delta\simeq30\%$.

\begin{figure}[h!]
   \centering
\includegraphics[width=0.9\linewidth]{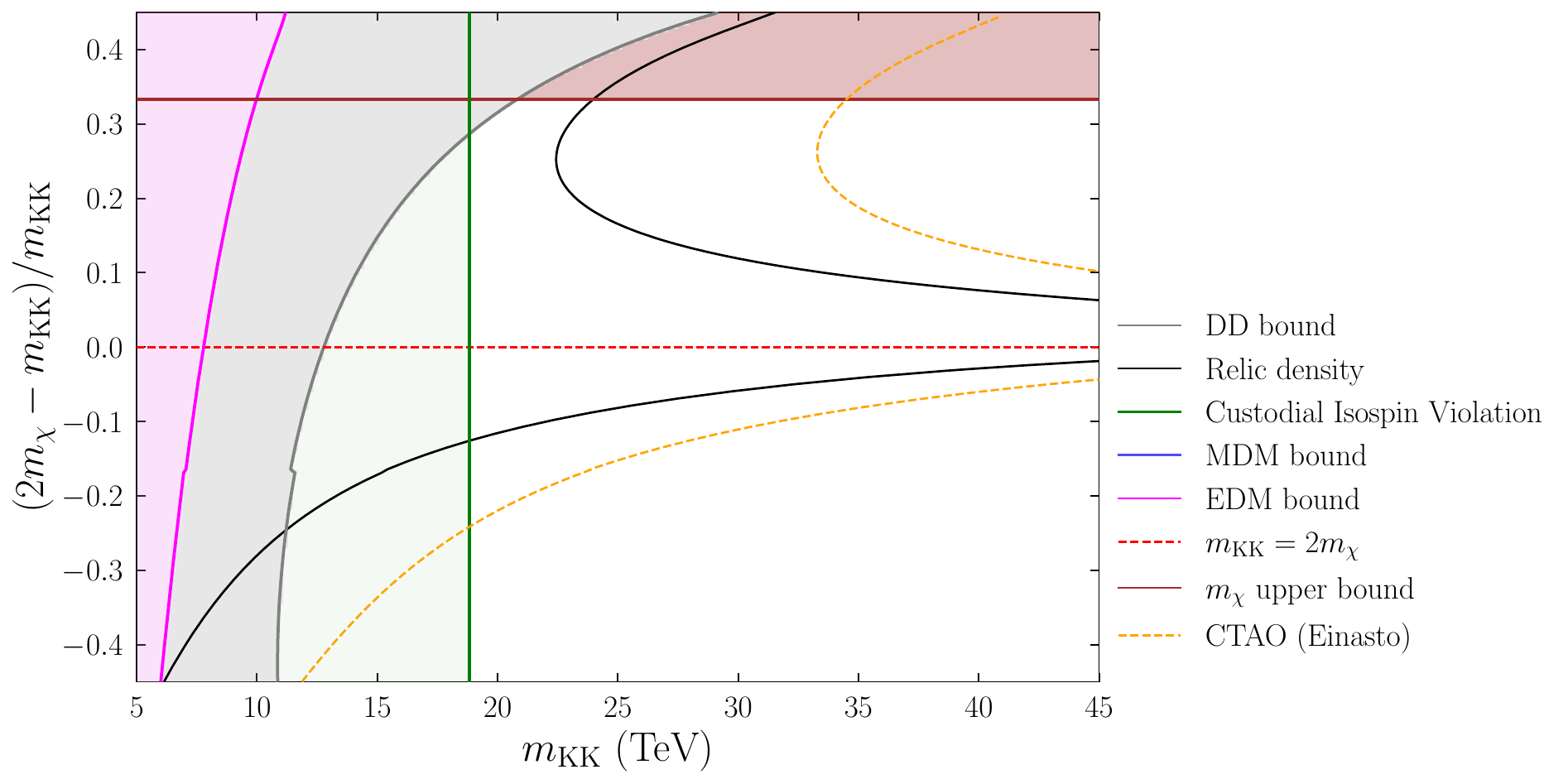}
   \caption{As in Fig.~\ref{fig:results_so10_unbroken}, here for $\mathcal{G}_3=SU(3)_c \times U(1)_Y \times SU(2)_L \times U(1)_B$ unbroken on the IR brane.}
   \label{fig:results_mm}
\end{figure}

\noindent \textbf{Variant III:} Finally, we consider the variant with $SO(10)$ broken to $SU(3)_c \times U(1)_Y \times SU(2)_L$ on the IR brane. 
The constraints on this model are shown in Fig.~\ref{fig:results_mm}. 
Now, both $X_s$ and $Z'$ have $(--)$ BCs, and therefore their couplings to $\chi$ are suppressed by an $\mathcal{O}(1)$ factor. 
Furthermore, both have the heavier mass $m_{(--)}$ instead of $m_{\rm KK}$. 
As a result, the $X_s$ and $Z'$ contributions to the annihilation cross sections are significantly reduced. 
We also find that the bound from direct detection gets even more relaxed as couplings of $\chi$ to both Higgs and $Z$ are suppressed. 
The suppression in the $\chi$-Higgs coupling is the same as Variant II. 
As discussed in Sec.~\ref{subsec:quark_masses}, the decrease in the $\chi$-$Z$ coupling is due to the decrease in the Higgs-$Z'$ coupling, and the increase in the $Z'$ mass. 
These two effects also weaken the $\rho$ parameter bound to $m_{\rm KK} \gtrsim 19\,\TeV$, which yet again, is the strongest bound. 
We thus obtain the required thermal relic density with a tuning $\Delta\simeq 30 \%$.

Akin to the results of Sec.~\ref{subsec: results_unbroken_so10}, the constraints from collider searches remain weaker than the bounds above, and we assume flavour and CP symmetries to protect the model from electron EDM bounds. 
With the $\rho$ parameter being the most stringent bound for all the variants, indirect detection is the leading signal. 
Once again, the points giving the correct DM relic abundance remain well within the reach of CTAO. 
The expected photon spectra from DM annihilation, which can be observed by CTAO, are shown in Fig.~\ref{fig:CTAO_spectra_all_models}. 
While the spectra for these variants are peaked at energies lower than the spectral peak of the $Z'$-portal model from Sec.~\ref{sec:IRZprime}, they are practically indistinguishable from the spectrum of the original model. 
Thus, while the model specifics and existing bounds change, the message for future signals at CTAO remains the same for these strongly coupled WIMP DM models.

\section{Dark Matter from Warped SUSY GUT}
\label{sec:warped_susy}

\begin{figure}[b]
    \centering
    \includegraphics[width=0.9\linewidth]{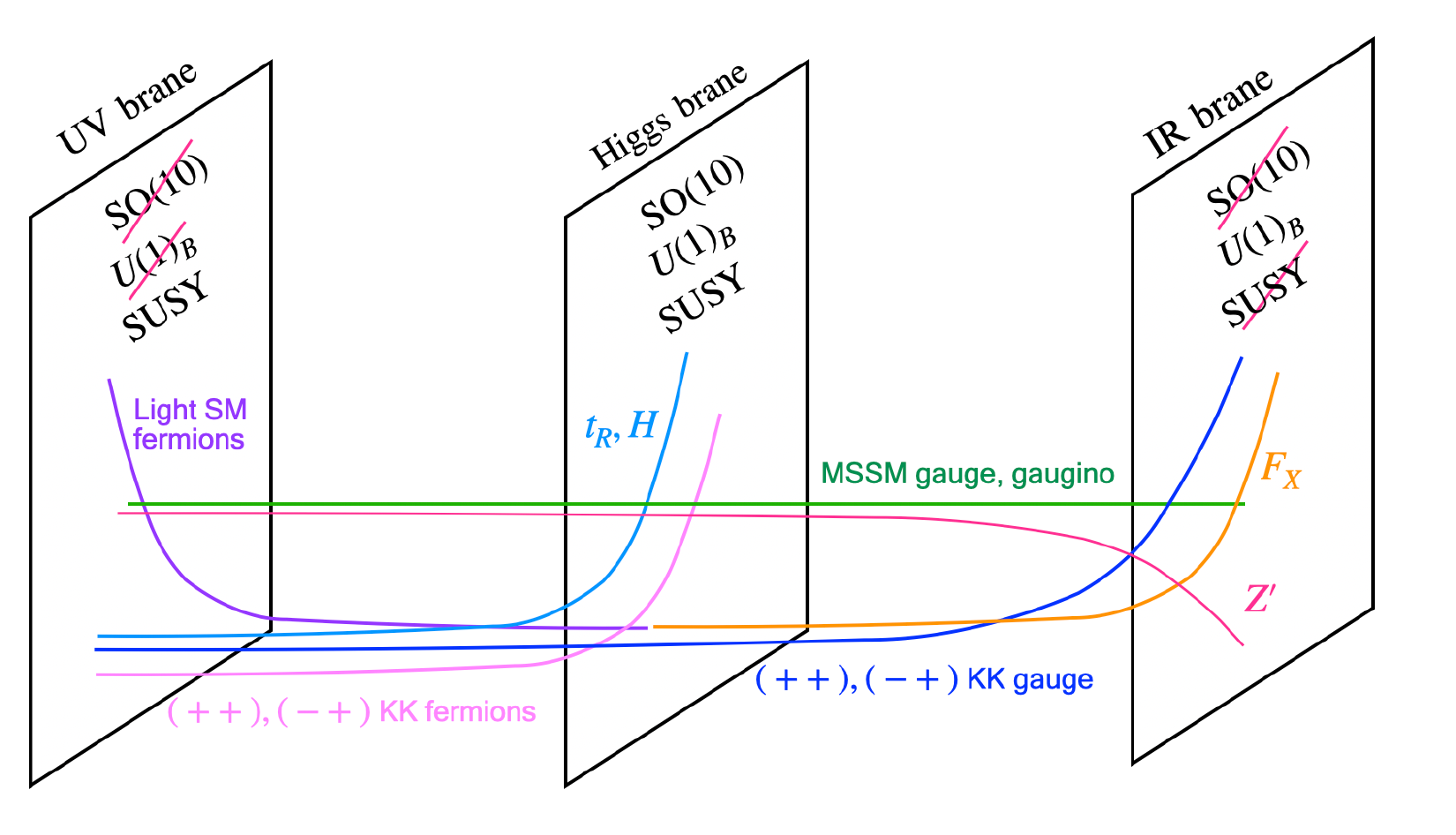}
    \caption{Basic set-up of the warped SUSY $SO(10)$ model. The curves show the 5D profiles of the various fields.}
    \label{fig:susy_setup}
\end{figure}

From the analysis performed in the previous section, it is evident that warped $SO(10)$ provides a plethora of UV completions for DM.
One problem that is common to all the models discussed, however, is that there remains a little hierarchy between the KK scale, which is pushed to $O(15$--$20)$\,TeV by various bounds, and the Higgs mass. 
Furthermore, if we consider flavour and CP violating effects in these models, the problem is even more severe. 
For instance, mixing between KK leptons can induce an electron EDM of
\begin{equation}
    d_e \simeq \frac{g^2_*} {16\pi^2} \frac{m_e}{m_{\rm KK}^2} , 
\label{eq:electron_edm}
\end{equation}
where we assume $\mathcal{O}(1)$ complex phases. 
Requiring this to satisfy the experimental bounds \cite{Roussy:2022cmp,Agashe:2026ggy} pushes $m_{\rm KK} \gtrsim \mathcal{O}(1000)\,\TeV$, giving yet another hierarchy between $m_{\rm KK}$ and the Higgs mass. 
There are other bounds on $m_{\rm KK}$ similar in magnitude from flavour physics, such as $K-\bar{K}$ mixing~\cite{Agashe:2026ggy}. 

Broadly speaking, there are two approaches to ameliorate these issues. 
The first is to impose CP and flavour symmetries. This has been an implicit assumption in all the previous sections, in which we did not discuss bounds from these effects. 
A second approach is to introduce SUSY to the warped model~\cite{Agashe:2026ggy}, and extend the setup to include three branes. 
The three branes are the UV, Higgs and the IR brane, with the SM Higgs peaked on the Higgs brane. We denote the scale of the Higgs brane by $m_{*}$, while the IR brane remains at the scale $m_{\rm KK}$. With this notation, we clarify that the CP and flavour bounds, such as Eq.~\eqref{eq:electron_edm}, constrain $m_{*}$ and not $m_{\rm KK}$. 
This is because the KK modes (such as KK electrons) in the electron EDM diagram have masses of order $m_*$. 
Introducing SUSY brings down the bounds on $m_*$ from flavour and CP violating effects, to $m_* \gtrsim \mathcal{O}(100) \,\TeV$.

In this model, SUSY is broken at a scale lower than $m_*$, and the corrections to the Higgs mass are proportional to the SUSY breaking parameters of the Minimally Supersymmetric Standard Model (MSSM) such as the stop mass, and not $m_*$. 
Therefore, as we assume stop masses of $\mathcal{O}(\TeV)$, the little hierarchy problem is ameliorated.
For breaking SUSY, we consider gaugino mediated supersymmetry breaking (GMSB) which provides flavour universal masses to all squarks and sleptons. 
This keeps any additional bounds from flavour violation under control.

Thus, the setup of Ref.~\cite{Agashe:2026ggy} accommodates low-scale GMSB, a natural Higgs mass, along with stop masses at the edge of discovery. 
Here, as shown in Fig.~\ref{fig:susy_setup}, we study an $SO(10)$ GUT extension of this model. 
Bulk fermions only propagate until the Higgs brane, with light fermions peaked near the UV brane. 
The KK/GUT fermions, including the DM $\chi$, along with $t^c_R$ and the SM Higgs are peaked near the Higgs brane. All gauge fields propagate till the IR brane, meaning the gauge KK modes still have masses of order $m_{\rm KK}$. 
We choose the IR brane scale $m_{\rm KK}\simeq \mathcal{O}(10)$\,TeV, and the Higgs brane scale $m_* \simeq \mathcal{O(}100)$\,TeV. 
As we show in Sec.~\ref{subsec:couplings_susy_model}, this setup can harbour a neutral gauge boson with electroweak strength couplings to both DM and the MSSM. 
Sparticles in this model get their masses primarily from GMSB. As SUSY breaking originates from the IR brane, the $F$-terms, as well as the masses of the GMSB messenger particles are both given by $\sqrt{F_X} \simeq M_{\rm mess} \simeq m_{\rm KK}$. 
The primordial SUSY breaking scale is then given by $\Lambda_{\rm SUSY }=F_X/M_{\rm mess} \simeq m_{\rm KK}$. 
The MSSM gauginos get a mass proportional to $g^2 \Lambda_{\rm SUSY}$, where $g$ is the corresponding gauge coupling. 
The breaking of SUSY on the IR brane is finally communicated to the rest of the MSSM particles via the MSSM gaugino fields, as their profiles span the entire fifth dimension. 
Furthermore, the $F$-term $F_X$ can itself have a profile, peaked on the IR brane. 
As shown in App.~\ref{app:mssm_masses}, this profile combined with GMSB can give squark and slepton masses consistent with LHC bounds. 
In order to obtain these masses, we assume that $SO(10)$ is broken on the final IR brane, and thus, the primordial SUSY breaking scale $\Lambda_{\rm SUSY}$ can be different for the three SM gauge groups. 
We use the masses calculated in App.~\ref{app:mssm_masses} for the rest of our analysis.

Details of this model, such as the validity of the 5D EFT and changes to the gauge coupling unification (with the addition of SUSY), will be presented in a future work~\cite{KAGGRRRS}. 
For the present work, we focus on the basic particle content and couplings of the framework that are relevant for the DM phenomenology. 
In this sense, our approach here is a simplified model of the full warped SUSY GUT scenario.

\subsection{Couplings of Dark Matter}\label{subsec:couplings_susy_model}

The DM candidate $\chi$ in this model is again the fermion $\nu'_R$ with $(-+)$ BCs, from the GUT multiplet of $t_R^c$. 
Since $\chi$ is now peaked near the Higgs brane instead of the IR brane, its couplings to KK gauge bosons change significantly. 
Instead of the enhanced coupling $g_*=g_{10}\sqrt{k\pi R}$ to $(-+)$ gauge bosons, $\chi$ now has a coupling 
\begin{equation}
g_{\rm \chi ,(-+) KK} \simeq g_{10}\sqrt{k\pi R} \left(\frac{m_{\rm KK}}{m_*}\right)^2\!.
\end{equation}
For $m_{\rm KK}\simeq 10 \rm \,TeV$ and $m_* \simeq 100 \rm \,TeV$, couplings of DM to $(-+)$ KK modes are diluted by a factor of 100 when compared to the two brane models in previous sections. 
This results in a highly suppressed annihilation cross section for $\chi$, unless its mass is tuned to be $\mdm = m_{\rm KK}/2$. 

To solve the problem of the overabundance of $\chi$, we consider a new type of BCs for $Z'$, namely $(+-)$, while still retaining $(-+)$ BCs for all other GUT gauge bosons. 
This $Z'$ has an almost flat profile (see Fig.~\ref{fig:susy_setup}). 
The flat profile can be understood by considering a scalar, denoted by $\phi_{\rm IR}$, localised on the final IR brane. 
This scalar obtains a vev and breaks $U(1)_{\zp}$. 
This mimics the $(-)$ BC for $Z'$ on the final IR brane, in the limit $\langle\phi_{\rm IR} \rangle\to \infty$. 
When $\langle\phi_{\rm IR} \rangle\to 0$, the zero mode is massless, and has a flat profile. 
A non-zero $\langle\phi_{\rm IR} \rangle$ leads to a mass-mixing between the zeroth and first KK modes of $Z'$. 
As $\langle\phi_{\rm IR} \rangle\to \infty$, the lowest lying mode still retains the zero mode profile far from the final IR brane. 
However, as the profile must vanish on the final IR brane, we are left with an almost flat profile.

Due to its flat profile between the UV and Higgs branes, $Z'$ couples to a majority of the MSSM. 
Hence, we essentially have a weakly coupled $Z'$-portal model from Sec.~\ref{sec:IRZprime}, but now with additional sfermion and Higgs couplings 
\begin{align}
\mathcal{L_{\rm gauge,Z'}}\supset g_{\zp}Z'_\mu \bigg[Q^{\chi}_{\zp} \bar{\chi}\gamma^\mu \chi 
+\!\sum_{\psi = f, \tilde{H}_{u,d}}\! Q_{\zp}^\psi{\psi}^\dagger\bar{\sigma^\mu} \psi 
+\!\sum_{\phi = \tilde{f}, H_{u,d}} \! Q_{\zp}^\phi ({\phi}^\dagger D^\mu {\phi}-D^\mu {\phi}^\dagger {\phi} )\bigg].
\end{align}
Here, $f$ and $\tilde{f}$ are the MSSM fermion and sfermions, whereas $\tilde{H}_{u,d}$ are the Higgsinos. 
The charges $Q_{\zp}^\phi$ and $Q_{\zp}^\psi$ are equal if $\psi$ and $\phi$ are superpartners. 
The mass, coupling, and charges for the $Z'$ are given by
\begin{equation}
   m_{Z'} \simeq 0.6 \frac{m_{\rm KK}}{\sqrt{k \pi R}},\qquad g_{Z'} =0.8,\qquad Q_{Z'}=T_{3R}-Y\sin^2\swp,
\end{equation}
where the coupling $g_{Z'}$ is chosen to be $0.8$ since it is a free parameter. 
All the heavy KK fermions and sfermions (including the DM) have a vectorial coupling to this $Z'$. 
Furthermore, the almost zero-mode nature of $Z'$ leads to the running of $\sin^2 \swp$, making it deviate from the GUT value $\sin ^2\swp = 3/5$. 
Since the hypercharge coupling $g_Y$ is fixed, the mixing angle is also fixed through the relation $g_Y=g_{Z'} \sin \swp \cos \swp $ to be $\sin^2\theta'=0.28$.

As seen previously, the DM can also have induced couplings to the SM $Z$ and the Higgs boson. 
As shown in Eq.~\eqref{eq:Higgs_dm}, the coupling to the Higgs boson, $y_\chi$ depends on the mass of $\mathcal{N}$ as $y_\chi \propto 1/m_\mathcal{N}$. 
As $\mathcal{N}$ is peaked on the Higgs brane, it has a large mass $m_{\neu}\simeq 100\,$TeV. 
This implies $y_\chi$ is now highly suppressed. 
Similarly, from Eq.~\eqref{eq:g_nul_nur}, the couplings $g^{\chi_R}_{Z,\rm mix},g^{\chi_L}_{Z,\rm mix} $ are also suppressed. 
The coupling due to $Z-Z'$ mixing in Eq.~\eqref{eq: g_z_zprime_mixing} is still sizeable, as the $Z'$ continues to couple with the Higgs, albeit with an electroweak-sized coupling. 
Thus, we obtain the DM couplings to SM $Z$
\begin{align} 
g^{ \chi_R }_{ Z-Z' }=g^{ \chi_L}_{ Z-Z' } \simeq
g_Z \left(\frac{ g_{\zp}^2v^2 \cos^2 \theta^{ \prime } Q^H_{Z'} Q^{\chi}_{ Z^{ \prime } } }{ 4 m^2_{ Z^{ \prime } } }\right)\!.
\label{eq: susy_g_z_zprime_mixing}
\end{align}
%

\subsection{Results for Warped SUSY Model}
\label{subsec:results_susy}
\label{subsec}

\begin{figure}[t!]
\centering
\includegraphics[width=0.95\linewidth]{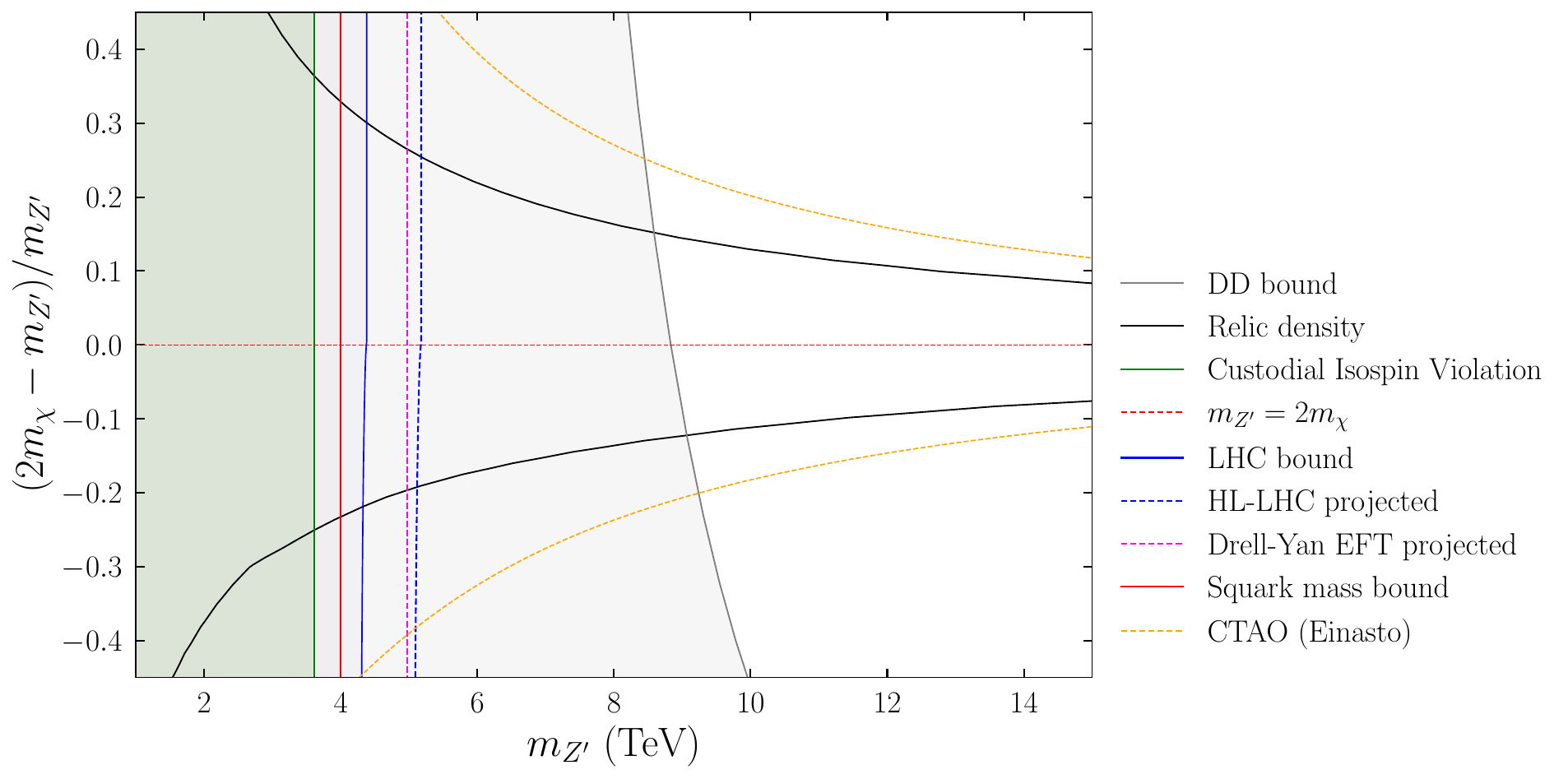}
\caption{The analogue of Fig.~\ref{fig:results_so10_unbroken} for a warped SUSY, 3 brane model, with a $(+-)$ $Z'$ mediating DM annihilation. For this plot, we choose $g_{Z'}=0.8$ and $\sin^2 \theta'=0.28$. DD stands for Direct Detection. The curves showing the CTAO projections are conservative as they only include the DM annihilations to SM final states; including superpartner final states would only improve the reach of CTAO.}
\label{fig:results_susy}
\end{figure}

We present the results for the DM phenomenology for the warped SUSY $SO(10)$ model in Fig.~\ref{fig:results_susy}. 
In what follows, we break down the various contributions to the result.

\textbf{Annihilation and direct detection:} As opposed to the models without SUSY, DM can now annihilate to all the MSSM states, including light fermions and their superpartners. 
This way, the model is equivalent to a generic $Z'$-portal model, such as the one discussed in Sec.~\ref{sec:IRZprime}. 
Even though the DM annihilates to a plethora of final states, a tuning of roughly $10 \%$ is required to get the correct annihilation cross section. 
This is because of a stronger direct detection bound, which now comes from two contributions. 
As seen in the two brane models, a sizeable contribution to direct detection is from the coupling of DM to SM $Z$ (see Eq.~\eqref{eq: susy_g_z_zprime_mixing}). 
However, akin to the weakly coupled $Z'$ in Sec.~\ref{sec:IRZprime}, $Z'$ couples directly to $u$ and $d$ quarks, thus giving another contribution to the direct detection cross section. 
These two contributions finally lead to a direct detection bound of $m_{Z'}\gtrsim 8.5\,\TeV$, or equivalently $m_{\rm KK} \gtrsim 80 \,\TeV$. 
This bound is stronger than the KK scale bounds for any of the two brane models discussed in previous sections.

\textbf{Dipole interactions of DM:} For this model, the bounds from the dipole interactions are considerably weak, and are therefore not shown in Fig.~\ref{fig:results_susy}. 
To understand this, we note that the dipole moments depend on the couplings of $X_s$ and $W^{\pm}_R$ with $\chi$. 
For a three brane setup, both of these couplings are diluted by a factor $(m_{\rm KK}/m_*)^2$, leading to a $(m_{\rm KK}/m_*)^4$ dilution of the dipole moments. 

\textbf{Custodial isospin violation:} Following the argument above for the $X_s$ and $W^{\pm}_R$ couplings, the couplings of SM Higgs to KK modes of $W^{\pm}$, KK modes of $Z$ as well as $W_R^{\pm}$, are also diluted by $(m_{\rm KK}/m_*)^2$. 
Thus, the only non-zero contribution to the $\rho$ parameter comes from $Z-Z'$ mixing, leading to the bound $m_{Z'}\geq 3.6 \,\TeV$, or equivalently, $m_{\rm KK} \geq 34\,\TeV$.

\textbf{Collider bounds on sparticles and $Z'$:} The mass of squarks is bounded by LHC searches to be $m_{\tilde{q}}\geq 1.75$\,TeV \cite{CMS:2017brl}. 
Due to the GMSB mass relations (see App.~\ref{app:mssm_masses}), this translates to a bound
\begin{equation}
    m_{\rm KK}/g_*\geq 12.3 \,\TeV\,,
\end{equation}
where $g_*$ is the strong coupling between the gauge KK modes. 
Taking $g_*\simeq 3$, we obtain $m_{\rm KK} \gtrsim 37 \rm \,\TeV$ or $m_{Z'} \gtrsim 4$\,TeV. 
The coupling of $Z'$ to light fermions being SM strength puts it within the reach of collider searches. 
For the parameter space considered here, we have a narrow width, $\Gamma_{Z'}/m_{Z'} \lesssim 6 \%$. 
Therefore, bounds from the LHC di-lepton searches can be recast for this model in a straightforward manner. 
We obtain LHC di-lepton bounds for this model by recasting the bounds for a sequential $Z'$~\cite{CMS:2021ctt}. As seen in Fig.~\ref{fig:results_susy}, this sets a bound of $m_{Z'}\gtrsim 4.3$\,TeV. 
The High-Luminosity LHC is only expected to reach $m_{Z'} \simeq 5.2\,\TeV$. 
Turning to off-shell searches for $Z'$, we see that EFT analyses of Drell-Yan cross sections are also only able to probe $O(5)$\,TeV masses. 
As highlighted in Ref.~\cite{Panico:2021vav}, this limited reach of the Drell-Yan analysis is due to the relative sign between the $Y$ and $(B-L)$ charges in the relation $Q_{Z'}= Y\cos^2\theta'-\frac{1}{2}(B-L)$.

We thus see that both on-shell and off-shell searches are subsumed by the direct detection bound. 
This again reinforces CTAO's position as a promising option to probe the model for the allowed region of parameter space in Fig.~\ref{fig:results_susy}.
For the CTAO curves in Fig.~\ref{fig:results_susy}, we neglect the DM annihilation to sparticle final states for simplicity. 
Nevertheless, due to sparticles and particles having equal $Z'$-charges, and being significantly lighter than $\chi$ and $Z'$, the annihilation cross section increases only by an $\mathcal{O}$(1) factor upon including sparticle final states. Therefore, compared to Fig.~\ref{fig:results_susy}, we do not expect a qualitative change in the CTAO reach if we do include the additional annihilation channels. 
As including more states increases the annihilation cross section, Fig.~\ref{fig:results_susy} is a conservative estimate of the CTAO reach. 
Due to the multitude of SM states into which the DM annihilates, the resulting photon spectrum is expected to be similar to the spectrum from a generic $Z'$, as shown in Fig.~\ref{fig:CTAO_spectra_all_models}. 
A calculation of the photon spectra in the full model, including all the MSSM states, is beyond the scope of this paper and is left for future work.

\section{Conclusion}
\label{sec:conclusion}

The coming generation of DM searches, with direct detection reaching the neutrino fog and indirect detection reaching TeV-scale thermal relics with CTAO, makes this a decisive period for the WIMP paradigm. 
This motivates the question of which WIMP scenarios lie at the edge of current bounds and remain discoverable. 
In this paper, we approached this question from both a bottom-up and top-down perspective.
From the IR, we considered the $Z'$-type portal as a benchmark, updated its direct-detection, collider and electroweak-precision constraints, and computed its reach at CTAO. 
We then provided a UV completion in a warped extra-dimensional GUT, which fixes both the mass scales and the couplings. 
The DM and its mediators are KK states with masses set by the scale that addresses the Planck-weak hierarchy. 
At the same time, unlike what emerges from the bottom-up, their couplings are largely fixed by the $SO(10)\times U(1)_B$ gauge structure and by where the fields are localised in the extra dimension.

Our study resulted in three main findings.
First, a $Z'$-type portal with electroweak-strength couplings survives only where the tuning between the DM and $Z'$ masses, measured by $\Delta\equiv(2m_\chi-m_{Z'})/m_{Z'}$, is at most 10--20\%. 
This tuning is relaxed if the $Z'$ couples strongly to DM, but a strongly coupled portal calls for a UV completion. 
Notably, the $Z'$ coupling to the Higgs dictates the reach of direct detection. 
When the $Z'$ couples to the Higgs sequentially, $Z$--$Z'$ mixing suppresses the spin-independent rate by the fourth power of the ratio of the momentum transfer to the $Z$ mass, which makes direct detection irrelevant, while electroweak precision tests, in this case the $\rho$ parameter, set the leading constraint together with LHC dilepton searches. 
Second, the warped $SO(10)\times U(1)_B$ GUT, in which the DM is stable for the same reason as the proton, provides a UV completion of a generalised, top-philic, strongly coupled $Z'$ portal. 
Including the baryon-number gauge boson in the annihilation for the first time, we obtain the correct relic abundance for KK modes as light as roughly $\sim 10$\,TeV, with a 40\% tuning.
We consider the various scenarios with $SO(10)$ broken on the IR brane which is allowed by such large KK masses. 
In one such variation, only the baryon-number gauge boson couples strongly to DM, consequently suppressing the direct-detection rate while preserving the relic abundance. 
Third, by augmenting the model with supersymmetry and a third brane on which supersymmetry is broken, we ameliorate the little hierarchy problem associated with a high KK scale. 
Because the SM fermions, the Higgs and the DM are sequestered near the Higgs brane, the electron EDM and other CP-violating observables constrain the scale of the Higgs brane, which sets the masses of the fermion KK modes, rather than the masses of the lighter gauge KK modes, and require it only to exceed $\sim 100$\,TeV.  Even with a Higgs brane scale of $\sim 100$ TeV, it turns out that the DM mass can still be $\sim \TeV$ due to profile suppression. 
In this new setup, DM is sequestered from the gauge KK modes. Hence, its gauge couplings are highly suppressed if the gauge KKs retain their BCs from the two-brane models above. This can lead to an overabundance of DM. Remarkably, a change of the $Z'$ boundary condition on the UV brane comes to the rescue, and turns the model into a UV completion of the weakly coupled $Z'$ portal with which we started, with a $Z'$ heavier than $\sim$$8.5\,$TeV and a 10\% tuning.

The above analysis demonstrates that WIMP DM and its variants remain an interesting avenue from both a model-building and detection perspective. 
For example, our analysis of the $Z'$ coupling to the Higgs calls for a systematic study of how UV completions fix this coupling, since it can change the phenomenology drastically. 
On the 5D side, the validity of the 5D EFT and gauge-coupling unification need to be worked out for the supersymmetric composite-Higgs setup with the geometry considered here~\cite{KAGGRRRS}. 
Improved electroweak precision tests are equally important for mapping the viable parameter space, since the $\rho$ parameter already provides the leading bound in several of the models we studied. 
On the signal side, we have shown that in every model we studied, the thermal-relic parameter space that survives current bounds lies within the projected reach of CTAO, and parts of it within reach of next-generation direct detection.
Indeed, since top-philic annihilation gives photon spectra that peak at lower energies than a generic $Z'$ portal, a CTAO signal could also indicate an extra-dimensional origin of the DM.
All CTAO sensitivities assumed the Einasto profile, however, and should a signal not emerge, understanding the implications of the resulting constraints will require an improved handle on the Galactic-centre $J$-factor.
Ultimately, it remains possible that the push for DM may yet also uncover a hint of extra dimensions.

\begin{acknowledgments}
We are grateful to Manuel Buen-Abad for valuable contributions in the early stages of this work. 
We would also like to thank Ben Lehman, Geraldine Servant, Tracy Slatyer and Raman Sundrum for discussions.
The research of KA, AB and KP is supported by NSF grants PHY-2210361, 
PHY-2514660 and PHY-2609893 and the Maryland Center for Fundamental Physics.
The research of AB is supported partly by the Deutsche Forschungsgemeinschaft under Germany’s Excellence Strategy EXC 2121 ``Quantum Universe" – 390833306. 
The research of NLR was supported by the Office of High Energy Physics of the U.S. Department of Energy under contract DE-AC02-05CH11231.  
\end{acknowledgments}

\appendix

\section{MSSM Masses in Warped SUSY}
\label{app:mssm_masses}

In this section, we derive the masses of the MSSM particles used for the calculations in the main text. 
As described in Sec.~\ref{sec:warped_susy}, there are two SUSY breaking mechanisms in play for the warped SUSY model. 
The first is GMSB. SUSY is broken on the final IR brane. This breaking is transmitted to the gauginos, which span the entire fifth dimension. The resulting masses of the gauginos are given by
\begin{equation}
    m^{\rm MSSM}_{\rm gaugino} = \left(\frac{g}{g_*} \right)^2 \frac{F_X}{M_{\rm mess}}= \left(\frac{g}{g_*} \right)^2 \Lambda_{\rm SUSY}.
\end{equation}
Here, $g_*$ is the coupling between the (IR brane peaked) strongly coupled KK modes, and $g$ is the gauge coupling for the MSSM gaugino under consideration. The $F$-term vev $F_X$ is the source of SUSY breaking in the messenger sector. We expect $\sqrt{F_X}\sim M_{\rm mess} \sim m_{\rm KK}$ giving $\Lambda_{\rm SUSY} \sim m_{\rm KK}$.
Since $SO(10)$ may be broken on the final IR brane, the SUSY breaking on the IR brane may also not respect $SO(10)$. 
To be general, we thus associate two different breaking scales, $\Lambda^{(3)}_{\rm SUSY},\Lambda^{(2)}_{\rm SUSY} \sim m_{\rm KK}$ to the $SU(3)_c$ and $SU(2)_L$ gauginos. The squark and the $SU(2)_L$ doublet slepton masses are roughly given by
\begin{equation}
m_{\tilde{q}} \simeq \sqrt{2C^{(2)}_3}\left(\frac{g^2_s}{4\pi g_*} \right)\Lambda^{(3)}_{\rm SUSY},\qquad m_{\tilde{\ell}}\simeq \sqrt{2C^{(2)}_2}\left(\frac{g^2_W}{4\pi g_*} \right) \Lambda^{(2)}_{\rm SUSY},
\label{eq:msusy_gmsb}
\end{equation}
where $C^{(2)}_3=4/3$ and $C^{(2)}_2=3/4$ are the quadratic Casimirs for $SU(3)$ and $SU(2)$ respectively. For the analysis done in Sec.~\ref{subsec:results_susy}, we assume that the masses of the $SU(2)_L$ singlet sleptons are equal to the doublet slepton masses. 
This is a simplifying but pessimistic assumption, as the singlet sleptons will be lighter. 
Thus, our analysis is more conservative, as annihilation of the DM to MSSM sparticles is larger than what we calculate in Sec.~\ref{subsec:results_susy}.

Another feature of the warped SUSY model is that the $F$-term $F_X$ itself has a profile, which is peaked near the IR brane with a value ${F_X} \simeq m^2_{\rm KK}$. This profile is the second SUSY breaking mechanism in play. The profile has a tail near the Higgs brane, where it has a value $F^{\rm tail}_X \simeq O(1)\,\TeV^2$. This $F$-term breaking generates mass terms for Higgsinos as well as additional mass terms for $\tilde{t}_R,\tilde{b}_R$, which are peaked near the Higgs brane. Since the DM has no sizeable couplings to most MSSM gauginos, $F_X$ is the only source of SUSY breaking mass for its superpartner, $\tilde{\chi}$. 
Therefore, we have the mass relations,
\begin{align}
    m^2_{\tilde{t_R}, \tilde{b_R}} &\simeq 2C^{(2)}_3\left(\frac{g^2_s}{4\pi g_*}\right)^2 (\Lambda^{(3)}_{\rm SUSY})^2 + c_1F^{\rm tail}_X,\nonumber\\
    \mu &\simeq m_{\tilde{H}}\simeq c_2\sqrt{F^{\rm tail}_X }\nonumber,\\
    m^2_{\tilde{\chi}} &\simeq c_3F^{\rm tail}_X+ \mdm^2,
\label{eq:Higgs_mssm_masses}
\end{align}
where $c_1,c_2,c_3$ are $\mathcal{O}(1)$ numbers. Since co-annihilation decreases the total annihilation cross section, we are motivated to suppress it. The suppression can be achieved through the Boltzmann factor $\text{exp} (-\Delta m /T_{\rm f.o.})$, where $\Delta m = m_{\tilde{\chi}}-m_{\chi}$ and $T_{\rm f.o.}$ is the freeze-out temperature. A rough estimate of the freeze-out temperature for TeV scale DM gives $T_{\rm f.o.}\simeq m_\chi/25$, meaning we need about an $8\%$ difference in the masses of DM and its superpartner to suppress co-annihilation.
Taking $c_3=1$, $F^{\rm tail}_X = 2-4\,\TeV^2$ is sufficient for suppressing co-annihilation for all of the parameter space.

Now, we combine the expressions in Eqs.~\eqref{eq:msusy_gmsb} and~\eqref{eq:Higgs_mssm_masses} to calculate the MSSM masses. We can take $c_2 \simeq 0.5-0.7$, coupled with $F^{\rm tail}_X = 2-4\, \TeV^2$ to get the Higgsino masses 
\begin{equation}
    m_{\tilde{H}} \simeq 1 \,\TeV.
\end{equation}
Moving on to squark masses, the relevant LHC bounds for GMSB vary according to the choice of the next lightest supersymmetric particle (NLSP). The most stringent limit is from Ref.~\cite{CMS:2017brl}, reaching $m_{\tilde{q}}\geq1.75$\,TeV. We work with this bound in order to present the most conservative analysis. With $F_X$ as above, and $c_1\simeq0.25-0.5$, we fix
\begin{align}  &m_{\tilde{t}_R}=m_{\tilde{b}_R}\simeq 2.3\,\TeV,\\
&m_{\tilde{q}}\simeq 1.8 \,\TeV \qquad \tilde{q}\neq \tilde{t}_R,\tilde{b}_R.
\end{align}  
Thus, all squark masses except $m_{\tilde{t}_R,\tilde{b}_R}$ are on the edge of the bound. For slepton masses, we fix 
\begin{equation}
m_{\tilde{l}} \simeq  1\,\mathrm{TeV}, 
\end{equation}
which can be achieved by taking $\Lambda^{(2)}_{\rm SUSY} \simeq 2 m_{\rm KK}$. Therefore, we have fixed the MSSM mass spectrum for the model. Finally, there is a bound on the scale $m_{\rm KK}$, coming from the bound on the sparticle masses. For instance, the squark mass bound above translates to
\begin{equation}
m_{\rm KK}/g_* \gtrsim 12.3 \,\TeV,
\label{eq:nmess_mkk_relation}
\end{equation}
which is the constraint used in Sec.~\ref{subsec:results_susy}.

\section{Profiles and Couplings in RS1}
\label{app:5D_specifics}

\begin{figure}[t!]
    \centering
    \includegraphics[width=0.9\linewidth]{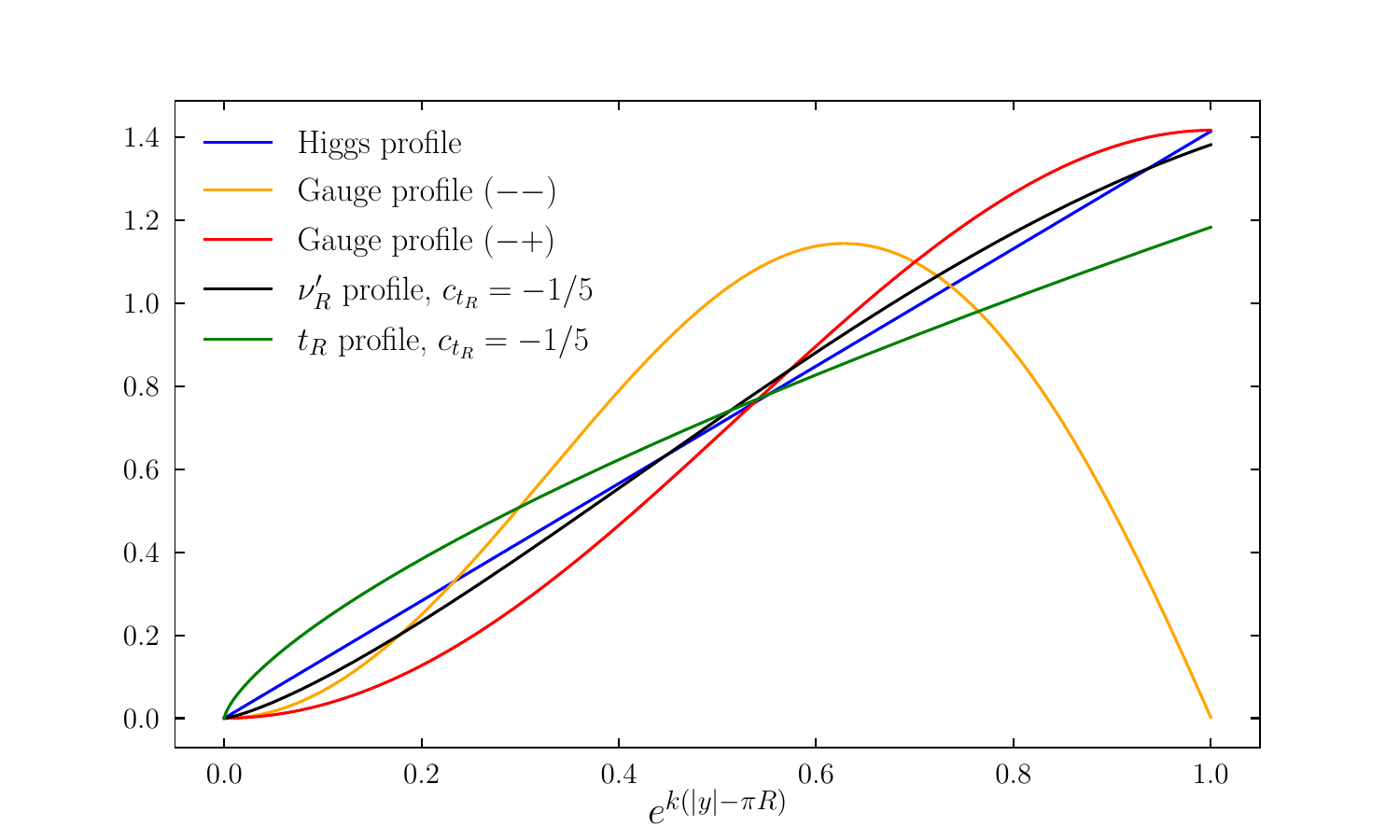}
    \caption{Canonically normalised profiles for 5D fields in RS1.}
    \label{fig:profiles}
\end{figure}

In this section, we review some of the calculations for 5D fields in a warped extra dimension. The extra dimension is an \(S^1/\mathbb Z_2\) orbifold given by
\begin{align}
    ds^2=e^{-2k|y|}dx_\mu dx^\mu-dy^2,
\end{align}
with $y=0$ and $y=\pi R$ being the UV and IR branes respectively. We define the redshift coordinate $z$
\begin{equation}
    z(y)= \frac{e^{k|y|}}{k} , ~~~z_h=z(0)=\frac{1}{k}, ~~~z_v =z(\pi R)= \frac{e^{k\pi R}}{k}~,
\end{equation}
so that the 5D metric now looks like
\begin{equation}
    ds^2= \frac{dx^\mu dx_\mu-dz^2}{k^2 z^2},
\end{equation}
with UV and IR branes at $z=z_h$ and $z=z_v$ respectively. We shall express the 5D profiles of various fields in terms of this coordinate. Below, we provide the profiles of various fields used in the main text. A plot of the profiles is shown in Fig.~\ref{fig:profiles}.\\
\underline{\it{5D Fermions}}: In five dimensions, fermions are Dirac and can be decomposed into left and right chiral fermions by assigning different $Z_2$ parities to the two components. This parity is defined by the operation $y \to -y$. The action for 5D fermions is then given by
\begin{align}
    \mathcal{S}=\int_{}^{}d^4x dy\sqrt{-g} \left[i\bar{\Psi}\Gamma^M\partial_M \Psi-ck~\text{sgn}(y)\bar{\Psi}\Psi\right]\!,
\end{align}
where $c$ is the 5D, $Z_2$ odd mass. We have the KK decomposition, 
\begin{equation}
    \Psi(x,y)= \sum_{n} f_n(y) \psi_n(x),
\end{equation}
where $f_n$ is the 5D profile of the $n^{th}$ KK mode. We can solve for the profiles for various boundary conditions. We focus on the Dirichlet $(-)$/Neumann$(+)$ type boundary conditions. Imposing $(++)$ BCs on the fermion profile gives a zero mass mode
\begin{equation}
    f_0(y)= \sqrt{\frac{1-2c}{z_h (e^{k\pi R(1-2c)}-1)}} \left(\frac{z}{z_h}\right)^{2-c}.
\end{equation}
Since the kinetic term for the fermion carries an additional $e^{-3k|y|}$ from gamma matrices and the determinant of the metric, we have an effective profile $\hat{f_0}(y)=e^{-3/2 ~k|y|} f_0(y)$. This gives the result of the profiles being UV peaked for $c > 1/2$ and IR peaked for $c<1/2$. The KK modes for the field $\Psi$ have two profiles for the two chiralities of the KK fermions. We can assign $(++)$ BCs to the LH chirality, which will mean $(--)$ BCs for the RH chirality. The LH chirality has the profile
\begin{align}
    f_n(y)&= N \sqrt{\frac{2}{z_h}}\left(\frac{z}{z_h}\right)^{5/2} \left[ J_\alpha (m_nz)+ b_n Y_\alpha (m_nz)\right] ,
    \label{eq:kk_fermion_profile}\\
    N&=\frac{1}{\sqrt{z_v^2 [J_{\alpha \mp 1}(m_n z_v)+b_n Y_{\alpha \mp 1}(m_n z_v)]^2 - z_h^2 [J_{\alpha \mp 1}(m_n z_h)+b_n Y_{\alpha \mp 1}(m_n z_h)]^2}},
\end{align}
where the coefficient $b_n$ is fixed by the BCs to be 
\begin{equation}
    -b_n=\frac{J_{\alpha\pm 1}(m_nz_h)}{Y_{\alpha \pm 1} (m_n z_h)}=\frac{J_{\alpha \pm 1}(m_nz_v)}{Y_{\alpha \pm 1} (m_n z_v)}.
\end{equation}
The relation above also gives the masses, $m_n$, of the KK modes. The profile of the LH chirality is  peaked near the IR brane.  The profile of the RH chirality goes to 0 at both branes and peaks at a small distance from the IR brane.

In the main text, we consider $(-+)$ BCs for the GUT fermions. Assigning $(-+)$ BCs to the LH chirality gives it the same profile as Eq.~\eqref{eq:kk_fermion_profile}, but with $b_n$ and $N$ now given by
\begin{align}
    -b_n&=\frac{J_{\alpha}(m_nz_h)}{Y_{\alpha} (m_n z_h)}=\frac{J_{\alpha \pm 1}(m_nz_v)}{Y_{\alpha \pm 1} (m_n z_v)},\\
    N&=\frac{1}{\sqrt{z_v^2 [J_\alpha(m_n z_v)+b_n Y_\alpha(m_n z_v)]^2 - z_h^2 [J_{\alpha \mp 1}(m_n z_h)+b_n Y_{\alpha \mp 1}(m_n z_h)]^2}}.
\end{align}
Using the small and large $x$ asymptotics of $J_\alpha(x)$ and $Y_\alpha(x)$, we can show that the first KK mode has the mass (with $\alpha=|c+1/2|$)
 \begin{equation}
    m \simeq \begin{cases}
    \frac{\pi}{2}(1+c)ke^{-k\pi R},~~~ &c \gtrsim-1/4,\\
2\sqrt{c+\frac{1}{2}} ke^{-k\pi R}, &-1/2 \lesssim c \lesssim -1/4, \\    2\sqrt{|c+\frac{1}{2}|\left(|c+\frac{1}{2}|+1\right)}\left(e^{-k\pi R}\right)^{|c+\frac{1}{2}|}ke^{-k\pi R}, ~~&c \lesssim -1/2~.
    \end{cases}
\label{eq:mp_fermion_mass_app}
\end{equation}

Therefore, we can get a very light KK fermion for $c$ close to $-1/2$. The equation above is key in postulating the existence of the LZP, that is $\nu'_R$, residing in the $t_R^c$ GUT multiplet, which is our DM candidate. For $(-+)$ BCs discussed above, the LH chirality is peaked near the IR brane. For $c<-1/2$, the RH chirality is peaked near the UV brane. However, for $c>-1/2$, the RH chirality is also peaked near the IR brane.

\underline{\it{5D gauge fields:}} The discussion can be generalised in a straightforward manner to 5D gauge fields. We have the gauge field action
\begin{equation}
    \mathcal{S}=\int_{}^{}d^4x dy\sqrt{-g} \left(-\frac{1}{4}F_{MN}F^{MN}\right)~~~~M,N =1,2,\ldots,5,
\end{equation}
where, working in the gauge where $A_5(x,y)=A^{(0)}_5(x)f_5(y)$, we can write the KK decomposition for the 4D gauge field
\begin{equation}    A_\mu(x,y)=\sum_nf_n(y)A^{(n)}_\mu(x).
\end{equation}
The fields $A_\mu$ and $A_5$ have opposite boundary conditions. Imposing $(++)$ BCs on the 4D gauge field gives a massless zero mode $A^{0}_\mu$, which can be identified with a SM gauge field. The zero mode has the flat profile
\begin{equation}
    F_0(y)= \frac{1}{\sqrt{\pi R}}.
\end{equation}
The KK modes for $(++)$ BCs have the profiles
\begin{align}
     F_n(y)&=\sqrt{\frac{2}{z_h}} N \left(\frac{z}{z_h}\right)\left[ J_1 (m_nz)+ b_n Y_1 (m_nz)\right] 
    \label{eq:kk_gauge_profile}\\
    -b_n&=\frac{J_{0}(m_nz_h)}{Y_{0} (m_n z_h)}=\frac{J_{0}(m_nz_v)}{Y_{0} (m_n z_v)} \\
    N&=\frac{1}{\sqrt{z_v^2 [J_1(m_n z_v)+b_n Y_1(m_n z_v)]^2 - z_h^2 [J_1(m_n z_h)+b_n Y_1(m_n z_h)]^2}},
\end{align}
that are all peaked near the IR brane. The mass of the first KK mode is found from above, to be
\begin{equation}
    m_{\rm KK} \simeq 2.4 z_v^{-1} = 2.4 k e^{-k \pi R}.
\end{equation}
As discussed in the main text, we can break the gauge symmetry on the UV or IR brane by imposing a $(-)$ BC. The profiles for gauge fields with $(-+)$ and $(--)$ BCs are identical to Eq.~\eqref{eq:kk_gauge_profile}, but with different coefficients,
\begin{align}
  -b^{(-+)}_n&=\frac{J_{1}(m_nz_h)}{Y_{1} (m_n z_h)}=\frac{J_{0}(m_nz_v)}{Y_{0} (m_n z_v)},\quad -b^{(--)}_n=\frac{J_{1}(m_nz_h)}{Y_{1} (m_n z_h)}=\frac{J_{1}(m_nz_v)}{Y_{1} (m_n z_v)}, \\
    N^{(-+)}&=\frac{1}{\sqrt{z_v^2 [J_1(m_n z_v)+b_n Y_1(m_n z_v)]^2 - z_h^2 [J_0(m_n z_h)+b_n Y_0(m_n z_h)]^2}},\\
     N^{(--)}&=\frac{1}{\sqrt{z_v^2 [J_0(m_n z_v)+b_n Y_0(m_n z_v)]^2 - z_h^2 [J_0(m_n z_h)+b_n Y_0(m_n z_h)]^2}}.
\end{align}
This gives the mass of the lightest modes 
\begin{align}
    m_{(-+)}&\simeq 2.4z_v^{-1}= m_{\rm KK},\\
    m_{(--)}&\simeq 3.8 z_v^{-1}\simeq1.6m_{\rm KK}.
\end{align}
Using the profiles derived above, we can calculate the couplings between different fields using the overlap of their profiles. For instance, given the 5D gauge coupling $g_5$, the coupling of a zero mode of a gauge field to any mode of a 5D fermion is given by:
\begin{equation}
    \mathcal{L} \supset g_5\bar{\Psi}\Gamma^\mu \Psi A_\mu \supset \left[\int_{}dy \sqrt{-g} e^{k|y|}~g_5 (f^{(n)})^2F^{(0)}\right]\bar{\Psi}^{(n)} A^{(0)}_\mu \gamma^\mu \Psi^{(n)}= \frac{g_5}{\sqrt{\pi R}}\bar{\Psi}^{(n)} A^{(0)}_\mu \gamma^\mu \Psi^{(n)}, 
\end{equation}
where the extra $e^{k|y|}$ is the f\"unfbein factor coming from the gamma matrix $\Gamma^\mu$. Therefore, we get $g_{\rm SM }=g_5/\sqrt{\pi R}$. The couplings of gauge KK modes are non-universal. The coupling of the $n^\mathrm{th}$ gauge KK mode to the $i^\mathrm{th}$ and $j^\mathrm{th}$ fermion mode is given by
\begin{equation}
    g_{4}= g_5 \int_{} dy \sqrt{-g} e^{k|y|} F^{(n)}(y) f^{(i)}(y)f^{(j)} (y).
\end{equation}
For couplings of $(-+)$ and $(++)$ gauge KK modes to fermionic KK modes, this expression can be approximated as $g_4 =g_* \simeq g_5 \sqrt{k}= g_{\rm SM}\sqrt{k\pi R}$. This is also true for IR peaked zero modes of fermions such as the $t_R^c$. Thus, as seen in the main text, all the IR peaked fields interact with a typical strong coupling $g_*$.

\underline{\it{The Higgs field:}} While faithful estimates can be constructed with an IR brane-localised SM Higgs, we discuss the more realistic case, where the Higgs has a profile. For most of the discussion in the main text, the Higgs was treated as a 5D scalar. However, in order to choose its profile, we turn to the standard picture of Higgs being the fifth component of a gauge field, $A^{(0)}_5$. Thus, while the main text discussed an $SO(10)$ GUT gauge group, in principle the model has an enlarged $SO(11)$ gauge symmetry, with the breaking $SO(11) \to SO(10)$ on both the branes yielding the Higgs field as an $A_5$ \cite{Frigerio:2011zg}.

Upon assigning $(++)$ boundary conditions to $A_5$ and $(--)$ BCs to $A_\mu$ for any 5D gauge field, we obtain a 4D scalar which we identify as the Higgs, with a profile
\begin{equation}
    f_H(z)= \sqrt{\frac{2}{z_h}} ~\frac{(z/z_h)^2}{\sqrt{e^{2 k\pi R}-1}}.
\end{equation}
The gauge coupling of the Higgs with the $n^{\rm th}$ gauge mode, and its Yukawa coupling with $i^{\rm th},j^{\rm th}$ fermion modes are then given by
\begin{align}
    g^{n}_{H}= g_5 \int_{} dy \sqrt{-g} e^{k|y|} F^{(n)}(y) (f_{H}(y))^2,\\
    y^{i,j}_{H}= g_5 \int_{} dy \sqrt{-g}  f^{(i)}(y) f^{(j)}(y) f_{H}(y).\
\end{align}

\section{Bounds from Electroweak Observables}\label{app:ew_obs}

\subsection{Custodial isospin violation}

\begin{figure}[t]
    \centering
    \begin{tikzpicture}
    \begin{scope}[xshift=2cm]
    \draw[dashed] (-0.5,0) -- ++(110:1.8);
    \draw[dashed] (-0.5,0) -- ++(70:1.8);

    \draw[dashed] (1.7,0) -- ++(110:1.8);

    \draw[dashed] (1.7,0) -- ++(70:1.8);
    
    \draw[thick,photon] (-2,0) to [out=0,in=-180] (3.4,0); 
    
    \node[] at (-1.7,-0.6) {$Z$};
    \node[] at (0.55,-0.6) {${\rm KK}~ Z/Z'$};
    \node[] at (3.2,-0.6) {$Z$};
    \end{scope}
    
    \begin{scope}[xshift=10cm]
    \draw[dashed] (-0.5,0) -- ++(110:1.8);
    \draw[dashed] (-0.5,0) -- ++(70:1.8);

    \draw[dashed] (1.7,0) -- ++(110:1.8);

    \draw[dashed] (1.7,0) -- ++(70:1.8);
    
    \draw[thick,photon] (-2,0) to [out=0,in=-180] (3.4,0); 
    
    \node[] at (-1.7,-0.6) {$W^{\pm}$};
    \node[] at (0.55,-0.6) {${\rm KK}~ W^{\pm}/W^{\pm}_R$};
    \node[] at (3.2,-0.6) {$W^{\pm}$};
    \end{scope}
    \end{tikzpicture}
    \caption{Possible mass mixings between gauge KK and SM $W$ and $Z$ that contribute to custodial isospin violation.}
    \label{fig:rho}
\end{figure}
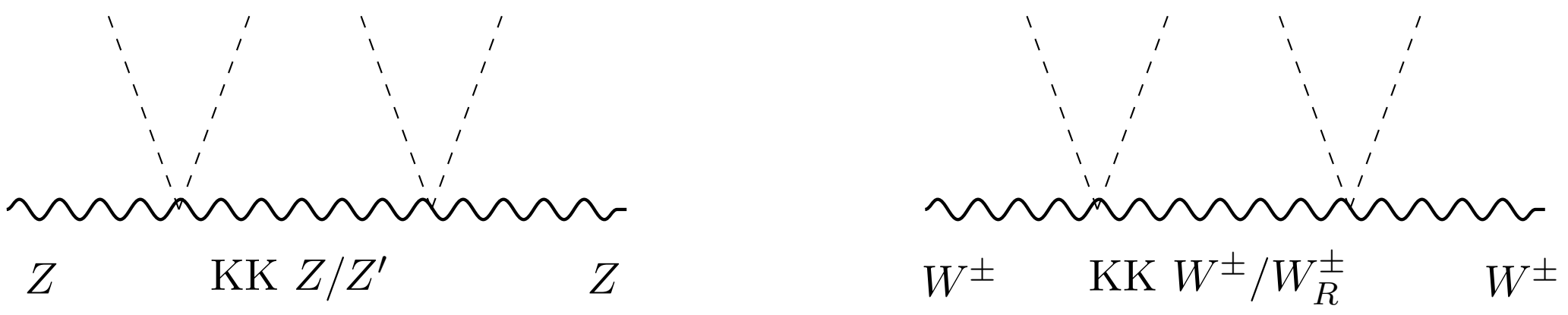

As discussed in the main text, breaking $SO(10)$ on the IR brane involves breaking the $SU(2)_R$ subgroup in order to reproduce the SM quark masses. In this section, we estimate this violation of custodial isospin through the $\rho$ parameter, and present the resulting bounds.
The $\rho$ parameter is defined as 
\begin{align}
    \rho = \frac{ m_W^2}{m_Z^2\cos^2\theta_W}.
\end{align}
We then define $\delta \rho= \rho -1$. In the standard model, at tree level, the mass difference of the W and Z bosons is only due to the mixing between hypercharge and the $SU(2)_L$ gauge boson $W_{3L}$, giving $\delta\rho =0$. However, loops of top quarks can give non-zero, albeit small contributions to $\delta \rho$. The experimentally measured value of the $\rho$ parameter is
\cite{ParticleDataGroup:2024cfk}
\begin{equation}
    \rho =  1.00031~\pm~0.00019.
\label{eq:rho_number}
\end{equation}
In the case of warped extra dimensions, heavy gauge KK modes can mix with the SM $W$ and $Z$, as shown in Fig.~\ref{fig:rho}, shifting their masses. The contribution to $\delta \rho$ from the KK modes of $W$ and $Z$ mixing with the zero modes is given by
\begin{align}
    \delta \rho_1&= \frac{g^2_Z (g^{H}_{\text{KK},Z})^2 v^4}{4^2 m^2_{\rm \text{KK},Z} m_Z^2}-\frac{g^2_W (g^{H}_{\text{KK},W})^2 v^4}{4^2 m^2_{\rm KK,W} m_W^2} \nonumber \\
    &= \left(\frac{g^{H}_{KK,Z}}{g_Z}\right)^2 \frac{m_Z^2}{m_{\text{} KK}^2} -\left(\frac{g^{H}_{\text{KK},W}}{g_W}\right)^2 \frac{m_W^2}{m_{\rm KK}^2} \nonumber\\
    &=\left(\frac{g^{H}_{KK,W}}{g_W}\right)^2 \frac{m_Z^2 \left(\frac{\cos^2 \w}{\cos^2 \gutw}\right)-m_W^2}{m_{\rm KK}^2} = \left(\frac{g_{10}}{g_W}\right)^2{k\pi R} \cdot \frac{m_W^2 \tan^2\gutw}{m_{\rm KK}^2}.
\label{eq:rho_1}
\end{align}
In the expressions above, $v=246\,\rm GeV$. We take the coupling of the KK Ws and Z to the Higgs, to be the GUT predicted value, i.e. $g^{H}_{KK,W}=g^{H}_{KK,Z} \cos \gutw$, where $\sin ^2\gutw=3/8$. This is because the KK modes are peaked near the IR brane, unlike zero modes, which are flat and whose couplings run, leading to deviations from the GUT predictions. We also have a contribution to $\delta \rho$ from the $SU(2)_R$ gauge bosons $W^{\pm}_R$ and $Z'$. Using $Q^{H}_{Z'}=1/2 \cos^2 \swp$, and noting that we are working in a 2 Higgs doublet model (see Sec.~\ref{subsec:quark_masses}) with $v_u/v_d=\tan \beta $ and $ v_u^2+v_d^2 =v^2/2$, this is given by
\begin{align}
    \delta \rho_2&= \frac{g^2_Z (g^{H}_{Z'})^2 \cos^4 \theta' v^4}{4^2 m^2_{Z'} m_Z^2}-\frac{g^2_W (g^{H}_{W_R})^2 v^4}{4^2 m^2_{W_R} m_W^2} \sin^2 2\beta \nonumber\\
    &=  \left(\frac{g^{H}_{Z'}}{g_Z}\right)^2\frac{m_Z^2}{\mzp^2}\cos^4 \theta' -\left(\frac{g^{H}_{W_R}}{g_W}\right)^2 \frac{m_W^2}{m_{W_R}^2} \sin^2 2\beta \nonumber\\
    &=\left(\frac{g_{10}}{g_W}\right)^2 k\pi R~m_W^2 \left(\frac{\cos^2\theta'}{\mzp^2}-\frac{\sin^2 2\beta}{m^2_{W_R}}\right)\!.
\label{eq:rho_2}
\end{align}
For the case where $SO(10)$ is unbroken on the IR brane, we have $\tan\beta =1$, and the expression above simplifies to
\begin{equation}
    \delta \rho_2= - \left(\frac{g_{10}}{g_W}\right)^2{k\pi R} \cdot \frac{m_W^2 \tan^2\gutw}{m_{\rm KK}^2} = -\delta \rho_1,
\end{equation}
giving $\delta \rho=0$, as expected. For the cases where $SO(10)$ is broken on the IR brane to $\mathcal{G}_1$ or $\mathcal{G}_2$, $ W_R^{\pm}$ has $(--)$ BCs, whereas $Z'$ retains $(-+)$ BCs, we have
\begin{align}
    \delta \rho = \left(\frac{g_{10}}{g_W}\right)^2{k\pi R} \cdot \frac{m_W^2}{m_{\rm KK}^2} \left(1- \frac{\sin^22\beta~m^2_{\rm KK}}{m^2_{\rm KK, --}} b_H^2\right)\!,
\end{align}
where $m_{(--)}$ is the mass of a gauge KK mode with $(--)$ BCs and $b_H$ is an $\mathcal{O}(1)$ factor coming from the profile overlap of the Higgs with a $(--)$ gauge KK mode (see Table~\ref{tab:dm_couplings}). Finally, for the case where $SO(10)$ is broken to $\mathcal{G}_3$, $Z'$ has $(--)$ BCs which leads to
\begin{align}
    \delta \rho = \left(\frac{g_{10}}{g_W}\right)^2{k\pi R} \cdot \frac{m_W^2}{m_{\rm KK}^2} \left(\sin^2 \swp- \frac{(\sin^22\beta-\cos^2\swp) ~m^2_{\rm KK}}{m^2_{(--)}} b_H^2\right)\!.
\end{align}
Imposing the $2\sigma$ bound, defined by $|\delta\rho/\rho|\leq 2\sigma=3.8\times 10^{-4}$, for our choice of $g_{10} =g_W $ and $k\pi R \simeq 32.23$ results in
\begin{align}
    m_{\rm KK}& \gtrsim 22.7\,\TeV ,~~~~~~\text{$SO(10) \to \mathcal{G}_1, \mathcal{G}_2$ on IR brane,}\\
    m_{\rm KK} &\gtrsim 18.8\,\TeV ,~~~~~~\text{$SO(10) \to \mathcal{G}_3$ on IR brane}.
\end{align}
For the warped SUSY case, all the KK gauge bosons are peaked near the final IR brane, except $Z'$, which has an almost flat profile. Due to this, the only significant contribution to $\delta \rho$ is due to $Z'$ mixing with SM $Z$ and is given by
\begin{equation}
    \delta\rho_{~\rm Warped-SUSY}=\frac{\gzp^2}{g_Z^2}\frac{m_Z^2}{\mzp^2}\cos^4\theta'.
\end{equation}
With the value of $g_Z'=0.8$ and the resulting  $\cos^2\theta' \simeq 0.72$ from the main text, we finally get the $2\sigma$ bound
\begin{equation}
    m_{Z'} \gtrsim 3.6 \,\TeV~~\text{for warped SUSY.}
\end{equation}

\subsection{Bounds on the electric and magnetic dipole moments of dark matter}

While the DM particle $\nu'_R$ is neutral, it is coupled to various charged fermions and gauge bosons. Thus, we expect it to have electromagnetic dipole interactions. We denote the electric and magnetic dipole moment of DM by $d_{\chi}$ and $\mu_{\rm \chi}$ respectively. These dipole moments can lead to photon mediated interactions between DM and nucleons, thus contributing to the direct detection cross section. The rate of direct detection is parametrically given by
\begin{equation}
    R \propto n_{\rm \chi} \sigma_{\rm DD} \propto d ^2/m_{\rm \chi} \rm ~or ~~\mu^2/m_{\chi}.
\label{eq:edm_main}
\end{equation}
Requiring this contribution to satisfy the direct detection bound gives the upper limits (in the limit $m_{\rm \chi}\gg {m_{\rm nucleon}}$)
\begin{align}
    |d_{\rm \chi}| \leq & ~5.62\times 10^{-24}~  \sqrt{\frac{m_{\chi}}{100\,{\rm GeV}}}\,e\,\rm cm,
    \label{eq:edm_bound}\\
    |\mu_{\rm \chi}| \leq & ~8.57 \times 10^{-21} \sqrt{\frac{m_{\chi}}{100\,{\rm GeV}}}\,e\,\rm cm.
    \label{eq:mdm_bound}
\end{align}  
These two inequalities are obtained by updating the bounds from Ref.~\cite{Pospelov:2000bq} with the current direct detection data. The bound on the EDM comes from a spin-independent interaction of the electric dipole of DM with the charge of the nucleons. On the other hand, the bound on the magnetic dipole moment comes from a spin-dependent interaction between the magnetic dipoles of the DM and the nucleon.

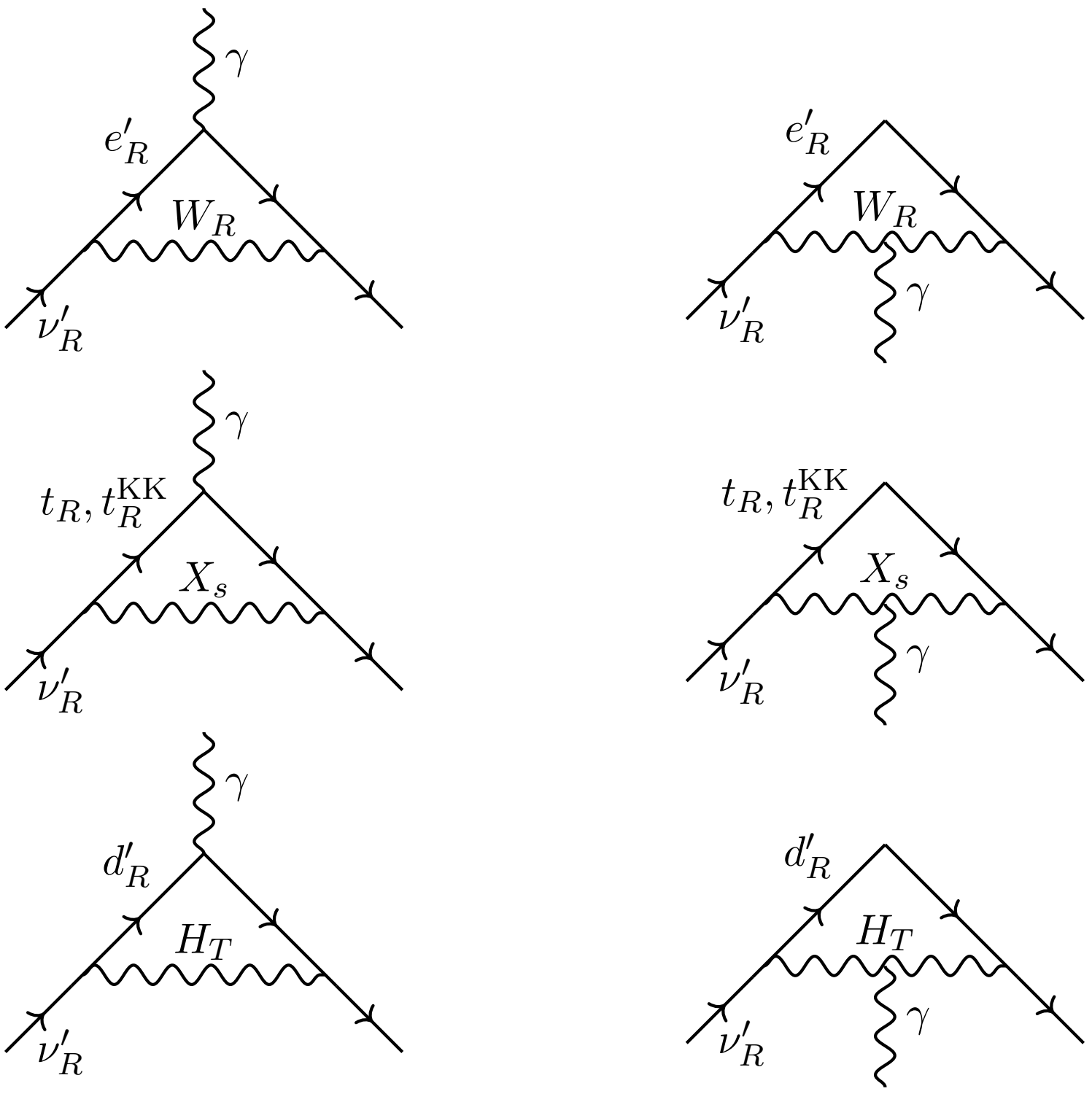
\begin{figure}[t!]
\centering
\begin{subfigure}{0.4\textwidth}
\centering
\begin{tikzpicture}
\begin{scope}
\draw[thick,electron] (0,0) to  [out=-45,in=135] (1.2,-1.2) ;
\draw[thick,electron] (1.2,-1.2) to  [out=-45,in=135] (1.8,-1.8);
\draw[thick,antielectron] (0,0) to [out=-135,in=45] (-1.2,-1.2);
\draw[thick,antielectron] (-1.2,-1.2) to  [out=-135,in=45] (-1.8,-1.8);
\draw[thick,photon] (0,0) to [out=90,in=-90] (0,1.1); 
\draw[thick,photon] (-1.1,-1.1) to [out=0,in=-180] (1.1,-1.1) ;
\node[] at (0,-0.8) {$W_R$};
\node[] at (0.3,0.6) {$\gamma$};
\node[] at (-0.7,-0.1){$e'_R$};
\node[] at (-1.3,-1.8){$\nu'_R$};
\end{scope}
\end{tikzpicture}
\end{subfigure}
\begin{subfigure}{0.4\textwidth}
\centering
\begin{tikzpicture}
\begin{scope}
\draw[thick,electron] (0,0) to  [out=-45,in=135] (1.2,-1.2) ;
\draw[thick,electron] (1.2,-1.2) to  [out=-45,in=135] (1.8,-1.8);
\draw[thick,antielectron] (0,0) to [out=-135,in=45] (-1.2,-1.2);
\draw[thick,antielectron] (-1.2,-1.2) to  [out=-135,in=45] (-1.8,-1.8);
\draw[thick,photon] (0,-1.1) to [out=-90,in=90] (0,-2.2); 
\draw[thick,photon] (-1.1,-1.1) to [out=0,in=-180] (1.1,-1.1) ;
\node[] at (0,-0.8) {$W_R$};
\node[] at (0.3,-1.6) {$\gamma$};
\node[] at (-0.7,-0.1){$e'_R$};
\node[] at (-1.3,-1.8){$\nu'_R$};
\end{scope}
\end{tikzpicture}
\end{subfigure}
\begin{subfigure}{0.4\textwidth}
\centering
\begin{tikzpicture}
\begin{scope}
\draw[thick,electron] (0,0) to  [out=-45,in=135] (1.2,-1.2) ;
\draw[thick,electron] (1.2,-1.2) to  [out=-45,in=135] (1.8,-1.8);
\draw[thick,antielectron] (0,0) to [out=-135,in=45] (-1.2,-1.2);
\draw[thick,antielectron] (-1.2,-1.2) to  [out=-135,in=45] (-1.8,-1.8);
\draw[thick,photon] (0,0) to [out=90,in=-90] (0,1.1); 
\draw[thick,photon] (-1.1,-1.1) to [out=0,in=-180] (1.1,-1.1) ;
\node[] at (0,-0.8) {$X_s$};
\node[] at (0.3,0.6) {$\gamma$};
\node[] at (-0.9,-0.1){$t_R,t^{\rm KK}_R$};
\node[] at (-1.3,-1.8){$\nu'_R$};
\end{scope}
\end{tikzpicture}
\end{subfigure}
\begin{subfigure}{0.4\textwidth}
\centering
\begin{tikzpicture}
\begin{scope}
\draw[thick,electron] (0,0) to  [out=-45,in=135] (1.2,-1.2) ;
\draw[thick,electron] (1.2,-1.2) to  [out=-45,in=135] (1.8,-1.8);
\draw[thick,antielectron] (0,0) to [out=-135,in=45] (-1.2,-1.2);
\draw[thick,antielectron] (-1.2,-1.2) to  [out=-135,in=45] (-1.8,-1.8);
\draw[thick,photon] (0,-1.1) to [out=-90,in=90] (0,-2.2); 
\draw[thick,photon] (-1.1,-1.1) to [out=0,in=-180] (1.1,-1.1) ;
\node[] at (0,-0.8) {$X_s$};
\node[] at (0.3,-1.6) {$\gamma$};
\node[] at (-0.9,-0.1){$t_R,t^{\rm KK}_R$};
\node[] at (-1.3,-1.8){$\nu'_R$};
\end{scope}
\end{tikzpicture}
\end{subfigure}
\begin{subfigure}{0.4\textwidth}
\centering
\begin{tikzpicture}
\begin{scope}
\draw[thick,electron] (0,0) to  [out=-45,in=135] (1.2,-1.2) ;
\draw[thick,electron] (1.2,-1.2) to  [out=-45,in=135] (1.8,-1.8);
\draw[thick,antielectron] (0,0) to [out=-135,in=45] (-1.2,-1.2);
\draw[thick,antielectron] (-1.2,-1.2) to  [out=-135,in=45] (-1.8,-1.8);
\draw[thick,photon] (0,0) to [out=90,in=-90] (0,1.1); 
\draw[thick,photon] (-1.1,-1.1) to [out=0,in=-180] (1.1,-1.1) ;
\node[] at (0,-0.8) {$H_T$};
\node[] at (0.3,0.6) {$\gamma$};
\node[] at (-0.7,-0.1){$d'_R$};
\node[] at (-1.3,-1.8){$\nu'_R$};
\end{scope}
\end{tikzpicture}
\end{subfigure}
\begin{subfigure}{0.4\textwidth}
\centering
\begin{tikzpicture}
\begin{scope}
\draw[thick,electron] (0,0) to  [out=-45,in=135] (1.2,-1.2) ;
\draw[thick,electron] (1.2,-1.2) to  [out=-45,in=135] (1.8,-1.8);
\draw[thick,antielectron] (0,0) to [out=-135,in=45] (-1.2,-1.2);
\draw[thick,antielectron] (-1.2,-1.2) to  [out=-135,in=45] (-1.8,-1.8);
\draw[thick,photon] (0,-1.1) to [out=-90,in=90] (0,-2.2); 
\draw[thick,photon] (-1.1,-1.1) to [out=0,in=-180] (1.1,-1.1) ;
\node[] at (0,-0.8) {$H_T$};
\node[] at (0.3,-1.6) {$\gamma$};
\node[] at (-0.7,-0.1){$d'_R$};
\node[] at (-1.3,-1.8){$\nu'_R$};
\end{scope}
\end{tikzpicture}
\end{subfigure}
\caption{Leading contributions to the magnetic dipole moment of DM.}
\label{fig:mdm_feynmann}
\end{figure}

Starting with the magnetic dipole moment, there are six diagrams shown in Fig.~\ref{fig:mdm_feynmann} that contribute to the magnetic dipole moment. We first consider the interaction
\begin{align}
   \mathcal{L_{\rm gauge}}\supset -\frac{g_{X_s}}{\sqrt{2}}X_s^\mu\bar{t_R}\gamma_\mu P_L\chi,
\end{align}
where $g_{X_s}$ can be calculated by profile overlaps. The magnetic dipole moment due to this operator is given by
\begin{align}
\mu_{t,X_s}=\frac{m_{\chi}}{16\pi^2} \left(\frac{g_{X_s}}{\sqrt{2}}\right)^2 N_c Q_{t} I_{\rm gauge} (M_{\rm gauge},m_f), 
\end{align}
where $N_c=3$ is the number of colours, and $Q_t=2/3$ is the electric charge of $t_R$ (and not $t_R^c$). Moving on to the $W_R,e'$ loop, we have  operators of the form
\begin{align}
    \mathcal{L_{\rm gauge}}\supset -\frac{g_{1}}{\sqrt{2}}W_R^\mu\bar{e'}\gamma_\mu P_L\chi~-\frac{g_{2}}{\sqrt{2}}W_R^\mu\bar{e'}\gamma_\mu P_R\chi,
\end{align}
with the hierarchy $g_1\gg g_2$ due to profile overlaps. From this, we obtain a dipole moment
\begin{align}
\mu_{e',W_R}\approx\frac{m_{\chi}}{16\pi^2} \left(\frac{g_{1}}{\sqrt{2}}\right)^2 Q_{e'} I_{\rm gauge} (M_{\rm gauge},m_{e'})+\frac{m_{e'}}{16\pi^2} \left(\frac{g_{1}}{\sqrt{2}}\right)\left(\frac{g_{2}}{\sqrt{2}}\right)  Q_{e'} J_{\rm gauge} (M_{\rm gauge},m_{e'}), 
\end{align}
where $m_{e'}\simeq m_{\chi}$. Here, the functions $I_{\rm gauge},J_{\rm gauge}$ are given by the integrals
\begin{align}
I_{\rm gauge}(M_{\rm gauge},m_f)=\frac{1}{2}\int_{0}^{1}dz\frac{z(2+z)}{\left[m_f^2(1-z)+M_{\rm gauge}^2 z -m_\chi^2 z(1-z)\right]},\\
J_{\rm gauge}(M_{\rm gauge},m_f)=\int_{0}^{1}dz\frac{3z(2z-1)}{\left[m_f^2(1-z)+M_{\rm gauge}^2 z -m_\chi^2 z(1-z)\right]}.
\label{eq:def_1_mudm}
\end{align}
Finally, the dipole moment due to a Yukawa interaction
\begin{align}
    \mathcal{L_{\rm Yukawa}}\supset-y_T H_u^T \bar{d}'P_R \chi,
\end{align}
(where $y_T$ is again calculated from profile overlap) takes the form
\begin{align}
\mu_{d',H^T}=-\frac{m_{\chi}}{16\pi^2} y_T^2 N_c Q_{d^c} I_{\rm yuk} (M_{\Phi},m_{d'}), 
\end{align}
with $Q_{d^c}=1/3$ and $I_{\rm yuk}$ given by the integral
\begin{align}
I_{\rm yuk}(M_{\Phi },m_f)=\frac{1}{2}\int_{0}^{1}dz\frac{z(1-z)(m_f/m_\chi+(2z-1))}{\left[m_f^2(1-z)+M_{\Phi}^2 z -m_\chi^2 z(1-z)\right]}.
\label{eq:def_2_mudm}
\end{align}
The total magnetic moment is given by the sum: $\mu_{\chi}=\mu_{t,X_s}+\mu_{e',W_R}+\mu_{d',H^T}$.

\begin{figure}
\centering
\begin{subfigure}{0.4\textwidth}
\begin{tikzpicture}
\begin{scope}[xshift=5cm]
\draw[thick,electron] (0,0) to  [out=-45,in=135] (1.2,-1.2) ;
\draw[thick,electron] (1.2,-1.2) to  [out=-45,in=135] (2.8,-2.8);
\draw[thick,antielectron] (0,0) to [out=-135,in=45] (-1.2,-1.2);
\draw[thick,antielectron] (-1.2,-1.2) to  [out=-135,in=45] (-2.8,-2.8);
\draw[thick,photon] (0,-1.1) to [out=-90,in=90] (0,-2.2); 
\draw[thick,photon] (-1.1,-1.1) to [out=0,in=-180] (1.1,-1.1) ;
\draw[dashed]
(0,0) to [out=45,in=-135]
(1,1);
\node[] at (0,0) {$\otimes$};
\node[] at (1,1) {$H$};
\node[] at (0,-0.8) {$X_s$};
\node[] at (0.3,-1.6) {$\gamma$};
\node[] at (-0.7,-0.1){$t_R$};
\node[] at (1,-0.1){$u^{KK, i}_L$};
\node[] at (1.7,-1.2){$ \nu^{i'}_L$};
\node[] at (1.8,-1.8) {$\otimes$};
\draw[dashed]
(1.8,-1.8,0) to [out=45,in=-135]
(3,-0.6);
\node[] at (3.1,-0.4) {$H$};
\node[] at (2.5,-2){$ \nu'_R$};
\node[] at (-1.3,-1.8){$\nu'_R$};
\end{scope}
\end{tikzpicture}
\end{subfigure}
\caption{The leading contribution to the electric dipole moment of DM.}
\label{fig:edm_feynmann}
\end{figure}
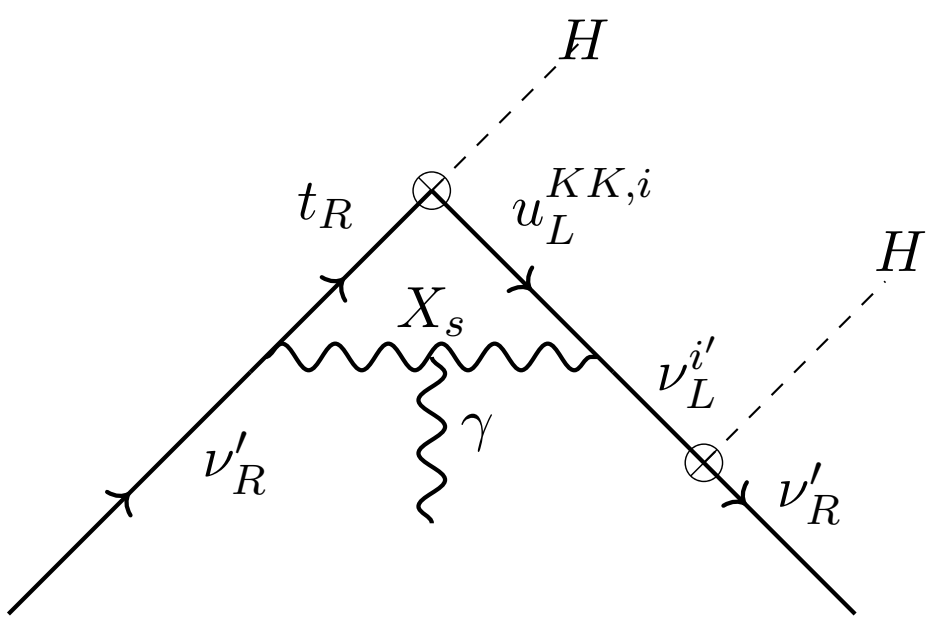

Moving on to electric dipole moments, we note that a non-zero electric dipole moment requires CP violating phases. These can be generated from mixing between KK fermions, as shown in Fig.~\ref{fig:edm_feynmann}. The mixing between different KK fermions is of the same order as $m_{A,B}$ in Eq.~\eqref{eq:mmix_nul_nur}. This gives rise to a DM electric dipole moment
\begin{align}
|d_\chi|=\delta \cdot \frac{m_{t} \sin \theta_1}{16\pi^2} \left(\frac{g_{X_s}}{\sqrt{2}}\right)^2 N_c Q_{t} I_{\rm EDM} (M_{X_s},m_t), 
\label{eq:edm_expression}
\end{align}
where $\delta$ is an $O(1)$ imaginary phase, coming from (possibly) complex Yukawa couplings of $\chi$, and $\sin\theta_1$ is defined in Eq.~\eqref{eq:basis_change}. The integral function $I_{\rm EDM}$ is defined as
\begin{align}
I_{\rm EDM}(M_{\Phi },m_f)=\int_{0}^{1}dz\frac{4z-7z^2}{\left[m_f^2(1-z)+M_{\Phi}^2 z -m_\chi^2 z(1-z)\right]}.
\label{eq:def_1_edm}
\end{align}
Even though the bound on $d_\chi$ is much stronger than that on $\mu_{\chi}$ (see Eq.~\eqref{eq:mdm_bound}), the resulting bounds on $m_\chi,m_{\rm KK}$ are comparable. This is due to the parametric suppression from $m_t$ in Eq.~\eqref{eq:edm_expression}.
It turns out that the bounds from the electromagnetic moments in this section are always weaker than the bounds from either (or both) direct detection or custodial isospin violation.

\section{Holographic Interpretation}\label{app:holography}
In this section, we provide the 4D picture dual to the 5D model given in the main text. The content of this section follows Refs.~\cite{Contino:2010rs,Gherghetta:2006ha,Panico:2015jxa,Agashe:2004bm,Contino:2004vy}.
The 4D dual consists of an elementary sector and a strong sector. The elementary sector contains the SM fermions and SM gauge bosons. The strong sector is a strongly interacting theory near a fixed point, which makes it a quasi-CFT. This quasi-CFT has a UV cutoff $\Lambda_{\rm UV}\simeq k\simeq M_{\rm pl}$, and it confines at a lower scale $m_{\rm KK}\simeq \mathcal{O}(10\,{\rm TeV})$. A fermion $f$ with $(++)$ BCs in the 5D theory, has the following dual interpretation. In the 4D theory, $f$ belongs to the elementary sector and couples to a composite operator $O_f$ of the strong sector
\begin{align}
    \mathcal{L}\supset \frac{1}{\Lambda^{\Delta_O-5/2}_{\rm UV}}\lambda_f~  f\mathcal{O}_f.
\label{eq:elem_CFT_mixing}
\end{align}
Upon confinement, $\mathcal{O}_f$ interpolates a fermion $\mathcal{O}_f \ket{0}\sim \psi_f \ket{0}$, leading to a mass-mixing
in the IR
\begin{align}
    \mathcal{L^{\rm mix}_{\rm IR}}\supset -m_{\rm KK}\left(\frac{m_{\rm KK}}{\Lambda_{\rm UV}}\right)^{\Delta_O-5/2}\bar{\psi} f-m_{\rm KK}\bar{\psi}\psi.
\label{eq:elem_comp_mixing}
\end{align}
The 5D mass parameter $c$ of $f$, and the scaling dimension $\Delta_O$ of the 4D theory are related by $\Delta_O=|c+1/2|+3/2$. Fermions with $\Delta_O >5/2$ (or $c>1/2$) have an irrelevant mixing with the CFT operators in the IR, and correspond to fermions peaked near the UV brane in the 5D dual theory. On the other hand, a fermion like $t^c_R$ has $\Delta_O <5/2$ (or $c<1/2$), which means it has a large, relevant mixing with composite operators in the IR. Correspondingly, its profile is peaked near the IR brane in the 5D theory.

A fermion with $(-+)$ BCs corresponds to a composite of the strong sector. Unlike $\mathcal{O}_f$ in Eq.~\eqref{eq:elem_CFT_mixing}, this composite is not sourced by an elementary field. In the main text, the fermionic GUT multiplets of $SO(10)$ have 15 fermions with $(-+)$ BCs, and one SM fermion with $(++)$ BCs. In the 4D theory, this corresponds to the following Lagrangian
\begin{align}
    \mathcal{L}\supset \frac{1}{\Lambda^{\Delta_O-5/2}_{\rm UV}}\lambda~  \bar{\mathcal{\bf O}}_{F} F,
\end{align}
where ${\mathcal{\bf O}_{F}}$ is a set of CFT operators forming a complete $\bf{16}$ of SO(10). These operators inside $\mathcal{\bf O}_F$ yield composite fermions at the scale $m_{\rm KK}$. The elementary source $F$ forms an incomplete multiplet, with only the SM field inside it. For instance, if we consider the $\bf{16_{t_R}}$ multiplet, $F$ is given by $F=(t^c_R,0,0,\ldots)$, while $\mathcal{\bf O}_F$ is a complete $\bf{16}$ of SO(10). In the IR, $\mathcal{\bf O}_F$ interpolates 16 composite fermions, including the DM $\chi$.

Having discussed the holography for fermions, we move to the correspondence of the gauge sectors of the two dual theories. The gauge symmetries of the UV brane of the 5D theory, correspond to the 4D gauge symmetries of the elementary sector. The bulk gauge group in 5D corresponds to the global symmetries of the strong sector. Finally, the symmetries of the IR brane correspond to the global symmetries that remain unbroken after confinement. Therefore, we have the following holographic interpretations of various BCs
\begin{itemize}
    \item A 5D gauge boson $A_\mu$ with $(++)$ BCs corresponds to a 4D elementary gauge field, with an unbroken gauge symmetry.
    \item A 5D gauge boson $A_\mu$ with $(+-)$ BCs corresponds to a 4D elementary gauge field. The gauge field receives a mass from the spontaneous breaking of the gauge symmetry by the CFT confinement.
    \item A 5D gauge boson $A_\mu$ with $(-+)$ BCs corresponds to a 4D spin-1 composite. The $(+)$ BC on the IR brane means that the strong sector has a conserved current $J_\mu$ that can interpolate a spin-1 composite upon confinement $J_\mu \ket{0}\sim m_{\rm KK}^2\rho_\mu \ket{0}$. This composite $\rho_\mu$ is dual to the 1st $(-+)$ KK gauge field of the 5D theory. The $(-)$ BC on the UV brane implies that the elementary sector does not respect this global symmetry.
    \item A 5D gauge boson $A_\mu$ with $(--)$ BCs again corresponds to a 4D spin-1 composite. The resulting $A_5$ with $(++)$ BCs corresponds to a pseudo-Nambu-Goldstone boson (pNGB). The SM Higgs, which can arise as an $A_5$ field in 5D, can therefore emerge as a pNGB of a strong sector in 4D.
\end{itemize}
Consider the 5D variant with the symmetry-breaking pattern
\[
SO(10)\to \mathcal{G}_{\rm SM}
=SU(3)_c\times U(1)_Y\times SU(2)_L,
\]
on the UV brane, and
\[
SO(10)\to \mathcal{G}_1
=SU(4)_c\times U(1)_R\times SU(2)_L,
\]
on the IR brane. In the 4D dual theory, the elementary sector respects only the \(\mathcal{G}_{\rm SM}\) gauge symmetry, while the CFT possesses a global \(SO(10)\) symmetry, of which the \(\mathcal{G}_{\rm SM}\) subgroup is gauged. Upon confinement, the \(SO(10)\) global symmetry is spontaneously broken to its \(\mathcal{G}_1\) subgroup. In the 5D description, the only gauge bosons with \((++)\) boundary conditions are the SM gauge bosons. Correspondingly, in the 4D dual description, these are identified with the elementary gauge fields. In contrast, KK gauge bosons such as \(X_s\) and \(Z'\), which have \((-+)\) boundary conditions, as well as \(W_R^\pm\), which have \((--)\) boundary conditions, correspond to spin-1 composite resonances of the 4D theory.

\bibliographystyle{JHEP}
\bibliography{dm_ch}

\end{document}